\documentclass[]{aa}  

\usepackage{siunitx}
\usepackage{lscape}
\usepackage{ulem}
\usepackage{xcolor}
\usepackage{subfig}

\definecolor{dark-red}{rgb}{0.9,0.0,0.0}
\definecolor{dark-blue}{rgb}{0.15,0.15,0.5}
\definecolor{dark-green}{rgb}{0.15,0.4,0.15}
\definecolor{medium-blue}{rgb}{0,0,0.9}

\usepackage{changepage} 
\usepackage{graphicx}
\usepackage{txfonts}
\begin{document} 

   \title{RVSPY - Radial Velocity Survey for Planets around Young stars}

   \subtitle{Target characterisation and high-cadence survey}

   \author{Olga V. Zakhozhay\inst{1,2}
          \and
          Ralf Launhardt\inst{1}
          \and
          Andre M\"uller\inst{1}
          \and
          Stefan S. Brems\inst{3}
           \and
           Paul Eigenthaler\inst{4,5,1}
            \and
           Mario Gennaro\inst{6}
           \and
           Angela Hempel\inst{7,1}
           \and
           Maren Hempel\inst{7,1}
           \and
           Thomas Henning\inst{1}           
           \and
           Grant M. Kennedy\inst{8,9}           
           \and
           Sam Kim\inst{5,1}           
           \and
           Martin K\"urster\inst{1}
           \and
           Régis Lachaume\inst{4,5,1}
           \and
           Yashodhan Manerikar\inst{10,1}
           \and
           Jayshil A. Patel\inst{11}           
           \and
           Alexey Pavlov\inst{1}           
           \and
           Sabine Reffert\inst{3}
          \and
          Trifon Trifonov\inst{1,12}           
          }
 \institute{Max-Planck-Institut f\"{u}r Astronomie,\ K\"{o}nigstuhl  17, 69117 Heidelberg, Germany \\ 
              \email{zakhozhay@mpia.de}
         \and
          Main Astronomical Observatory, National Academy of Sciences of the Ukraine, 03143 Kyiv, Ukraine\\ \vspace{-3mm}
          \and      
          Landessternwarte, Zentrum f\"{u}r Astronomie der Universit\"{a}t Heidelberg,\ K\"{o}nigstuhl 12, 69117 Heidelberg, Germany\\ \vspace{-3mm}
          \and
          Instituto de Astrof\'{i}sica, Pontificia Universidad Cat\'{o}lica de Chile, Av. Vicu\~{n}a Mackenna 4860, 7820436 Macul, Santiago, Chile\\ \vspace{-3mm}
          \and      
          Astro-engineering center (AIUC), Instituto de Astrof\'{i}sica, Pontificia Universidad Cat\'{o}lica de Chile, Santiago, Chile\\ \vspace{-3mm}
          \and
          Space Telescope Science Institute, 3700 San Martin Drive, Baltimore, MD 21218, USA\\ \vspace{-3mm}          
          \and
          Universidad Andrés Bello, Departemento de Ciencias Fisicas, Facultad de Ciencias Exactas, Campus Casona de Las Condes, Astronomia, Fern\'{a}ndez Concha 700, 7591538 Santiago de Chile, Chile\\ \vspace{-3mm}          
          \and
          Department of Physics, University of Warwick, Coventry CV4 7AL, UK\\ \vspace{-3mm}
          \and  
          Centre for Exoplanets and Habitability, University of Warwick, Gibbet Hill Road, Coventry CV4 7AL, UK\\ \vspace{-3mm}
          \and
          Indian Institute of Technology, Madras, India\\ \vspace{-3mm}
          \and
          Department of Astronomy, Stockholm University, AlbaNova University Center, SE-10691 Stockholm, Sweden\\ \vspace{-3mm}
          \and
          Department of Astronomy, Sofia University 'St Kliment Ohridski',5 James Bourchier Blvd, BG-1164 Sofia, Bulgaria\\ \vspace{-3mm}
}
   \date{Received June 08, 2022; accepted August 05, 2022\vspace{-3mm}}
% \abstract{}{}{}{}{} 
% 5 {} token are mandatory

\abstract
% context heading (optional)
% {} leave it empty if necessary  
{The occurrence rate and period distribution of (giant) planets around young stars is still not as well constrained as for older main-sequence stars. This is mostly due to the intrinsic activity-related complications and the avoidance of young stars in many large planet search programmes. Yet, dynamical restructuring processes in planetary systems may last significantly longer than the actual planet formation phase and may well extend long into the debris disc phase, such that the planet populations around young stars may differ from those observed around main-sequence stars.}
% aims heading (mandatory)
{We introduce our Radial Velocity Survey for Planets around Young stars (RVSPY), which is closely related to the \mbox{NaCo-ISPY} direct imaging survey, characterise our target stars, and search for substellar companions at orbital separations smaller than a few au from the host star.}
% methods heading (mandatory)
{We used the FEROS spectrograph, mounted to the MPG/ESO 2.2\,m telescope in Chile, to obtain high signal-to-noise spectra and time series of precise radial velocities (RVs) of 111 stars, most of which are surrounded by debris discs. Our target stars have spectral types between early F and late K, a median age of 400\,Myr, and a median distance of 45\,pc. During the initial reconnaissance phase of our survey, we determined stellar parameters and used high-cadence observations to characterise the intrinsic stellar activity, searched for hot companions with orbital periods of up to 10 days, and derived the detection thresholds for longer-period companions. In our analysis we, have included archival spectroscopic data, spectral energy distribution, and data for photometric time series from the TESS mission.
}
% results heading (mandatory)
{For all target stars we determined their basic stellar parameters and present the results of the high-cadence RV survey and activity characterisation. We have achieved a median single-measurement RV precision of 6\,m/s and derived the short-term intrinsic RV scatter of our targets (median 23\,m/s), which is mostly caused by stellar activity and decays with an age from >100\,m/s at <20\,Myr to <20\,m/s at >500\,Myr. 
We analysed time series periodograms of the high-cadence RV data and the shape of the individual cross-correlation functions. We discovered six previously unknown close companions with orbital periods between 10 and 100 days, three of which are low-mass stars, and three are in the brown dwarf mass regime. We detected no hot companion with an orbital period <10 days down to a median mass limit of $\sim$1\,M$_{\rm Jup}$\ for stars younger than 500\,Myr, which is still compatible with the established occurrence rate of such companions around main-sequence stars.
We found significant RV periodicities between 1.3 and 4.5 days for 14 stars, which are, however, all caused by rotational modulation due to starspots. We also analysed the data for TESS photometric time series and found significant periodicities for most of the stars. For 11 stars, the photometric periods are also clearly detected in the RV data. 
We also derived stellar rotation periods ranging from 1 to 10 days for 91 stars, mostly from the TESS data. 
From the intrinsic activity-related short-term RV jitter, we derived the expected mass-detection thresholds for longer-period companions, and selected 84 targets for the longer-term RV monitoring.\looseness=-4}
{}
   \maketitle

%
%%%%%%%%%%%%%%%%%%%%%%%%%%%%%%%%%%%%%%%%%%%%%%%%%%%%%%%%%%%%%%%
%\clearpage

\section{Introduction}
\label{sec:intro}

%%%%%%%%%%%%%%%%%%%%%%
\begin{figure}[htb]
\includegraphics[width=8.0cm,angle=0,clip=true]{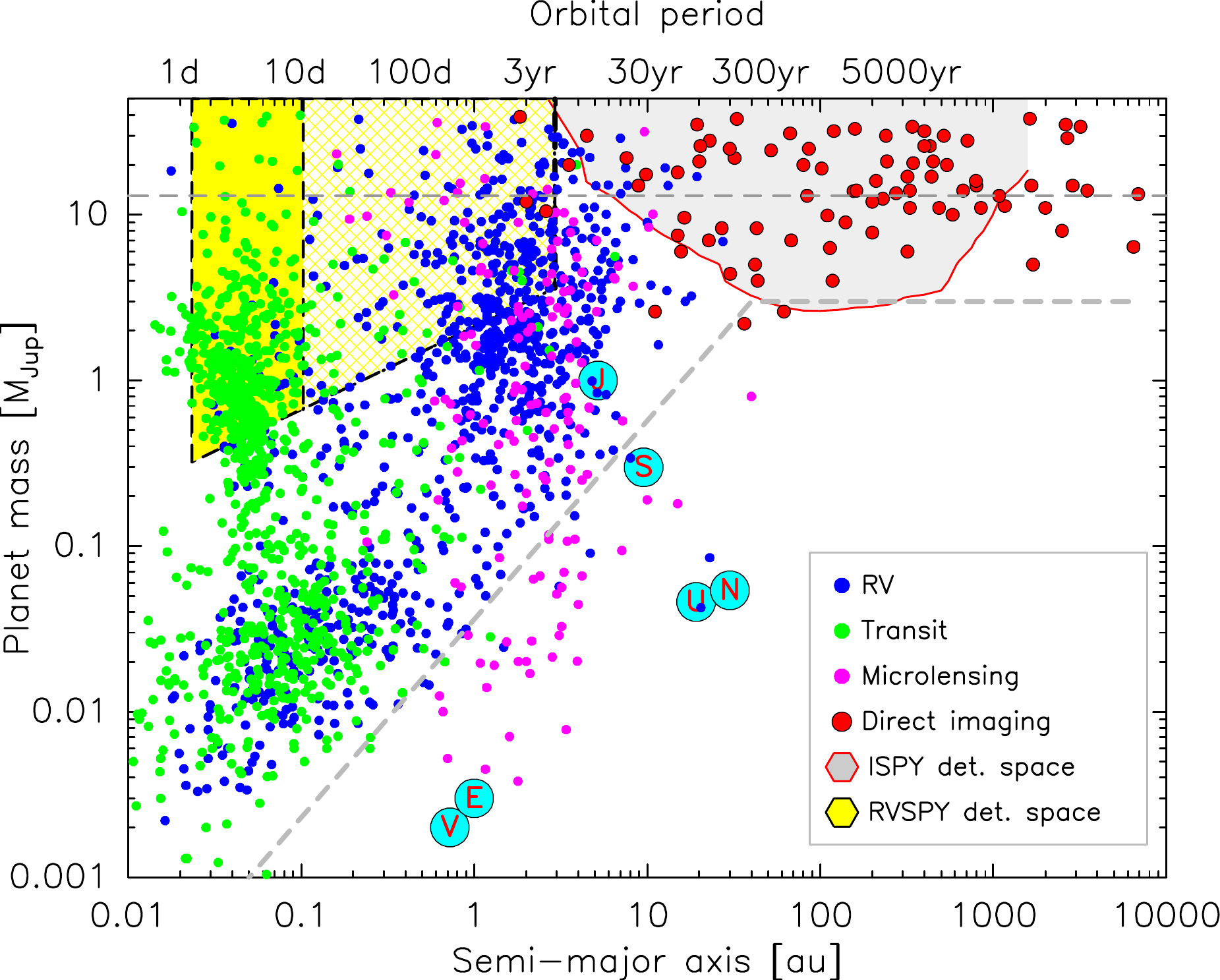}
\centering
\caption{\label{fig:mass-sma}
Distribution of planet mass vs.\ orbital separation of confirmed exoplanets as listed on exopolanet.eu \mbox{\citep[January 2022,][]{schneider2011}}. Labelled on top are the corresponding orbital periods for $M_{\ast}=1\,{\rm M}_{\odot}$\ and $M_{\rm p}<<M_{\ast}$. The main detection methods are marked by different colours. Solar System planets are represented by cyan circles and red letters. The horizontal dashed line marks the approximate deuterium burning mass limit. The solid yellow-shaded area marks the parameter space probed by our high-cadence RVSPY survey, assuming a conservative mean 3\,$\sigma$-sensitivity of 60\,m/s (Sect.\,\ref{ssec:res:rv}) and a mean stellar mass of 1\,M$_{\odot}$. The yellow-hatched area marks the extended detection space probed by our RVSPY survey assuming a long-term monitoring duration of 5 years. 
The grey-shaded area marks the parameter space probed by the NACO-ISPY survey (10\% detection probability, \citealt{launhardt2020}). 
The light-grey dashed line marks the approximate detection threshold of current-day exoplanet searches.
}
\end{figure}
%%%%%%%%%%%%%%%%%%%%%%

Within only a few million years, protoplanetary discs dissipate their gas due to accretion onto the star and newly formed planets, but also via disc winds and photoevaporation \mbox{\citep[e.g.][]{Williams2011, Pascucci2022}}. By this time, most of the primordial dust has coagulated to pebbles and planetesimals or has been accreted onto forming planets. Collisional cascades within the disc then lead to the formation of debris dust. The size, mass, and brightness of the debris discs depend on the properties of the preceding protoplanetary discs \mbox{\citep[e.g.][]{najita2021}}. Relatively bright \mbox{($L_{\rm d}/L_{\ast}\sim10^{-4}-10^{-2}$)} and large debris discs with radii of up to a few 100\,au are observed around \mbox{$\sim$20\,--\,25\%} of low-mass (FGK) class\,III T\,Tauri stars with ages between 5 and $\sim$100\,Myr \mbox{\citep{hughes2018}}. During this time, that is to say after the initial planet formation, the planetary systems still evolve due to dynamical interactions and migration, but the central stars already experience much less chromospheric activity than at younger ages during the protoplanetary phase, such that planet searches become possible, albeit still challenging (e.g. \mbox{\citealt{Gregory17}}). However, many stars retain their debris discs much longer, albeit with slowly fading fractional luminosities reaching \mbox{$L_{\rm d}/L_{\ast}\sim10^{-6}-10^{-5}$} at few Gyr \mbox{\citep[][]{najita2021}}.

Several specific properties of debris discs such as the large inner gaps, the co-existence of cold outer and hot inner dust belts, or outer dust belts that are much larger than predicted by collisional cascade models, for example, may be explained by the action of newly formed planets (e.g. \mbox{\citealt{Moor2015}}). Thus, the observable properties of debris discs could be indicative of embedded and still evolving planetary systems. 
Yet, the relation between debris disc properties and the existence and properties of planets is still poorly understood. There are indications, albeit still debated, that the frequency of giant planets in young debris discs might be significantly higher than around main-sequence (MS) stars \mbox{\citep[e.g.][]{Meshkat2017,Yelverton2020}}.

To systematically investigate the relation between debris disc properties and the occurrence of giant planets, the large direct Imaging (DI) Survey for Planets around Young stars (NaCo-ISPY) was initiated in 2015 \mbox{\citep{launhardt2020}}. Since DI is only sensitive to companions at large separations \mbox{($>$5-10\,au)}, we have launched a complementary systematic radial velocity (RV) survey for planets in closer orbits around these debris disc stars in 2017, using a Fibre-fed Extended Range Optical Spectrograph, mounted at the 2.2\,m MPG/ESO telescope at ESO's La Silla Observatory in Chile, \mbox{\citealt{Kaufer99}} (Sect.\,\ref{sec:obs}). Figure\,\ref{fig:mass-sma} illustrates the detection spaces of these two complementary surveys in terms of planet masses and semi-major axes. 

This paper introduces our systematic RV Survey for Planets around Young stars (RVSPY) with debris discs, mainly targeting the giant planets and brown dwarfs in orbits shorter than seen by DI. The paper is organised as follows: 
In Sect.\,\ref{sec:motivation} we outline the motivation, goals, and general survey strategy. 
The targets and their properties are presented in Sect.\,\ref{sec:targets}. 
The observations and complementary data are described in Sect.\,\ref{sec:obs} and the data reduction and analysis are described in Sect.\,\ref{sec:red}. Results from the initial high-cadence observations are presented in Sect.\,\ref{sec:res} and discussed in Sect.\,\ref{sec:discussion}. Section\,\ref{sec:summary} summarises the paper.

%%%%%%%%%%%%%%%%%%%%%%%%%%%%%%%%%%%%%%%%%%%%%%%%%%%%%%%%%%%%%

\section{Motivation, goals, and survey strategy}
\label{sec:motivation}

The main goal of the RVSPY survey is to help constraining the occurrence rate of Jovian-mass planets around low--mass stars with bright (detectable) debris discs. We aim to investigate the relation between the occurrence of such planets and the properties of the debris discs, which will provide important constraints for informing planet formation and evolution models that are used in planet population synthesis modelling. Since many main-sequence stars retain detectable debris discs for up to several Gyr, our survey is not restricted to young stars and thus contains age as an additional parameter to be tested. 

Different planet detection methods have different detection biases and each method alone can explore only a certain part of the parameter space we are interested in. While DI is only sensitive to planets with large enough projected angular separations to be spatially resolved from the bright host star, RV has the opposite detection bias, that is, it is most sensitive to planets orbiting their host stars at small separations and with short orbital periods (Fig.\,\ref{fig:mass-sma}). Thus, combining these two methods by designing this \mbox{FEROS-RVSPY} survey such that it has the largest possible target overlap with the NaCo-ISPY DI survey \mbox{\citep{launhardt2020}} will allow us to characterise the (giant) planet population around young debris disc stars over most orbital scales. 

However, the two methods do not harmonise easily in all aspects. DI works best for young stars, when the planets are still hot from the formation process, and thus bright. Young stars in general rotate fast and the interplay between stellar rotation, convection, and magnetic fields, in particular in cool stars, leads to enhanced photospheric (e.g. star spots) and chromospheric (e.g. emission lines) activity. Both can alter the effective shape of spectral lines such that rotational modulation of line shapes due to activity features in the stellar atmosphere leads to an effective jitter (random or quasi-periodic) of the measured mean RV, which can mask or even mimic the RV modulation caused by an orbiting (planetary) companion \mbox{\citep[e.g.][]{Hatzes2002}}. 

At very young ages, when a forming star efficiently accretes matter from the disc (i.e. for at most a few Myr), disc- and wind-locking mechanisms should keep its rotation speed low \citep[e.g.][]{camenzind1990,matt2005}. When the accretion rate drops below some critical value, the star decouples from the disc and starts spinning up, owing to its ongoing contraction. However, stars less massive than approximately 1.3\,M$_{\odot}$\ and cooler than \mbox{$T_{\rm eff}\approx6500$\,K} (corresponding to a main-sequence spectral type later than F5) soon develop and maintain a convective envelope, in which dynamo processes generate a permanent magnetic field that causes the stars to spin down again via magnetic braking on a timescale of about 100\,Myr (at 1\,M$_{\odot}$) to 1\,Gyr (at <0.3\,M$_{\odot}$; \citealt{lamm2005,irwin2009,weise2010,bouvier2014}). Consequently, such stars, when sufficiently slowed down, usually have many (because they are cool) and narrow (because they rotate slowly) absorption lines in their spectra and are thus well-suited for precision RV measurements. In stars more massive than about 1.3\,M$_{\odot}$, the dominating energy-efficient CNO cycle leads to the formation of a convective core and a radiation-dominated envelope in which no efficient dynamo can generate magnetic fields which could slow down the stellar rotation. These stars have not only fewer (because they are hotter), but also much broader spectral lines and are therefore less suited for precision RV measurements.

Hence, RV planet searches around young stars require one to carefully consider stellar ages, rotation velocity indicators (e.g. $v\sin(i)$), and stellar spectral types ($T_{\rm eff}$) when defining the target sample and tailoring the observing strategy and cadence of the RV measurements. An efficient synergy between DI and RV planet searches is most promising for stars older than about 10\,Myr, 

and for spectral types of F5 or later. Attention should be given to pre-main-sequence (PMS) stars younger than about 50\,Myr, which are still contracting and heating up and may still rotate relatively fast even if their mass is below 1.3\,M$_{\odot}$. 
With these considerations we define our target selection criteria (Sect.\,\ref{ssec:targets:sel}), draw a suitable sub-sample from the ISPY target list, and extend it with more debris disc targets that were not observed within ISPY. 

As outlined above, the target stars in question still exhibit stronger stellar activity than their more evolved MS counterparts. Rotational modulation of spectral line strengths and shapes due to activity-related stellar surface features causes signals that interfere with precise RV measurements. Therefore, it is mandatory to characterise the intrinsic stellar activity on the rotational time scales before attempting to search for RV variations that could be caused by orbiting companions. Rotational periods of young low- and intermediate-mass stars are well-explored and are typically constrained to periods between 1 and 10 days with a bimodal distribution peaking at $\sim8$\,days (slow rotators) and $\sim2$\,days (fast rotators; e.g. \mbox{\citealt{herbst2002}}). The orbital period distribution of 'hot Jupiters', which are thought to have formed at larger separations in the discs and migrated inwards (e.g. \mbox{\citealt{Lin1986,Dawson18}}), peaks in a similar range at 3--5\,days. 

Hence, with high-cadence RV measurement series covering about two weeks, with one to three spectra per night, we can both characterise the most significant activity-related variability of our targets as well as constrain the presence of hot companions (HC, which include hot Jupiters, with orbital periods $\le$10\,days). 
Although HC seem to be rare around 'normal' MS stars \citep[0.4-1.2\%; e.g.][]{santerne2016,Dawson18,zhou2019}, it is not clear whether this also applies to young stars and to stars with massive debris discs.
Since HCs induce much larger RV signals than longer-period planets, they are also easily detectable in Doppler data. Therefore, the initial phase of our FEROS-RVSPY survey is tailored to sample variability time scales of \mbox{$\sim1-10$\,days} to characterise stellar activity, while at the same time screening for the presence of HCs. The results and detection thresholds derived from this initial high-cadence survey are then used to identify those targets for which searching for longer-period companions (several months up to $\sim$5\,yrs) seems feasible. In the second phase of the RV survey, which is not the subject of this paper, we continue to monitor the stars from this down-selected sample with an adapted lower cadence for up to a few years.

%%%%%%%%%%%%%%%%%%%%%%%%%%%%%%%%%%%%%%%%%%%%%%%%%%%%%%%%%%
%\clearpage

\section{Targets} 
\label{sec:targets}

\subsection{Target selection criteria} 
\label{ssec:targets:sel}

In accordance with our science goals and the specific complications outlined in Sect.\,\ref{sec:motivation}, we restrict our target list to stars older than $\sim$10\,Myr, focus on young stars with ages of up to a few hundred million years, but do not apply a strict upper age limit, and restrict to stars that show a significant debris disc signature.
This lower age limit is chosen to avoid the most active phase of young stars and the overlap with the preceeding protoplanetary disc phase. \mbox{\citet{brems19}} show that the mean activity-induced RV jitter of young solar-type stars decreases from \mbox{$\gtrsim$500\,m/s} at 5\,Myr to $\lesssim$200\,m/s at 10\,Myr, thus making age a crucial parameter for the target selection. 

Spectral types are restricted to the range F6\,--\,M2, because earlier types have too broad and too few spectral lines (see Sect.\,\ref{sec:motivation}), and later spectral types have too many spectral lines that blend. Furthermore, the latter are known to have less massive debris discs and fewer giant planets. 
We also set a brightness limit at \mbox{V$\le$ 11\,mag} to limit individual exposure times. The declination range is naturally set to \mbox{$-75\degr \le {\rm DEC}\le +25\degr$} by the location of the La Silla observatory. For some particularly interesting targets, the limits on declination and spectral type were not strictly obeyed. In particular we included a few stars with earlier spectral types (up to A7) to verify their $v\sin(i)$\ and the achievable RV precision from test spectra before a decision was made to schedule them for high-cadence RV monitoring.

We also do not include in our target list stars for which a sufficiently large number of useable RV data was already available in archives. For this purpose, we have queried the ESO archive (FEROS, HARPS{\footnote{HARPS -- High Accuracy Radial velocity Planet Searcher, mounted at 3.6\,m telescope at La Silla observatory in Chile \mbox{\citep{Mayor2003}}.}}) and the Keck archive (HIRES{\footnote{HIRES -- High Resolution Echelle Spectrometer, mounted on the 10\,m Keck\,I telescope, USA, Hawaii \mbox{\citep{Vogt94}}.}}). 
Stars with less than ten archival spectra are still kept in our survey list. Stars with $\ge$120 spectra are excluded, but we analyse their archival data in the same way as our own FEROS spectra and include the results in our final survey analysis. For stars with an intermediate number of archival spectra, we verify case by case whether sequences of spectra with sufficiently high cadence (see Sect.\,\ref{ssec:obs:strat}) are available and include or reject them from our survey list.  

Also in accordance with our main science goals, we select our core target list for this RV survey from the target list of the NACO-ISPY survey \mbox{\citep{launhardt2020}}, with the RV-specific selection criteria described above. This resulted in a list of 54 targets. However, for statistical reasons, in order to put robust constraints on the frequency of giant planets in debris discs and their relation to debris disc properties, one would ideally have a total sample size of $\ge$\,100 stars.
We therefore enlarged our core target list by 57 additional targets selected from the Spitzer IRS catalogue of \mbox{\citet{chen2014}}, and from \cite{CottenSong2016} based on the criteria described above, although we verified the significance of the debris disc excess only later (see below, Sect.\,\ref{ssec:targets:list}). 
Some of these stars were actually already included in the NACO-ISPY master target list, but could not be imaged in the end. 

%%%%%%%%%%%%%%%%%%%%%%%%%%%%%%%%%
\subsection{Target properties} 
\label{ssec:targets:list}

%%%%%%%%%%%%%%%%%% Fig. 2
\begin{figure}[htb]
\includegraphics[width=9cm]{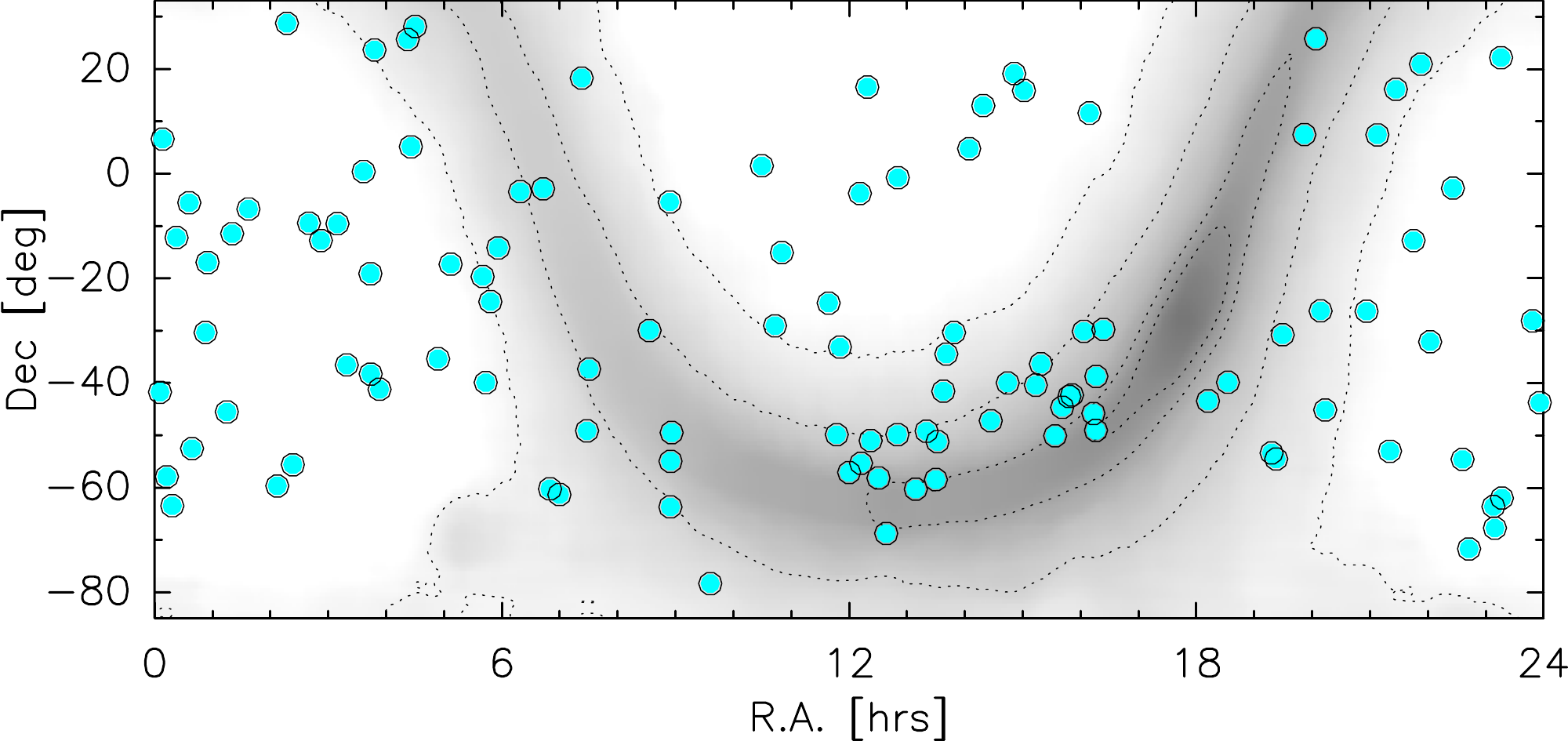}
\centering
\caption{\label{fig:sky_distribution}
Sky distribution of all 111 targets of this survey. The grey-shaded area and dotted contours outline the Milky Way disc and bulge as traced by the COBE-DIRBE band\,2 ($K$) zodi-subtracted all-sky map \mbox{\citep{hauser1989}}.
}
\end{figure}
%%%%%%%%%%%%%%%%%%

Our final target list for the survey observations consists of 111 stars, of which 54 stars were also imaged with NACO--ISPY \mbox{\citep{launhardt2020}}. Figure\,\ref{fig:sky_distribution} shows how the 111 targets are evenly distributed over the (southern) sky. Distances to all our stars are inferred from {\it Gaia}\,DR2 parallaxes \mbox{\citep{gaia_mission,gaia_dr2}} with the method described by \mbox{\citet{bailer2018}}. Spectral types and $V$\ and $K$\ magnitudes are compiled from SIMBAD, the Hipparcos and Tycho Catalogues (\mbox{\citealt{leeuwen2010}}; \mbox{\citealt{Tycho2}}), and from 2MASS \mbox{\citep{2mass}}. For the 54 stars overlapping with ISPY \mbox{\citep{launhardt2020}}, we adopt the stellar parameters from \mbox{\citep{pearce2022}}. For the other 57 non-ISPY stars, we derive stellar parameters in the same way as described in this paper.

The stellar effective temperature $T_{\rm eff}$, bolometric luminosity $L_{\ast}$\ of the stars, as well as the fractional luminosity and blackbody temperature $T_{\rm BB}$\ of the associated debris discs\footnote{In this initial stage of our survey, we do not use the actual disc parameters yet, but only evaluate the significance of the disc excess to select the targets.} are derived by fitting simultaneous stellar \mbox{\citep[PHOENIX;][]{husser2013}} and blackbody models to the observed photometry and spectra as described in \mbox{\citet{launhardt2020}} and \mbox{\citet{pearce2022}}. 
In this process, it turned out that 18 of the 57 non-ISPY stars selected from \mbox{\citet{chen2014}} and \cite{CottenSong2016} show only marginal or no significant debris disc emission\footnote{For most of these 18 stars, the issue appears to be related to the normalisation of the Spitzer IRS spectra as discussed by \citet[][their Sect. 4.3]{kennedy2014}}. The majority of these 18 stars are older than 1\,Gyr, and nearly all are older than 500\,Myr, as can be seen in Fig.\,\ref{fig:sky_distribution}.
Since this was done only after the survey observations had started, these 18 stars are still included in our initial high-cadence survey, but they will neither be scheduled for longer-term monitoring (Sect.\,\ref{ssec:obs:strat}), nor will we use them for our anticipated statistical analysis of the occurrence of GPs around debris disc stars. 

To derive stellar ages, we first checked each target for membership of known associations using the banyan\,$\Sigma$\ tool\footnote{http://www.exoplanetes.umontreal.ca/banyan/} \mbox{\citep{gagne2018}}. In addition, we checked if other youth indicators, such as $v\sin(i)$, are in agreement with the association age. If the membership probability was >\,80\%\footnote{For two stars with membership probabilities of 64 and 70\%, we also assigned the mean age of the association because these ages are widely used in the literature and our isochronal ages did not contradict the association ages. These have corresponding notes in Table\,~\ref{tab:RVspy_phys_pars}}, and neither our isochronal age estimate, nor the measured $v\sin(i)$\ contradict the association age, we assign the mean age of the association to the star. In total we assigned association ages to 43 of our 111 stars. For the remaining 68 field stars, ages are assigned by compiling various literature estimates as described in \citet{pearce2022}. 
To validate our adopted ages, and to derive stellar masses, we also performed Hertzsprung–Russell diagram (HRD) isochrone fits as described in detail in \cite{pearce2022}. 
The resulting stellar masses were adopted for each star. The corresponding ages and uncertainties (which are naturally large for stars close to the main sequence) were not adopted. They were only used to verify the consistency of our adopted ages with the isochrone fits.
Only for two stars, for which we could not find a valid literature age, we adopted the isochrone age directly.
For most of the other stars, our adopted and isochronal ages agree to within 2$\sigma$. 
The uncertainties of our age estimates are discussed in Sect.\,\ref{ssec:dis:stellpar}.

%%%%%%%%%%%%%%%%%%%%%%%% Fig. 3
\begin{figure}[htb]
\includegraphics[width=9cm,angle=0,clip=true]{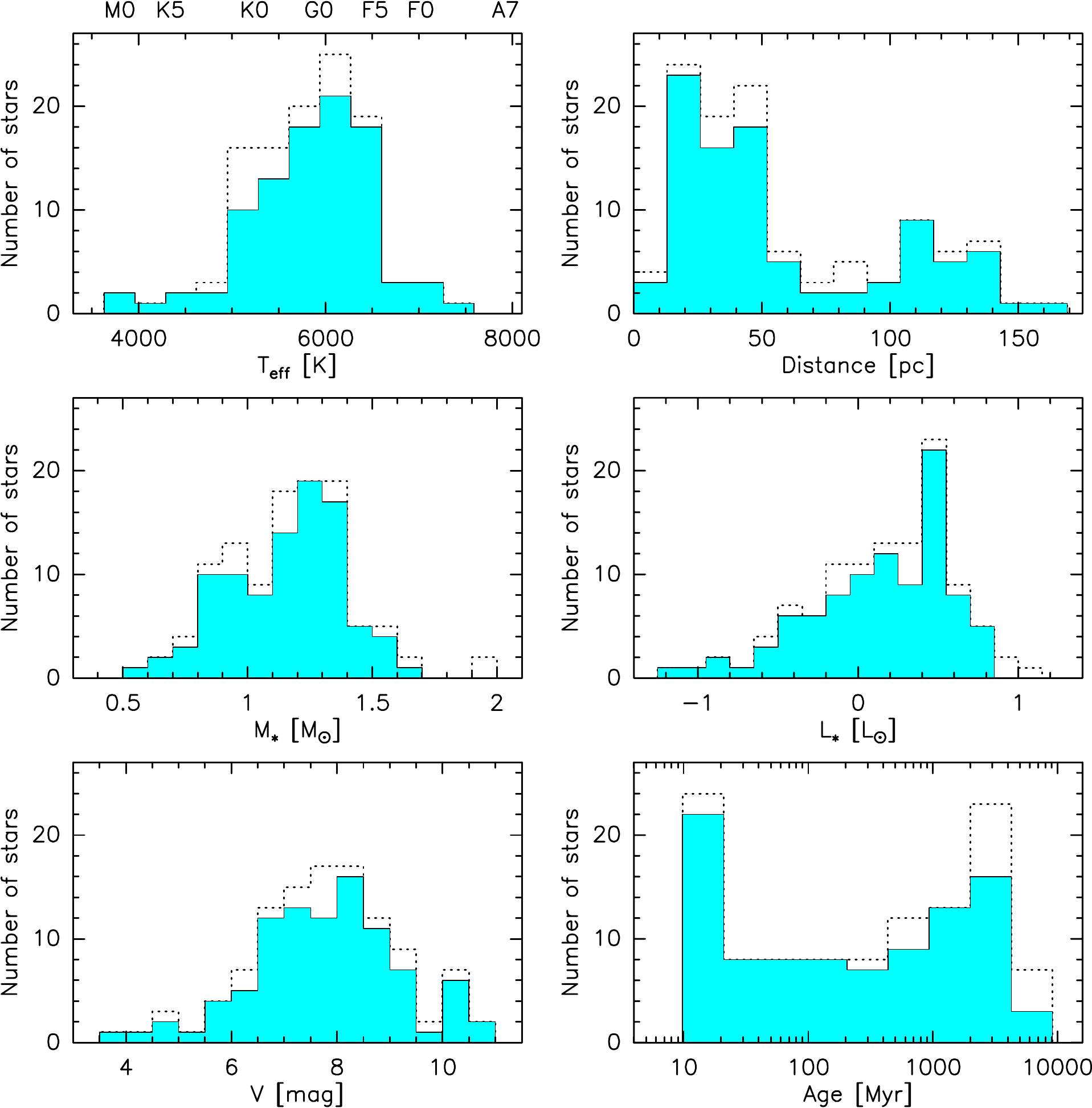}
\centering
\caption{\label{fig:distribution_param}
Histograms showing the distributions of stellar effective temperatures (corresponding main-sequence spectral types marked on top), distances, masses, luminosities, $V$\,magnitudes, and ages of our survey targets. Dotted histograms show all 111 targets, while the blue filled histograms account only for targets with confirmed significant debris disc signal.
}
\end{figure}
%%%%%%%%%%%%%%%%%%%%%%%%

Figure\,\ref{fig:distribution_param} shows the distribution of $T_{\rm eff}$, distances, stellar masses, luminosities, $V$\,magnitudes, and ages of all 111 survey targets. Effective temperatures range from 3800 to 7480\,K (corresponding to spectral types A9 to M0, median 5840\,K or G1). Stellar masses range from 0.56 to 2.34\,M$_{\odot}$\ (median 1.18\,M$_{\odot}$), $V$\ magnitudes from 3.7 to 10.6\,mag (median 7.8\,mag), distances range from 6 to 160\,pc (median 45\,pc and one outlier with uncertain distance of 335\,pc), 
ages range from 10\,Myr to 7.6\,Gyr (median 400\,Myr), and bolometric luminosities range from 0.06 to 60\,L$_{\odot}$\ (median 1.7\,L$_{\odot}$). The distance distribution is bimodal with a pronounced peak at 20-50\,pc accounting for the most nearby stars within the local bubble, and a second peak 120-140\,pc accounting for young stars associated to the nearest star-forming regions.
Further target properties are derived in this paper from the FEROS spectra (Sect.\,\ref{ssec:res:par}).
Target designations, coordinates, and most physical parameters mentioned above of all 111 survey stars are listed in Tables\,\ref{tab:RVspy_phys_pars} (debris disc stars) and \ref{tab:RVspy_phys_pars_nodeb} (stars without a significant debris disc signal).

%%%%%%%%%%%%%%%%%% Fig. 4
\begin{figure}[htb]
\includegraphics[width=9cm]{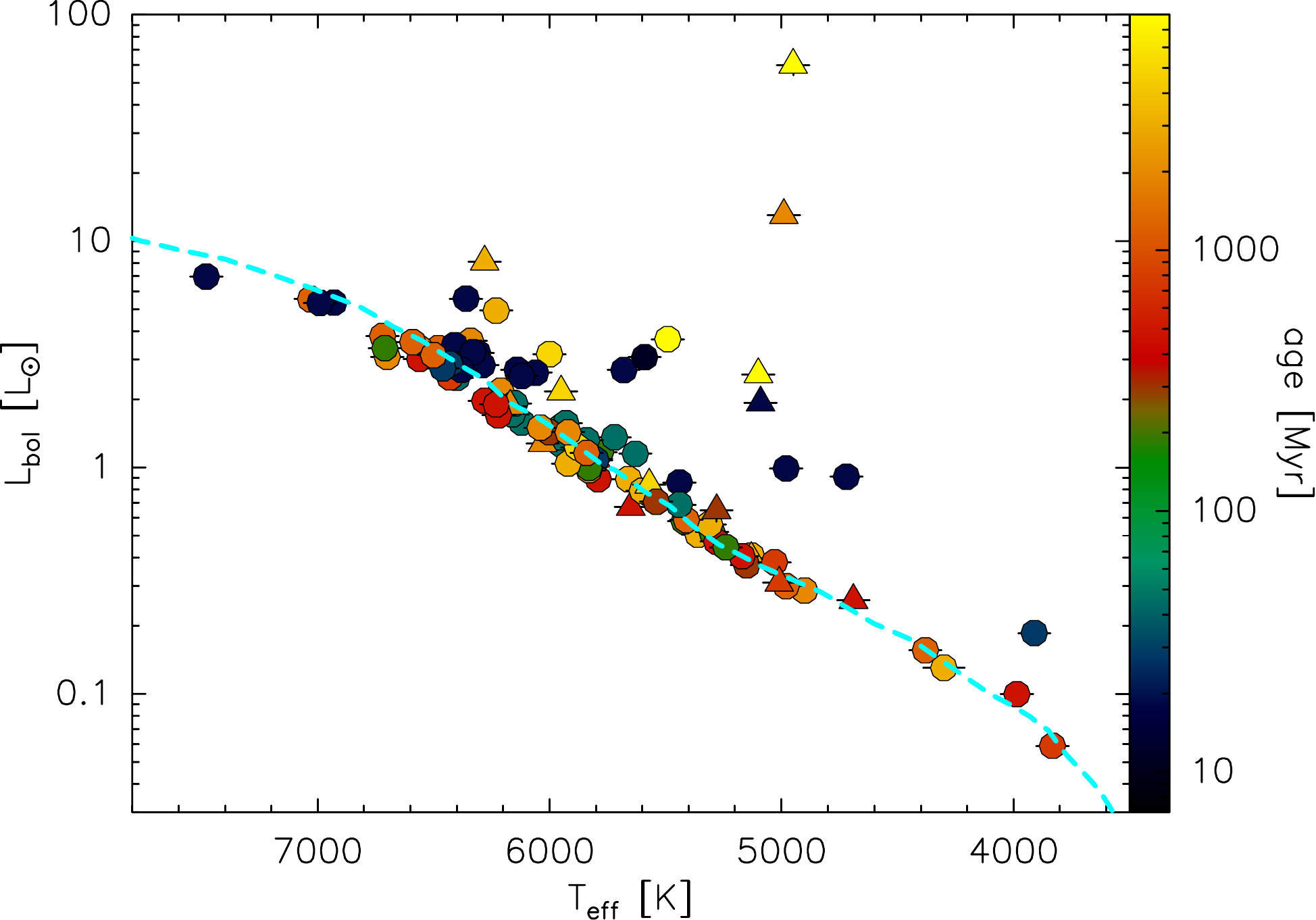}
\centering
\caption{\label{fig:hrd}
HRD ($L_{\ast}$\ vs. $T_{\rm eff}$) of single RVSPY target stars. Stellar ages are colour-coded. Stars marked as triangles do not have significant IR excess (see Sect.\,\ref{ssec:targets:sel}). 
To guide the eye, an updated version of the main sequence from \citet{pm2013} is marked by a dashed light-blue curve.
}
\end{figure}
%%%%%%%%%%%%%%%%%%

Figure\,\ref{fig:hrd} shows the HRD of our target stars with the stellar ages colour-coded. The main sequence is clearly visible. Also evident is a larger number of slightly overluminous pre-main sequence stars with ages below 25\,Myr. Nearly all of these stars have significant infrared (IR) excess from debris discs. Furthermore, it can be seen that the very overluminous stars are actually old stars (giants) without significant IR debris disc excess.

%%%%%%%%%%%%%%%%%%%%%%%%%%%%%%%%%%%%%%%%%%%%%%%%%%%%%%%%%%%%%%%%%

\section{Observations}
\label{sec:obs}

\subsection{Observing strategy}
\label{ssec:obs:strat}

As outlined in Sect.\,\ref{sec:motivation}, the first phase of our survey is designed to characterise the stellar activity jitter of all targets on the timescales of their rotational periods and to search for HCs. For this purpose, we scheduled every target star for one period of high-cadence observations consisting of 14\,--\,15 consecutive nights in which we obtained one spectrum per night during at least 13, ideally consecutive nights, plus 2\,--\,3 spectra per night during one to two nights. The order in which the stars during one such block were observed was changed from night to night and within their visibility time to avoid exact 24 hr window function peaks in the periodograms, thereby probing the frequencies (or periodicities) relevant for both activity-related rotational modulation and hot companions.

Many of the more quiet stars in our sample exhibit an intrinsic  RV scatter of <10\,m/s (see Fig.\,\ref{fig:rmsrv_teff}), which is close to the instrumental limit of FEROS for stars with small $v\sin(i)$\ and a sufficient number of narrow spectral lines, that is, cool stars. Such a precision can typically be achieved with an signal-to-noise ratio\,(S/N)\,\mbox{$\gtrsim100$}. In practice however, the achievable RV precision also depends on the stellar spectral type, $v\sin(i)$, and the atmospheric conditions. Since the RV amplitude induced by a 1\,$M_{\rm Jup}$-mass planet in a 3-day orbit around a solar-mass star is about 50\,m/s, we aimed to identify and monitor only stars for which an intrinsic RV precision of $\sigma_{RV}\leq50$\,m/s can be reached.
In addition to the high-cadence observations, we took individual 'test' spectra for stars that did not have any archival spectra available to assess their basic stellar parameters and the achievable RV precision (Sect.\,\ref{ssec:targets:list}) and to decide whether or not they will be scheduled for high-cadence observations. 
The results and detection thresholds derived from this initial high-cadence survey were then used to identify those targets for which searching for longer-period companions (months up to \mbox{2\,--\,3\,yrs}) seems feasible. In the second phase of the survey, we will continue to monitor the stars from this down-selected sample with an individually adapted lower cadence for up to a few years (see Sect.\,\ref{sec:discussion}).

%%%%%%%%%%%%%%%%%%%%%%%%%%%%%%%%%%%%

\subsection{High-cadence observations}
\label{ssec:obs:obs}

%%%%%%%%%%%%%%%%%% Table 1
\begin{table*}[htb]
\caption{Observing campaigns}
\label{tab:obs}
\centering          
\begin{tabular}{lccccc}     % 7 columns 
\hline\hline       
ESO Period, Program ID & Time period & Nights allocated\tablefootmark{a} & Time allocated &Time used\tablefootmark{b} & N of spectra\tablefootmark{c} \\
\hline
 101, 0101.A-9012(A)  & Apr. 2018 - Sept. 2018 & 29 & 165~h & 143~h 45~m & 685 \\
 102, 0102.A-9008(A)  & Oct. 2018 - Mar. 2019 & 29 & 130~h & 120~h 44~m & 635 \\
 103, 0103.A-9010(A)  & Apr. 2019 - Sept. 2019 & 27 & 140~h & 110~h 54~m & 539\\
 104, 0104.A-9003(A)  & Oct. 2019 - Mar. 2020 & 29 & 150~h & 131~h 45~m & 654 \\
 105, 0105.A-9010(A)  & Apr. 2020 - Sept. 2020  & 23 & 140~h &  0~h 0~m\tablefootmark{d} & 0\tablefootmark{d} \\
 106, 0106.A-9015(A)  & Oct. 2020 - Mar. 2021 & 21 & 92~h & 73~h 37~m & 346 \\
 107, 0107.A-9004(A)  & Apr. 2021 - Sept. 2021 & 18 & 68~h & 51~h 25~m & 260 \\
\hline                  
\end{tabular}
\tablefoot{
\tablefoottext{a}{Number of nights officially dedicated to the RVSPY program. We not specify the observing dates here, since many additional spectra were taken during the other nights: as a weather or technical loss compensation during DDT and due to the time exchange agreement with other large programs (observing semesters p103-p107).}
\tablefoottext{b}{Total exposure time dedicated to the program during the observing semester according to RAW ESO Science Archive \footnote{\url{http://archive.eso.org/eso/eso_archive_main.html}}.}
\tablefoottext{c}{Total number of spectra taken during the run.}
\tablefoottext{d}{The telescope was shut down due to the world lock down restrictions.}
}
\end{table*}
%%%%%%%%%%%%%%%%%%

During the first 3.5 observing years, between April 2018 and March 2022 (see Table\,\ref{tab:obs}), we obtained high-cadence time series of spectra for all our 111 survey targets. The observing campaigns typically consisted of about 14-15 consecutive nights.

All observations were carried out with the FEROS instrument, which provides a spectral resolution of $R=48000$, high efficiency ($\sim20\%$), and covers the wavelength range 350\,--\,920\,nm. All spectra were taken in the 'object-calibration' mode of FEROS with one fibre on the star and the second fibre observing simultaneously the 'ThAr+Ne' calibration lamp. Standard calibration files (BIAS, FLATS, and lamp spectra for wavelength calibration 'WAVE') were taken before each observing night during the afternoon and reduced with the CERES pipeline (see Sect.\,\ref{ssec:red:tools} and \mbox{\cite{Brahm2017a}} for details). To monitor the long-term stability of the spectrograph, we also observed a few RV standard stars during each campaign.

The required \mbox{S/N$\geq100$} (Sect.\,\ref{ssec:obs:strat}) was achieved with integration times between 5\,min for stars with $V\le6$\,mag and 20\,min for stars with $V=10-11$\,mag. When the seeing exceeded $1.8\arcsec$, the exposure time was increased. To ensure that the required S/N\ was actually achieved, and to identify possible spectroscopic binaries early on, we verified the data quality during the observations and analyse the RV time series after the first few observing nights of each run. 
During April - September of 2020 (observing semester p105) and April - July 2021 (observing semester p107), no observations were carried out due to the pandemic lockdown. 
The weather was good during most of the observing nights, with clear to thin cloud conditions and seeing mostly in the range \mbox{0\farcs6\,--\,1\farcs7} (median $1\arcsec$). No observations were taken when the seeing exceeded $2.5\arcsec$. About 10\% of time was lost due to bad weather conditions (thick clouds or strong wind). Table\,\ref{tab:obs} summarises the observing semesters.

%%%%%%%%%%%%%%%%%%%%%%%%%%%%%%%%%%%%

\subsection{Complementary data}
\label{ssec:obs:compl}

To extend the temporal baseline of the RV data of our targets and check if other high-cadence time series already existed for our targets, we looked for available archival spectra or already processed RVs that qualify for our purpose and can be combined with our own FEROS data. Archival FEROS spectra, which were not taken during our program, were downloaded from the RAW ESO Science Archive\footnote{\url{http://archive.eso.org}} and reduced and analysed with the same procedure as we use for our own observing data.
For 21 targets, we found precise Keck/HIRES RV data in the HIRES archive published in \citet{Tal-Or2019}\footnote{Originally, this is the HIRES archive published by \citet{Butler2017}, but corrected by \citet{Tal-Or2019} for small systematic errors in the RVs.}.
In addition, we found precise HARPS RV data and activity indicators for 38 targets in the
{\sc HARPS-RVBank}\footnote{\url{www.mpia.de/homes/trifonov/HARPS_RVBank.html}} \citep{Trifonov2020}.
However, since we found no existing high-cadence time-series dataset that qualifies for our purpose for any of our targets in the archives, we do not use these data here, but keep them for the longer-term follow-up survey, which is subject to a subsequent paper.

Furthermore, we used photometric time series data to derive additional constraints on stellar rotation periods and activity cycles. The majority of our targets are in the TESS\footnote{TESS -- Transiting Exoplanet Survey Satellite \mbox{\citep{Ricker2015}}.} observing list. For 91 of the 111 targets (82\%), the TESS light curves are already publicly available and the data are retrieved, and included in our analysis.

To search for known indications of close companions to our targets, we evaluate the ninth catalogue of spectroscopic binary orbits \citep[SB9;][]{pourbaix2004} and the Washington Visual Double Star Catalog \citep[WDS;][]{mason2020}. Close visual (i.e. not necessarily spectroscopically verifiable) companions are relevant for our study insofar as they can affect the photometry from which we derive certain stellar parameters (see Sect.\,\ref{ssec:targets:list}).
We also evaluated if our targets have significant ($>3\,\sigma$) proper motion anomalies (PMa) between the long-term HIP\,--\,{\it Gaia} proper motion vector and the GDR2 measurements \citep[PMaG2;][]{kervella2019}, which could hint at the presence of (known and still unknown) close companions.

%%%%%%%%%%%%%%%%%%%%%%%%%%%%%%%%%%%%%%%%%%%%%%%%%%%%%%%%%%%%%%%%%%%%%%%%%%%%%%%%%%%%%%%%%%

\section{Data reduction and analysis}
\label{sec:red}

\subsection{Tools}
\label{ssec:red:tools}

The basic reduction of the spectra and the extraction of RVs were done in a semi-automatic fashion with the CERES pipeline \mbox{\citep{Brahm2017a}}. CERES computes the RV\ of an observed spectrum by cross-correlating it with a binary mask for the stellar spectral lines. CERES provides three default masks for spectral types G2, K5, and M2. We used the mask that comes closest to the spectral type of a star (Tables\,\ref{tab:RVspy_phys_pars} and \ref{tab:RVspy_phys_pars_nodeb}).

The shape of the resulting cross-correlation function (CCF), which is also provided by CERES, is an important carrier of information about the origin of the measured RV\ variations (see below). 
For the computation of periodograms from time series data of RVs and various other quantities, we used the Generalised Lomb-Scargle (GLS) periodogram tool \mbox{\citep{Zechmeister2009}}, with the eccentricity fixed to zero and sine frequency (or period) output.
For the extraction of stellar parameters, we used the ZASPE pipeline \mbox{\citep{Brahm2017b}}.
To fit the RV data and derive the orbital parameters of the hypothetical companions we used The Exo-Striker fitting toolbox \mbox{\citep{Trifonov2019_es}}.

%%%%%%%%%%%%%%%%%%%%%%%%%%

\subsection{Activity indicators}
\label{ssec:red:activity}

Stars generally exhibit higher levels of activity when they are young as compared to the rather quiet stage on the main sequence (e.g. \citealt{Gregory17,brems19}). Hence, possible planet signals in the RVs will always be accompanied by some degree of activity-related rotational modulation of the spectra, such that planet signals do not stand out as clearly. Their recovery requires, besides an adapted observing strategy (Sect.\,\ref{ssec:obs:strat}), both a special analysis strategy and a case-by-case treatment. These difficulties are also reflected in the unofficial nickname of our high-cadence survey, 'Hot planets and rubble - let's face the trouble'.

Rotational modulation of spectral line shapes due to activity features on the stellar surface can induce large RV scatter that can mask a planetary RV signal or even mimic it, when the spots are stable. To characterise the stellar activity and disentangle possible planetary signals from activity-induced RV\ variations, we derive  activity-related parameters for all our observed spectra. In particular, H$\alpha$ line indices \citep{2003AA...403.1077K,2009AA...495..959B}, Ca\,II H\&K S$_{\rm index}$\ \citep{1991ApJS...76..383D}, line profile variations of Ca\,II H\&K, H$\alpha$, H$\beta$, and H$\gamma$ \citep{1995AJ....109.2800J},  line depth ratios of temperature-sensitive lines using line pairs \citep{2007AN....328..938B}, bisector shape, span (BS), displacement, and curvature \citep{2001AJ....121.1136P} and full width at half maximum (FWHM) of the cross-correlation function derived by CERES. We determine stellar rotation periods ($P_{\rm rot}$) from time series of photometric data from TESS \mbox{\citep{tess}}.

To distinguish periodic RV variations caused by a physical companion from those that are related to rotational modulation of line shapes, we calculated both GLS periodgrams and linear correlations of all activity indicators with the RVs. In addition, to detect periodic variations of the line profiles or small parts of the lines, we calculated two-dimensional GLS periodograms for the Ca\,II\,H\&K, H$\alpha$,H$\beta$, and H$\gamma$\ lines (see Fig.\,\ref{fig:2dgls}).

Depending on spectral type and age of the star, as well as geometrical aspects, chromospheric and photospheric activity can occur at different levels and might not be detectable equally in all indicators. 
For example, a well-established indicator to identify periodic rotational modulations of line shapes by dark spots on the stellar surface is the correlation between bisector span (BS) and RV. A strong anti-correlation between  BS and RV is a clear indication of a dark spot \mbox{\citep[e.g.][]{Queloz01}}. However, depending on the inclination of the star, the spot latitude, and $v\sin(i)$, the strength of the correlation and the amplitude of the BS can vary significantly \mbox{\citep[e.g.][]{desort2007}}. It is therefore necessary to rely not only on one indicator, but to measure a variety of indicators of different origin and different sensitivities, especially in the case of a large RV survey. 

We did not undertake this full activity analysis routinely for all targets, but only when we found a periodicity in the RV data that cannot be easily explained with fewer analyses. In Appendices\,\ref{sec:activity:HD38949} and \ref{sec:activity:CPD-722713}, we demonstrate as two examples our described activity analysis for targets HD\,38949 and CPD-72\,2713, for which most activity indicators show clear and unambiguous signals.

In addition, we used GLS periodograms of the photometric time series data from TESS (Sect.\,\ref{ssec:obs:compl}), where available, to independently derive photometric variability periods and draw conclusions on the most likely stellar rotational periods (Sect.\,\ref{ssec:res:phot}). Knowing the stellar rotation periods helps us to interpret RV periodograms and to distinguish rotational modulation from companion-related RV variability (Sect.\,\ref{ssec:res:rv}).

%%%%%%%%%%%%%%%%%%%%%%%%%%%%%%%%%%%%%%%%%%%%%%%%%%%%%%%%%%%%%%%%

 \section{Results}
 \label{sec:res}

We have obtained high-cadence time series of 19 spectra (median) over 12-20 days for all 111 stars of our survey. From the spectra, we determined the basic stellar physical parameters $T_{\rm eff}$, $[{\rm Fe/H}]$, and $\log(g)$), and measured the rotation-dominated line broadening, $v\sin(i)$, and the activity-dominated short-term RV jitter, \mbox{${\rm rms}_{\rm RV}(\tau\,14\,{\rm d})$}. We searched for RV periodicities on the 14-d timescale, and, where we found periodicities, analysed the correlation between RV and BS and compared the RV variability with the photometric variability in the TESS data where available. In addition, we used archival spectra where available to verify the longer-term RV variability.

%%%%%%%%%%%%%%%%%%%%%%%%%%

\subsection{Spectroscopic stellar parameters}
 \label{ssec:res:par}

In addition to the target properties presented in Sect.\,\ref{ssec:targets:list}, we derived in this paper from our FEROS data spectroscopic $T_{\rm eff}$, metallicity \mbox{$[{\rm Fe/H}]\equiv\log_{10}[({\rm Fe/H})]/({\rm Fe/H})_{\odot}$}, surface gravity $\log_{10}(g)$, and projected rotational line broadening $v\sin(i)$, using the open source ZASPE pipeline \mbox{\citep{Brahm2017b}}.
Since the derivation of uncertainties for individual stars within ZASPE is computationally very expensive, we computed them only for selected stars and extrapolate the uncertainties to the entire sample\footnote{In the high-signal-to-noise regime, the uncertainty of the ZASPE-derived parameters is mainly determined by the mismatch between the observed and modelled spectra.}.
For $T_{\rm ]eff}$, we adopted a general 1\,$\sigma$\ uncertainty of 124\,K (see Sect.\,\ref{ssec:dis:stellpar}), with the exception of a few stars for which the rms scatter of the  $T_{\rm eff}$\ derived from the individual spectra was larger.
For $v\sin(i)$, we selected seven stars with $v\sin(i)$ coming closest to 3, 5, 10, 20, 30, 40, and 50\,km\,s$^{-1}$\ to compute individual 1\,$\sigma$\, uncertainties. Based on the computed uncertainties for these example cases, we derived an empirical fit of the form \mbox{$\sigma_{v\sin(i)}=\sqrt{v\sin(i)}$}, with a lower limit of 5\% towards larger $v\sin(i)$, for all our targets, with the following exception. For values of \mbox{$v\sin(i)<3$\,km\,s$^{-1}$}, we can no longer distinguish between instrumental and rotational broadening of the spectral lines and adopted therefore a lower limit of 3\,km/s \mbox{\citep{2012AJ....143...93R}}.

%%%%%%%%%%%%%%%%%%%%%%%% Fig. 5
\begin{figure}[htb]
\includegraphics[width=9cm,angle=0,clip=true]{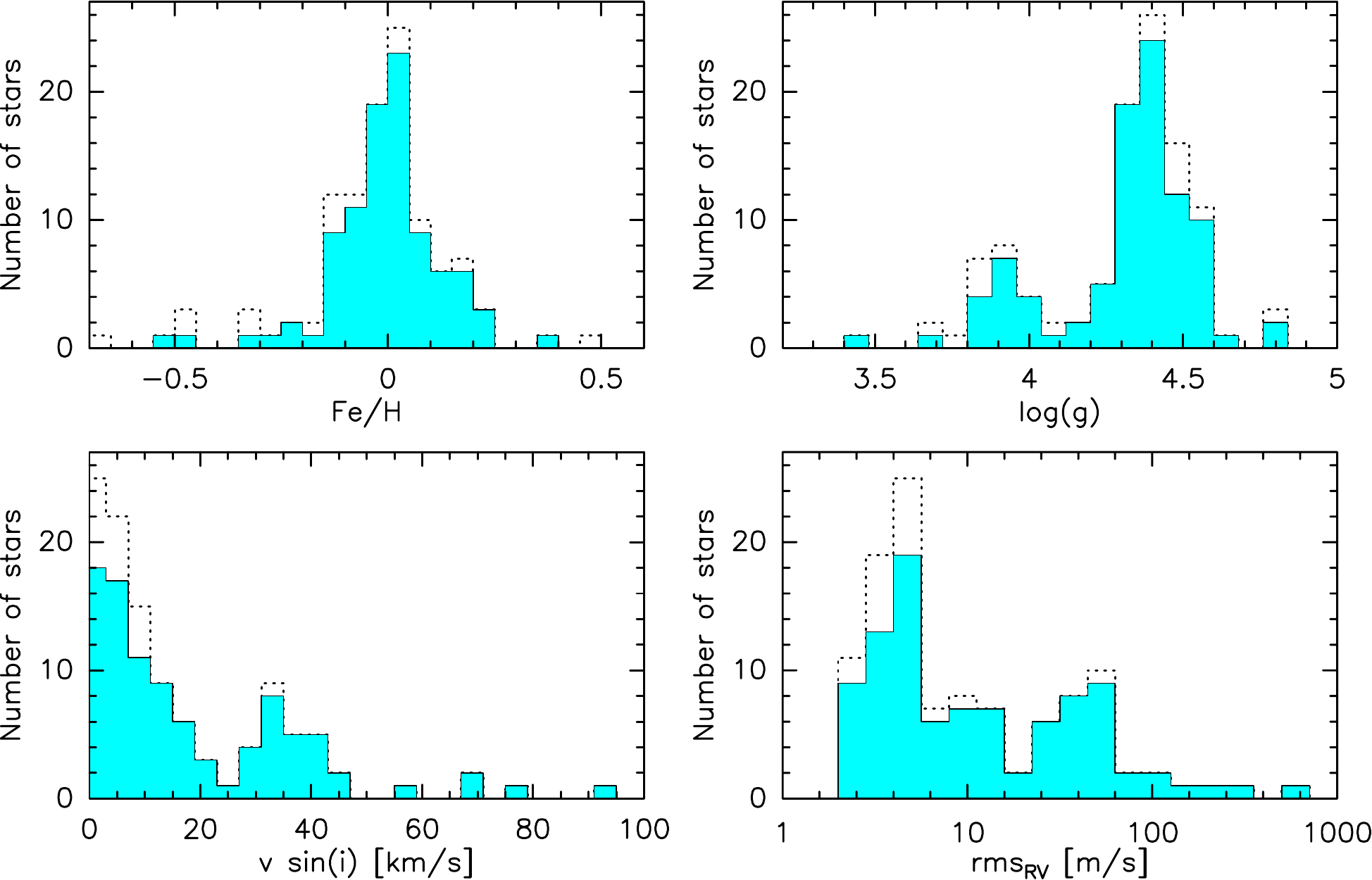}
\centering
\caption{\label{fig:distribution_param2}
Histograms showing the distributions of stellar metallicities $[{\rm Fe/H}]$, surface gravity $\log(g)$, $v\sin(i)$, and intrinsic rms scatter of the RVs in the 2-week high-cadence data (see Sect.\,\ref{ssec:res:rv}). Dotted histograms show all 111 targets, while the blue filled histograms account only for targets with confirmed significant debris disc signal.
}
\end{figure}
%%%%%%%%%%%%%%%%%%%%%%%%

Figure\,\ref{fig:distribution_param2} shows the distributions of stellar metallicities $[{\rm Fe/H}]$, surface gravity $\log(g)$, and $v\sin(i)$\ as dervived from our FEROS spectra.
The spectroscopically derived $T_{\rm eff}$\ mostly agree with the photometrically derived ones (Sect.\,\ref{ssec:targets:list}) within 3$\sigma$, and the $T_{\rm eff}$\ distributions (Fig.\,\ref{fig:distribution_param}) are indistinguishable. There are only four stars for which the relative difference is 3\,--\,5\,$\sigma$, which are discussed in Sect.\,\ref{ssec:dis:stellpar}.
Our FEROS/ZASPE-derived metallicities, $[{\rm Fe/H}]$, agree mostly within 3\,$\sigma$\ with those derived by \citet{gaspar2016} where those are available. There are only three stars for which the relative difference is larger (3.3\,--\,5\,$\sigma$, see Sect.\,\ref{ssec:dis:stellpar}). We verified our results on ten or more different individual spectra for each of these stars, but obtained very consistent numbers. Our metallicities cluster around solar (zero, as expected) with only eight stars having $|[{\rm Fe/H}]|>0.4$. These 'outliers' are discussed in Sect.\,\ref{ssec:dis:stellpar}.
The distribution of surface gravities clearly shows the main-sequence peak around \mbox{$\log(g)\sim4.4$} and a secondary peak around \mbox{$\log(g)\sim3.9$}, which encompasses both the youngest PMS stars with ages between 10 and 30\,Myr as well as the few old (sub-)giant stars which mostly have no significant debris disc excess (see Sect.\,\ref{ssec:targets:list}). The very few 'outliers' are discussed in Sect.\,\ref{sec:discussion}.
The distribution of $v\sin(i)$\ is also bimodal, with the first peak between $v\sin(i)\sim0.5$\ and 20\,km/s encompassing those stars for which magnetic braking is efficient, and the second peak between $v\sin(i)\sim30$\ and 45\,km/s encompassing stars for which magnetic braking does not act, that is, stars more massive than $\sim1.3$\,M$_{\odot}$\ (see Sect.\,\ref{sec:motivation} and discussion in Sect.\,\ref{ssec:dis:Teff_vsini_age}). The five very fast rotating stars ($v\sin(i)>50$\,km/s) are all young with ages between 15 and 200\,Myr.
The distribution of ${\rm rms}_{\rm RV}(\tau\,14\,d)$\ is discussed in Sect.\,\ref{ssec:res:rv}.
The individual values for $T_{\rm eff}$(sp), $[{\rm Fe/H}]$, $\log(g)$, $v\sin(i)$, and ${\rm rms}_{\rm RV}(\tau\,14\,d)$\ are listed in Tables\,\ref{tab:RVspy_phys_pars} and \ref{tab:RVspy_phys_pars_nodeb}.

%%%%%%%%%%%%%%%%%%%%%%%%%%

\subsection{RV variability in the high-cadence data}
 \label{ssec:res:rv}
 
%%%%%% Fig. 6
\begin{figure}[htb]
\includegraphics[width=9.0cm]{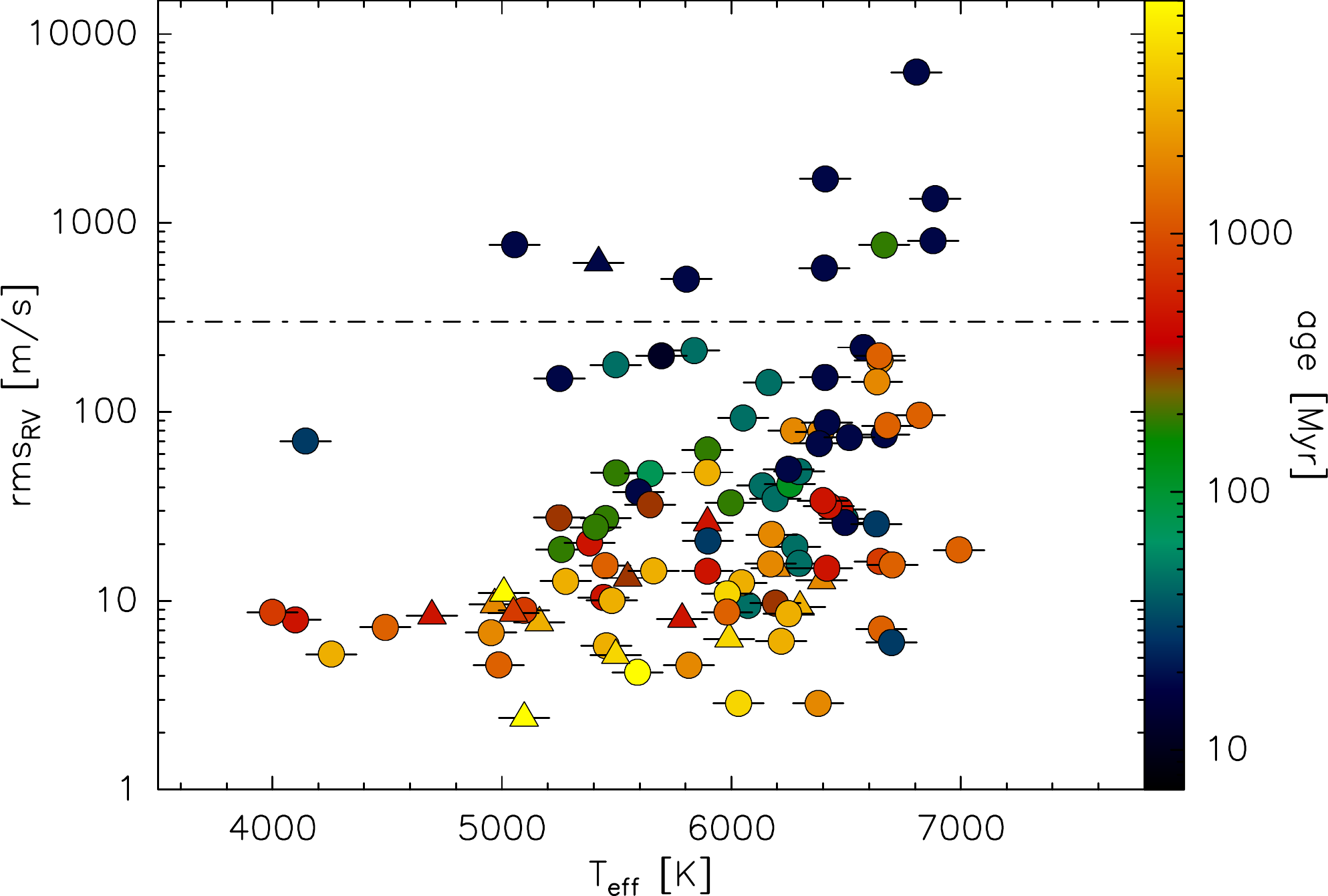}
\centering
\caption{\label{fig:rmsrv_teff}
Intrinsic rms scatter of the RVs in the 2-week high-cadence data vs. (spectroscopic) effective stellar temperature, $T_{\rm eff}$, with age encoded in colour. Stars marked as triangles do not have significant IR excess (see Sect.\,\ref{ssec:targets:sel}). 
The horizontal dashed-dotted line marks the approximate boundary above which we consider RV exoplanet searches not feasible (although we use a mass detection threshold in the end; see Sect.\,\ref{sec:dis:prospects} and Fig.\,\ref{fig:Mp_vs_Teff}).
}
\end{figure}
%%%%%%

To quantify the RV variability of our targets on a 2-week time scale, we first looked at the intrinsic rms scatter of the RVs, irrespective of whether there are periodicities in the RVs or not. We achieved a median single-measurement RV precision of \mbox{$\langle \sigma RV\rangle\,=\,5.75$\,m/s} (4.45\,m/s for stars younger than 500\,Myr and 13.1\,m/s for older stars), and measured a median short-term intrinsic RV scatter, 
\mbox{${\rm rms}_{\rm RV}(\tau\,14\,d)=\sqrt{{\rm rms}_{\rm obs}^2-\sigma RV^2}$} of 23\,m/s (44\,m/s for stars younger than 500\,Myr and 10\,m/s for older stars).
Figure\,\ref{fig:rmsrv_teff} shows ${\rm rms}_{\rm RV}(\tau\,14\,d)$ vs.\ the effective temperature, $T_{\rm eff}$, for all targets, except the SB's (see Sect.\,\ref{ssec:res:sb}). 
We observed three stars with \mbox{${\rm rms}_{\rm RV}(\tau\,14\,d)\gtrsim1000$\,m/s}: HD\,115820, HD\,141011, and HD\,145972. All three stars are very young (15-16\,Myr) and none of them shows any significant periodicity in the high-cadence RV data, although the dominant photometric TESS periods (see Sect.\,\ref{ssec:res:phot}) in HD\,141011 (0.95\,d) and HD\,145972 (1.5\,d) might be marginally seen in the RVs. However, both stars also show several smaller TESS periodgram peaks at periods between 0.2 and 0.7\,d, which could be indicative of week pulsations that could explain the large RV jitter, since these periods are not temporarily resolved by the sampling of our RV data. This goes along with the notion by \citet{Grandjean2021} and the analysis of \citet{lagrange2009} that the RV jitter in young A--F5 stars is dominated by pulsations.

For HD\,115820 (A7\,V, 15\,Myr), the star with the largest \mbox{${\rm rms}_{\rm RV}(\tau\,14\,d)$} of $\sim$6.2\,km\,s$^{-1}$, the TESS data indicate several strong and sharp periods between 42 and 68\,min, plus numerous fainter periodgram peaks at longer periods up to a few days. These can most likely be attributed to g-mode pulsations, which typically occur in $\delta $\,Scuti stars \citep[e.g.][]{Breger2000,Murphy2015}. 
These strong and very short-period pulsations appear as random 'noise' in our RV data and explain the large \mbox{${\rm rms}_{\rm RV}(\tau\,14\,d)$} of HD\,115820.

Excluding these three young and very active stars with signs of pulsations, we observed intrinsic RV rms scatter values between 2.4 and $\sim$800\,m/s, with a general trend of increasing scatter with increasing $T_{\rm eff}$\ and with youth. Using the stellar masses (Tables\,\ref{tab:RVspy_phys_pars} and \ref{tab:RVspy_phys_pars_nodeb}), we also derived the corresponding mass-detection limits for HC corresponding to 3$\times{\rm rms}_{\rm RV}(\tau\,14\,{\rm d})$\ and list the respective values for $P=10$\,d in Table\,\ref{tab:periods}. For $P=3$\,d, we achieved a median mass-detection limit for HC around stars younger than 500\,yr of $\sim$1.4\,M$_{\odot}$\ ($\sim$0.3\,M$_{\odot}$\ for older stars).
RV searches for longer-period planets become unfeasible if the short-term RV scatter rms exceeds a value of $\sim$300\,m/s (conservative estimate), but we actually used these \mbox{${\rm rms}_{\rm RV}(\tau\,14\,d)$} values to calculate mass detection limits on longer-period companions and select the targets suitable for longer-term RV monitoring (Sect.\,\ref{sec:dis:prospects}).

To further evaluate the nature of the RV variablity, we analysed the time series periodograms\footnote{We use the Generalised Lomb-Scargle (GLS) periodogram tool by \mbox{\citet{Zechmeister2009}}, with the eccentricity fixed to zero.}, the correlation between BS and RV \mbox{\citep{Queloz01}}, and the shape of the cross-correlation functions (CCF) for each star. We found a strong anti-correlation ($r_P\le-0.6$)\footnote{$r_P=$ Pearson correlation coefficient} between BS and RV in the high-cadence data of 40 stars, all marked in Table\,\ref{tab:periods}, indicating that stellar spots dominate the observed RV variability. We also found significant periods between 1.3 and 4.5\,d in the RV data of 14 stars, which we list in Table\,\ref{tab:periods}. Of these, the RV periods of 12 stars agree with photometric periods seen in the TESS data (Sect.\,\ref{ssec:res:phot}). Nine of these show the aforementioned strong anti-correlation between BS and RV, suggesting that we have indeed detected the stellar rotation period and not an HC. 

%%%%%% Fig. 7
\begin{figure}[htb]
\includegraphics[width=9.0cm]{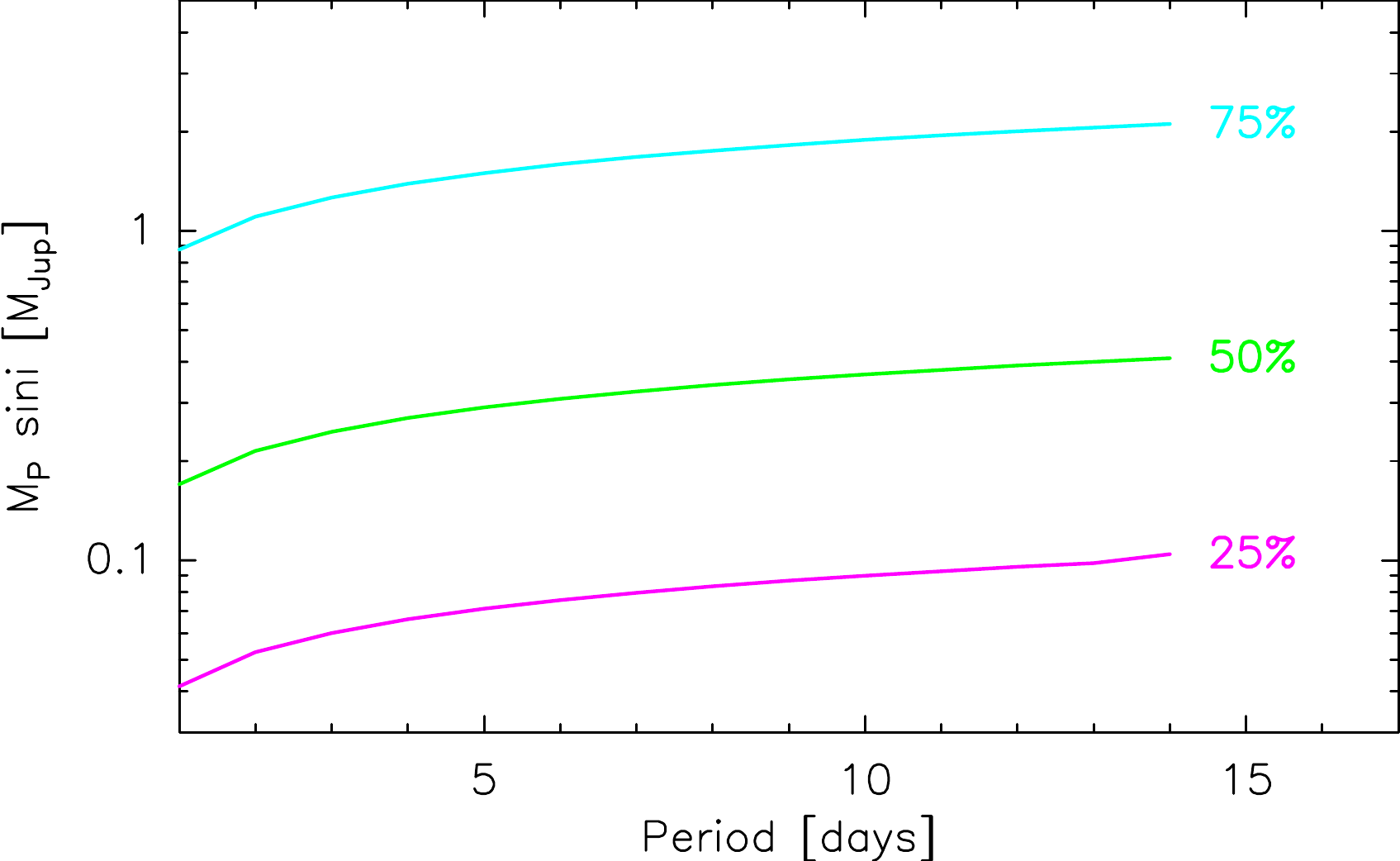}
\centering
\caption{\label{fig:mpdetlim_survey}
Planet mass detection limits and search completeness for our high-cadence survey, corresponding to intrinsic rms scatter of the RVs in the 2-week high-cadence data.
}
\end{figure}
%%%%%%

Two stars with significant RV periods of 3.6\,d (HD\,107146) and 1.56\,d (HD\,219498) have no TESS data available (yet), but the corresponding BS/RV correlation factors of \mbox{($r_P\le-0.79$)} suggest that we also see spot-dominated RV variability and may have detected the stellar rotation period. Two more stars without TESS data show only marginally significant RV periods of 3.1\, (HD\,131156) and 6\,d (HD\,23340), respectively. Since they also show BS/RV correlation factors of \mbox{($r_P\le-0.7$)}, we interpret their RV variability as tentative detection of the stellar rotation period.
In 18 more stars, the main TESS period is also marginally detected in the RVs, indicating spot-dominated rotational modulation of the RVs.
In none of the remaining 69 targets we detect a significant RV periodicity in the range 1\,--\,14\,d, that is, we do not find an HC among our 111 target stars. The planet mass detection limits and search completeness for our high-cadence survey are shown in Fig.\,\ref{fig:mpdetlim_survey}.

%%%%%%%%%%%%%%%%%%%%%%%%%%

\subsection{Photometric variability}
 \label{ssec:res:phot}

TESS photometric time series data are available and retrieved for 91 of our 111 target stars. We employed GLS periodograms to search for significant photometric variability periods between 0.1 and 20\,days. If more than one TESS observing sector was available (for 67 out of 91 stars), we analysed the individual sectors separately. If similar periodicities were detected in all sectors of a given star, we averaged the respective period values and GLS power and derived the uncertainty from the scatter between the individual sectors. In Table\,\ref{tab:periods} we list the up to three most significant periods together with their GLS periodogram power. All listed photometric periods are highly significant and have a false alarm probability (FAP) of $<<10^{-4}$. We also list in Table\,\ref{tab:periods} our best estimate of the most likely stellar rotation period and assume a minimum uncertainty of 0.1\,d.
For stars that show large ($>$1\,km/s) unexplained \mbox{${\rm rms}_{\rm RV}(\tau\,14\,d)$}, we also searched the TESS data for frequencies up to 300\,d$^{-1}$, corresponding to periods down about five minutes. These short periodicities are not related to the stellar rotation period, but to possible pulsations (see Sect.\,\ref{ssec:res:rv}).

For 30 stars, we identified one single dominant photometric period of which we are confident that it is caused by a single dominant spot (group) and represents the stellar rotation period. For 29 stars, we identified two dominant periods with an approximate 2:1 period ratio, which we interpret as the result of two dominant spot groups, such that the longer of the two periods represents the most likely stellar rotation period. For many of these stars, a visual inspection of the light curves confirms that the shapes and strength of the clearly visible wiggles are indeed alternating. For 7 stars we identified two periods close together and adopt the mean as the most likely rotation period. For ten stars, we only found a range of significant periods, but cannot clearly identify a rotation period. For seven stars, we identified two dominant periods that are not in a 2:1 ratio and it is unclear which one represents the rotation period. Finally, five stars do not show any significant photometric periodicity. 

As reported in Sect.\,\ref{ssec:res:rv}, we further detected rotation periods for two more stars without TESS data based on their RV variability. In total, we thus derived confident estimates of stellar rotation periods for 59 stars, and uncertain guesses for 32 stars. Our derived rotation periods range from 0.9 to 11\,days.

%%%%%%%%%%%%%%%%%%%%%%%%%%

\subsection{Spectroscopic companions}
 \label{ssec:res:sb}

We identified nine spectroscopic binaries (SB) in our target list, of which only three were previously known.
An orbit solution for HD\,16673 (37-d nearly circular orbit with a $M_2\sin(i)\sim0.2$\,M$_{\odot}$\ secondary) has been published only recently by \citet{gorynya2018}, after our survey had started. 
HD\,27638\,B has a relatively massive (mass ratio 0.85) seconday in a 17.6-d orbit \citep{tokovinin2001}, which we had obviously missed in our target selection. 
HD\,141521 was previously identified as an SB2 by \citet{weise2010}, but no orbit solution has been published yet. We confirm the SB2 nature of this star, but cannot provide an orbit solution yet. None of these three stars shows a significant PMaG2.

In addition, we found six new SB, which were not reported previously. Since our high-cadence observations did not cover their full orbits, we can here only roughly estimate their orbital periods and secondary (minimum) masses and will publish the orbit solutions later. 
Three of these six stars have companions that are potentially in the brown dwarf mass regime with extrapolated orbital periods between 14 and 50\,days (HD\,102902, MML\,43, and HD\,129590). Of these, only HD\,102902 shows a significant (4.8\,$\sigma$) PMaG2 \citep{kervella2019} that would be consistent with a brown dwarf companion, albeit the orbital radius or period remain unconstrained. The orbital periods of the two other SB may be too short to detect such a PMa.
HD\,129590 has a bright nearly edge-on disc that was imaged in scattered light \citep{Matthews2017,Olofsson2022}. 

Three other stars are found to have companions in the low-mass stellar regime with extrapolated orbital periods between 20 and 80 days (HD\,20759, HD\,108857, and HD\,143811). Two of these also show a significant PMaG2 of 7.6\,$\sigma$\ (HD\,20759) and 3.5\,$\sigma$\ (HD\,108857), respectively, both consistent with low-mass stellar companions.
Since these SB are not resolved spatially, their photometrically derived luminosities and $T_{\rm eff}$\ are also affected, which in turn affects the age and mass estimates derived from HRD isochrone fits (Sect.\,\ref{ssec:targets:list}). We mark these stars in Tables\,\ref{tab:RVspy_phys_pars} and \ref{tab:RVspy_phys_pars_nodeb} and discuss them in Sect.\,\ref{ssec:dis:stellpar}.

In addition to the nine targets with known or newly detected close (spectroscopic) companions mentioned above, we found eight more targets with significant (3.3-30\,$\sigma$) PMaG2, but no other hints (yet) at the presence of a close companion. These stars are all marked in Tables\,\ref{tab:RVspy_phys_pars} and \ref{tab:RVspy_phys_pars_nodeb}. Since the PMaG2 has very little constraining power for orbital periods shorter than the Gaia observing time window (668\,d) due to observing window smearing \citep{kervella2019}, and the longer-period RV variability and related PMa constraints are subject to a subsequent paper, we do not further discuss these sources with PMa here. 
We also do not consider the nine SB's anymore in the following discussion since their photometric and spectroscpic properties are affected by the unresolved binarity.)

%%%%%%%%%%%%%%%%%%%%%%%%%%%%%%%%%%%%%%%%%%%%%%%%%%%%%%%%%%%%%%%%%%%%%%%%%%%%%%%%%%%%%%%%%%%%%%

\section{Discussion}
\label{sec:discussion}

%%%%%%%%%%%%%%%%%%%%%%%%%%%%%%%%%%%%

\subsection{Uncertainties of stellar parameters} 
\label{ssec:dis:stellpar}

% Teff:
The median (1\,$\sigma$) uncertainty of the photometrically derived $T_{\rm eff}$\ is 70\,K, that of the spectroscopically derived $T_{\rm eff}$\ is 110\,K. Systematic uncertainties such as, for example, those caused by unresolved binarity, are not accounted for by these model uncertainties.
The spectroscopically derived $T_{\rm eff}$\ are systematically higher by 124\,K (median) than the photometrically derived ones, but they mostly agree with each other to within 3$\sigma$. 
This systematic difference is most likely related to the different synthetic model spectra used: while the spectral energy distribution (SED) fitting uses the PHOENIX models from \citet{husser2013}, ZASPE uses its own library of synthetic spectra \citep{Brahm2017b}.
There are only three stars (not counting the SB) for which the relative difference is larger (3\,--\,5\,$\sigma$), all very young (15\,Myr) and with $T_{\rm eff}>6000$\,K. 
Of these, HD\,114082 has a visual companion at 1\farcs5, which might have affected the photometry. 
HD\,115820 shows a huge RV scatter rms of 6.3\,km/s, most likely owing to its $\delta$\,Scuti-like pulsations (see Table\,\ref{tab:periods}), which may also have affected derivation of $T_{\rm eff}$.
HD\,111520 (3.3\,$\sigma$\ difference) also has a relatively large RV scatter rms of 220\,m/s, but has no known close companion and the spectra provide no hint at a SB. These slight $T_{\rm eff}$\ discrepancies might also be related to the limited applicability of the respective stellar atmosphere models used for such young stars.

% ages:
Our literature-adopted and isochronal ages agree to within 2$\sigma$\ for all but four stars. 
For two supposedly young debris disc stars (HD\,117524 and HD141521, 15\,--\,16\,Myr), our upper limit on the isochronal age is slightly more than 2$\sigma$\ below our adopted association age, but still within a factor of two only. These slight discrepancies can be easily explained by the large model uncertainties at these young ages, which are not accounted for by the formal fitting uncertainties.
For two other stars (HD\,5349, HD\,102902, both without a significant debris disc signal), our isochronal fits suggest ages of below 5\,Myr, while other studies classify them as Gyr old giant stars \citep{Casa11,delgado2019}. Both stars have very small (unresolved) $v\sin(i)$, which suggests an old age. HD\,102902 turns out to be a SB, that is, the blended photometry can explain the wrong isochronal age. For HD\,5349, a closer look at the posterior distributions shows that both a 5\,Myr old 1.75\,M$_{\odot}$\ PMS star and an $\sim$10\,Gyr old 1.1\,M$_{\odot}$\ K\,giant would be consistent with its location in the HRD. Based on the low $v\sin(i)$\ and chemical (abundance rates) age derived by \citet{delgado2019}, the Gyr age is however the more likely one.

% Masses:
Our stellar masses given in Tables\,\ref{tab:RVspy_phys_pars} and \ref{tab:RVspy_phys_pars_nodeb} are derived from HRD isochrone fits to the MIST evolutionary models \citep{choi2016} and assuming a \citet{chabrier2003} single-star prior on mass \citep[see][]{pearce2022}. This, together with the fact that the photometry does not resolve SB, implies that the masses for SB and other close binaries (visual and those with a significant PMaG2) are not reliable. The respective stars are marked in Tables\,\ref{tab:RVspy_phys_pars} and \ref{tab:RVspy_phys_pars_nodeb}. Although we prefer our mass estimates over those of \citet{kervella2019}, because we use full SED evolutionary model fits and take metallicity into account, we compare our masses with those of \citet{kervella2019}, where available. We found on average good agreement with relative discrepancies rarely exceeding 20\%, and only three stars excceding the combined 3\,$\sigma$\ significance level on the mass difference. These are HD\,27638\,B and HD\,141521, which both are SB, and HD\,117524, a very young (15\,Myr) PMS star with a significant (20\,$\sigma$) PMaG2, indicating the likely presence of a still unknown close companion. 

% F[e/H]:
Our FEROS/ZASPE-derived $[{\rm Fe/H}]$\ are mostly consistent, within the error bars, with those derived by \citet{gaspar2016}. They differ by more than 3\,$\sigma$\ (3.3-4.4\,$\sigma$, assuming a lower limit for the uncertainty of 0.1\,dex) for only three stars. Two of these are actually SB, which means the quantities derived from the combined spectra might be wrong (HD\,102902 and HD\,141521). The third star is HD\,191849, the lowest-mass M-dwarf in our sample.
All three stars also belong to the group of eight stars with $|[{\rm Fe/H}]|>0.4$. The other five stars with $|[{\rm Fe/H}]|>0.4$\ are HD\,20759, which is also a SB, HD\,5349, HD\,101259, and HD\,213941, which are all Gyr-old giants and for which our and the Gaspar metallicities are consistent with ours, and CPD-72\,2713, a very young star (14\,Myr) which was not in the list of \citet{gaspar2016}.

% Surface gravity:
Our FEROS/ZASPE-derived surface gravities indicate only six stars with $\log(g)<3.7$. Three of these are Gyr-old giants which we already excluded from further observations (HD\,5349, HD\,101259, and HD\,138398). The three other stars $\log(g)<3.7$\ are very young PMS stars (HD\,115820, HD\,117214, and CPD-72\,2713), which explains their large radii and related low surface gravity.

%%%%%%%%%%%%%%%%%%%%%%%%%%%%%%%%%%%%

\subsection{RV precision as a function of $T_{\rm eff}$ and $v\sin(i)$} 
\label{ssec:dis:sRV_Teff_vsini}

%%%%%% Fig. 8
\begin{figure}[htb]
\includegraphics[width=9.0cm,angle=0,clip=true]{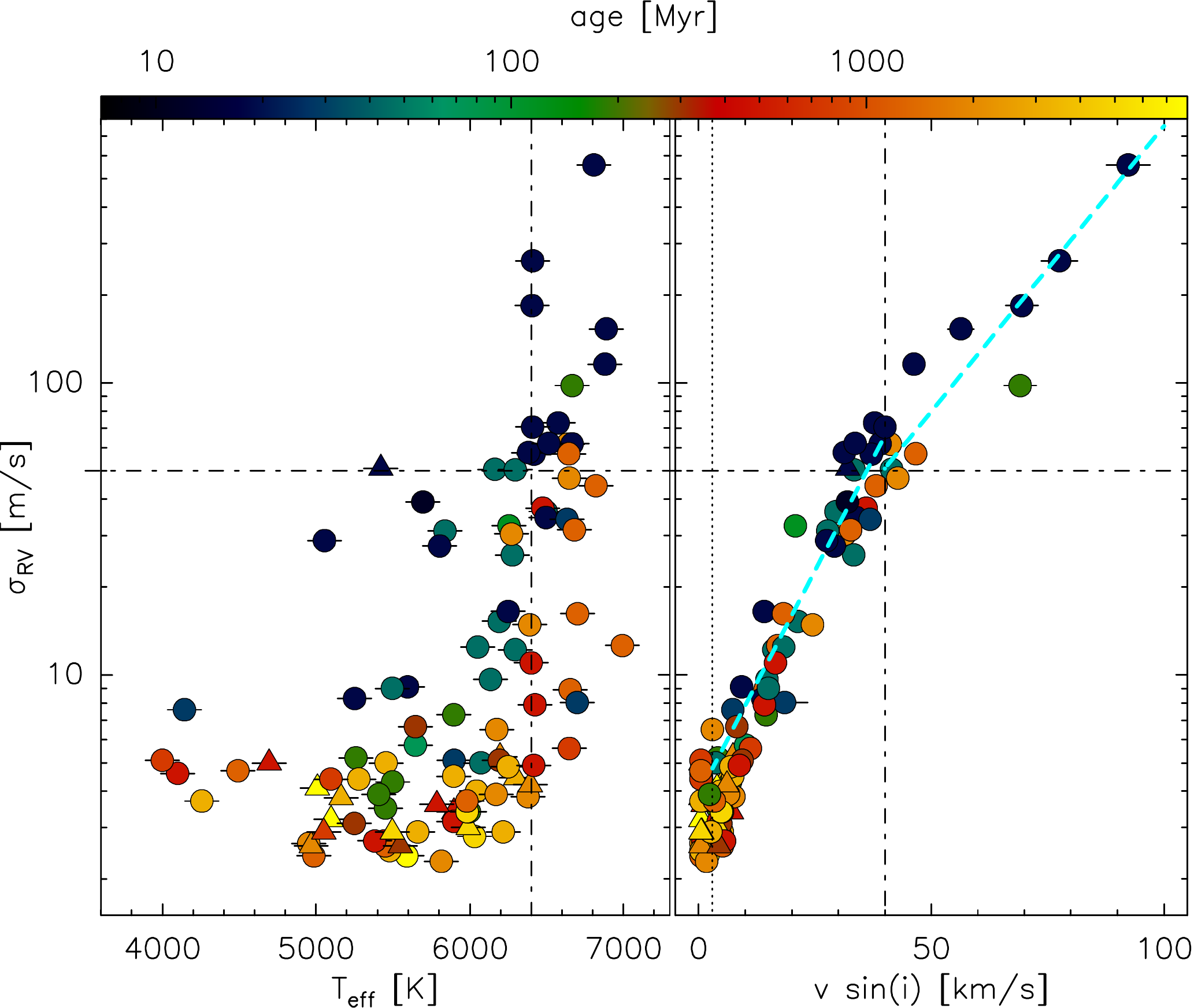}
\centering
\caption{\label{fig:vsini}
Relation between single-measurement RV\ precision, $\sigma_{RV}$, and $T_{\rm eff}$\ ({\it left}) and $v\sin(i)$\ ({\it right}). Stellar ages are colour-coded. Stars marked as triangles do not have significant IR excess (see Sect.\,\ref{ssec:targets:sel}). 
Horizontal dashed-dotted lines mark the anticipated threshold value for the RV precision of $\sigma_{RV}=50$\,m/s for our longer-term survey (Sect.\,\ref{ssec:obs:strat}). Vertical dashed-dotted lines indicate the values of $T_{\rm eff}=6400$\,K (left panel) and $v\sin(i)=40$\,km/s (right panel), above which many target are no longer compatible with our $\sigma_{RV}\leq50$\,m/s single-measurement RV-precision goal. The vertical dotted line on the right panel marks the lower sensitivity limit (3\,km/s) of our $v\sin(i)$\ measurements (Sect.\,\ref{ssec:res:par}). The light-blue dashed lines in the right panel show the best linear fits to the correlation between $v\sin(i)$ and $\log(\sigma_{RV})$\ for $v\sin(i)\leq40$\,km/s and $v\sin(i)\geq40$\,km/s.
}
\end{figure}
%%%%%% 

Here we investigate the relation between stellar effective temperature, $T_{\rm eff}$, projected rotational line broadening, $v\sin(i)$, and achievable RV precision, $\sigma_{RV}$. While $v\sin(i)$\ directly affects the achievable RV precision, $T_{\rm eff}$\ is indirectly related to $v\sin(i)$\ and $\sigma_{RV}$\ via the mechanisms discussed in Sect.\,\ref{sec:motivation}. 
Figure\,\ref{fig:vsini} shows the relation between $\sigma_{RV}$\ and $T_{\rm eff}$, and $v\sin(i)$\ for our target stars as derived from our FEROS spectra. 
There is a clear trend indicating that $\sigma_{RV}$\ increases with increasing $T_{\rm eff}$, although the scatter is large. All stars with \mbox{$T_{\rm eff}\leq6400$\,K} have $\sigma_{RV}\leq50$\,m/s, which is the precision we aim for in our survey (Sect.\,\ref{ssec:obs:strat}). Many stars with higher $T_{\rm eff}$\ have significantly larger $\sigma_{RV}$, and not all of them are very young. This transition is actually close to the $T_{\rm eff}$\ limit above which magnetic braking is no longer efficient (Sect.\,\ref{sec:motivation}).
We used this coarse figure already to pre-select and prioritise those targets for which no values of $v\sin(i)$\ (see below) were initially available from archival spectra or the literature, based on their $T_{\rm eff}$\ obtained from {\it Gaia}\,DR2 \mbox{\citep{gaia_dr2}} or from spectral energy distribution fits (Sect.\,\ref{ssec:targets:list}). 
While stars with \mbox{$T_{\rm eff}\leq6000$\,K} are in general safe targets, and stars with \mbox{$T_{\rm eff}\geq7500$\,K} were not considered at all, selected stars with $6000\,<T_{\rm eff}<7500\,K$\ are scheduled for obtaining test spectra to characterise their $v\sin(i)$\ and $\sigma_{RV}$\ before a decision is made about their inclusion in the survey.

The correlation between $v\sin(i)$\ and $\sigma_{RV}$ is significantly tighter and we derive a relation of \mbox{$\log(\sigma_{RV}[$m/s$]) = a\,(v\sin(i)[$km/s$])+b$} with \mbox{$a=0.071$} and \mbox{$b=1.35$} for \mbox{$v\sin(i)\leq40$\,km/s}, and \mbox{$a=0.045$} and \mbox{$b=2.13$} for \mbox{$v\sin(i)\geq40$\,km/s}. The right panel of Fig.\,\ref{fig:vsini} shows that stars with $v\sin(i)\leq30$\,km/s all have $\sigma_{RV}<50$\,m/s and are thus safe survey targets, while stars with $v\sin(i)>45$\,km/s all have $\sigma_{RV}>50$\,m/s are thus in general incompatible with our survey goals. Stars with intermediate values of $v\sin(i)$\ require a case-by-case inspection of test spectra before a decision is made about their inclusion in the survey for longer-period companions.

%%%%%%%%%%%%%%%%%%%%%%%%%%%%%%%%%%%%

\subsection{Dependence of $v\sin(i)$\ on age and $T_{\rm eff}$} 
\label{ssec:dis:Teff_vsini_age}

%%%%%% Fig. 9
\begin{figure}[htb]
\includegraphics[width=9.0cm]{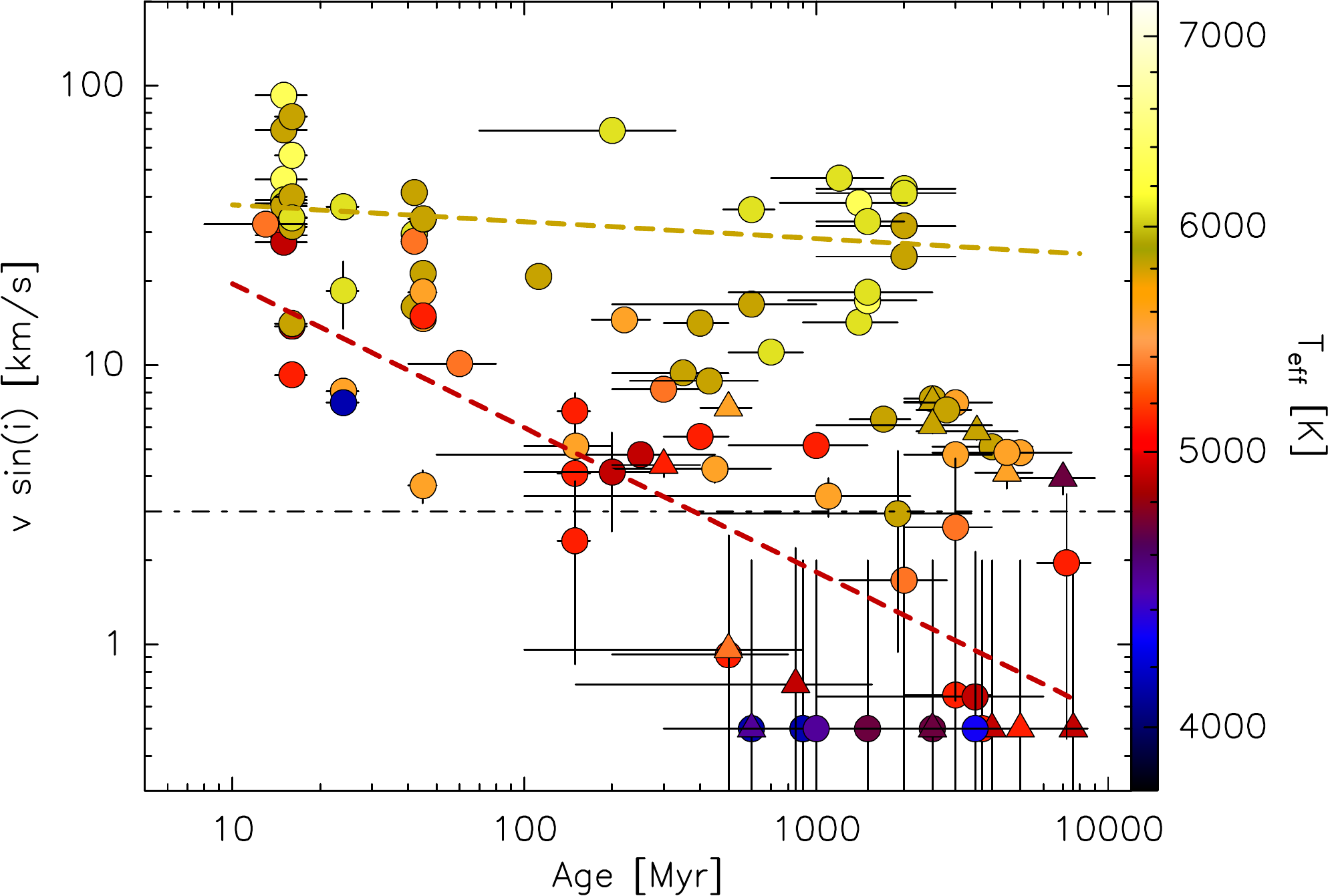}
\centering
\caption{\label{fig:vsini-age}
Relation between $v\sin(i)$, age, and $T_{\rm eff}$\ for stars observed in the RVSPY high-cadence programme. $T_{\rm eff}$\ is coded in colour. 
Stars marked as triangles do not have a significant IR excess (see Sect.\,\ref{ssec:targets:sel}). 
Coloured dashed lines show separate linear fits to the relation between $\log(v\sin(i))$\ and $\log$(age) for stars with \mbox{$T_{\rm eff}\leq6000$\,K} (dark red) and stars with \mbox{$T_{\rm eff}\geq6500$\,K} (dark yellow). The horizontal dashed-dotted line marks the lower sensitivity limit of our $v\sin(i)$\ measurements.
}
\end{figure}
%%%%%%

We use the derived values of $v\sin(i)$\ to analyse their mutual dependence on $T_{\rm eff}$\ and on the stellar age. Figure\,\ref{fig:vsini-age} shows the relation between $v\sin(i)$\ and age for our targets, with $T_{\rm eff}$\ coded in colour. There is a clear correlation between $T_{\rm eff}$, $v\sin(i)$, and age. The hottest and youngest stars have the largest $v\sin(i)$, and the oldest and coolest stars have the smallest $v\sin(i)$. In addition, there seems to be a general $v\sin(i)$\ offset between cool and hot stars. In accordance with the reasoning outlined in Sect.\,\ref{sec:motivation}, we assume that this $v\sin(i)$\ offset is related to the presence or absence of magnetic braking. We therefore perform separate fits to the relation \mbox{$\log(v\sin(i)[$km/s$]) = a\,\log(age[$Myr$])+b$}. For the cooler stars with \mbox{$T_{\rm eff}\leq 6000$\,K}, we derive \mbox{$a=-0.52$} and \mbox{$b=1.81$}, that is, they spin down with age much faster than stars with \mbox{$T_{\rm eff}\geq6500$\,K}, for which magnetic braking is inefficient and for which we derive \mbox{$a=-0.06$} and \mbox{$b=1.63$}.
A comparison between Figs.\,\ref{fig:vsini} and \ref{fig:vsini-age} shows that all targets with \mbox{$T_{\rm eff}<6000$\,K} down to ages of 10\,Myr are suited for RV measurements with the anticipated precision of 50\,m/s. Hotter stars with \mbox{$T_{\rm eff}>6000$\,K} may show too large $v\sin(i)$\ for precise RV measurements when they are younger than 50\,--\,100\,Myr, but each star must be checked individually since the scatter in the $v\sin(i)$ vs. age relation is large.

%%%%%%%%%%%%%%%%%%%%%%%%%%%%%%%%%%%%

\subsection{Dependence of short-term RV jitter on age} 
\label{ssec:dis:rmsrv-age}

%%%%%% Fig. 10
\begin{figure}[htb]
\includegraphics[width=9.0cm]{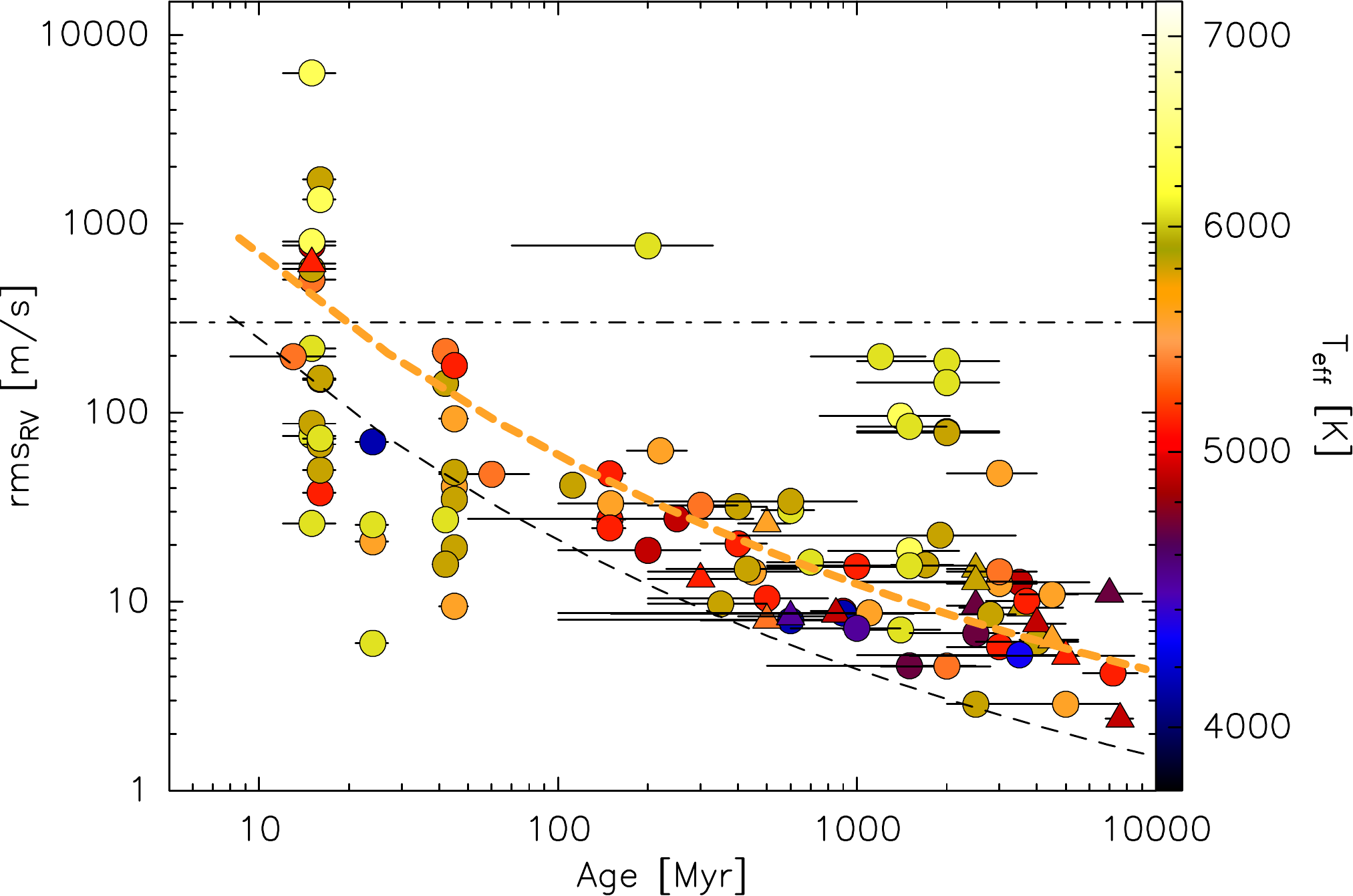}
\centering
\caption{\label{fig:rms-age}
Relation between the intrinsic RV scatter, ${\rm rms}_{\rm RV}(\tau$=14\,d), age, and $T_{\rm eff}$\ for stars observed in the RVSPY high-cadence programme. $T_{\rm eff}$\ is coded in colour. Stars marked as triangles do not have significant IR excess (see Sect.\,\ref{ssec:targets:sel}). 
The black dashed line shows the relation derived by \citet[][their eq.\,(5) with parameters for model (a) from Table\,3]{brems19} for $\tau$=14\,d. The thick orange dashed line shows the same relation with $\epsilon$\ modified from 1.382 to 0.93, adapted to our sample (see text). The horizontal dashed-dotted line marks the approximate boundary above which we consider RV exoplanet searches not feasible (although we use a mass detection threshold in the end; see Sect.\,\ref{sec:dis:prospects} and Fig.\,\ref{fig:Mp_vs_Teff}).
}
\end{figure}
%%%%%%

The median short-term intrinsic RV scatter, ${\rm rms}_{\rm RV}(\tau$=14\,d), of our survey targets is 23\,m/s (Sect.\,\ref{ssec:res:rv}). Figure\,\ref{fig:rms-age} shows the relation between ${\rm rms}_{\rm RV}(\tau$=14\,d), derived in Sect.\,\ref{ssec:res:rv} (see also Fig.\,\ref{fig:rmsrv_teff}), and the stellar age. As expected for young stars with activity-related rotational modulation of line shapes and activity decaying  with age, we find a relatively tight correlation, albeit with a number of outliers that show significantly larger RV jitter. 
These are mostly hot stars with $T_{\rm eff}>6000$\,K, which we have shown in Sect.\,\ref{ssec:dis:Teff_vsini_age} to not spin down but maintain a large $v\sin(i)$\ even at older ages. These stars are not considered here.

For the remaining stars, our ${\rm rms}_{\rm RV}$\,--\,age relation agrees at large with the relation derived for a smaller sample by \mbox{\citet{brems19}}. 

However, our targets exhibit a slightly larger (factor $\sim$2.6) ${\rm rms}_{\rm RV}(\tau$=14\,d) than described by the relation of \mbox{\citet{brems19}}, at least at ages $>50$\,Myr, where the scatter between individual targets is smaller than at younger ages. Adapting $\epsilon$\ in eq.\,(5) of \mbox{\citet{brems19}} from $1.382\pm0.041$\ to 0.93 would describe our sample better. 
This slightly larger mean RV jitter over 14\,d in our data could be related to the larger size of our sample as compared to \citet[][111 vs. 27]{brems19}, to the ages they use (with only two stars in overlap, a direct comparison is not possible), or also to the fact that our ${\rm rms}_{\rm RV}$\ is in many cases derived from more than one high-cadence period, albeit after taking out the mean RV of the individual periods. Furthermore, since all timescales and stellar ages are fit with the same functional form, it is not surprising that discrepancies arise for larger parameter ranges than the ones used to derive this relation originally.

%%%%%%%%%%%%%%%%%%%%%%%%%%%%%%%%%%%%

\subsection{Identifying targets for longer-period planet search}
 \label{sec:dis:prospects}

%%%%%% Fig. 11
\begin{figure}[h!]
\includegraphics[width=8.7cm,angle=0,clip=true]{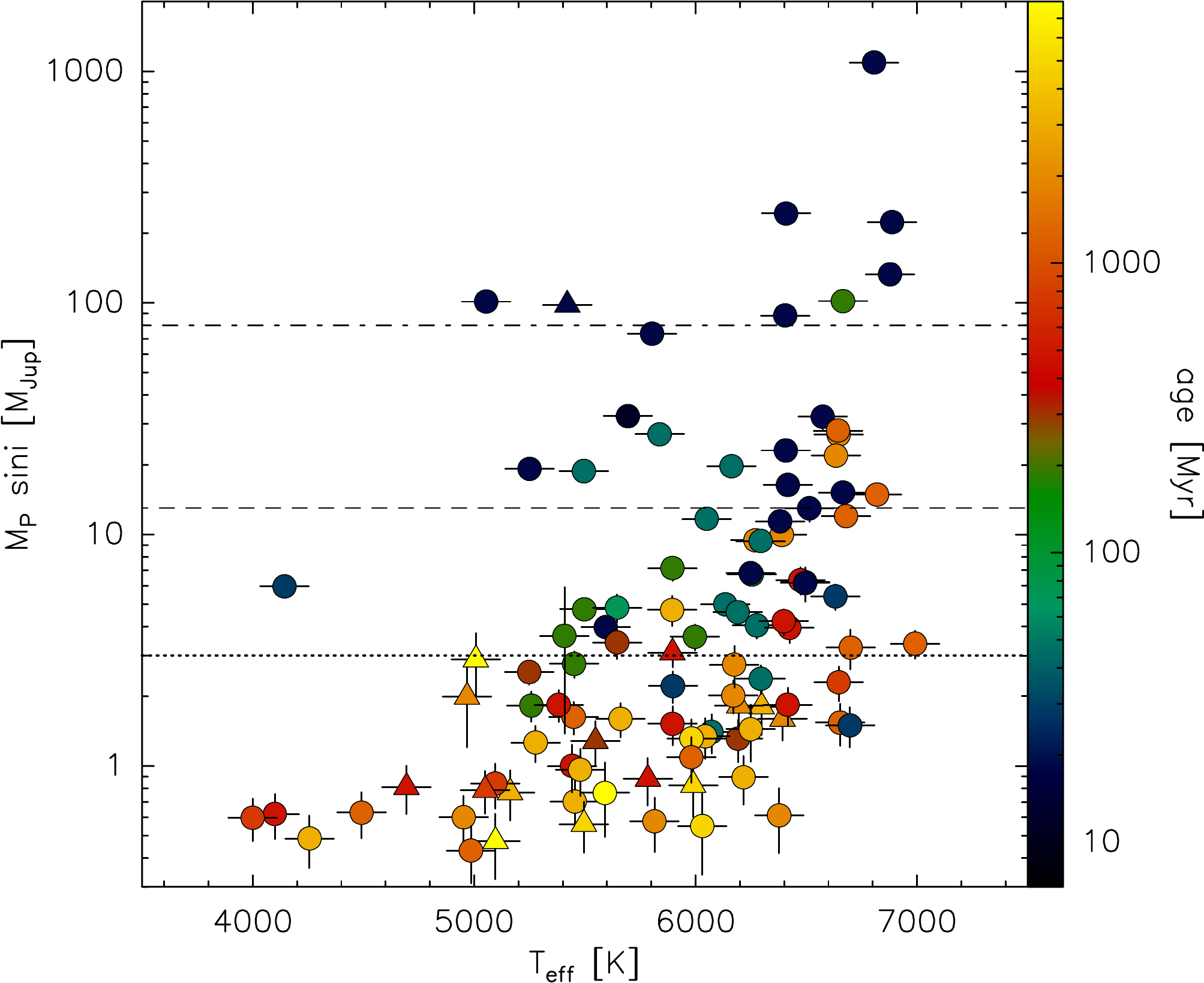}

\centering
\caption{\label{fig:Mp_vs_Teff}
Mass detection limits for hypothetical companions with $P=1$\,yr that induce an RV amplitude three times larger than the activity-jitter rms derived from the high-cadence observations, plotted vs. host star (spectroscopic) $T_{\rm eff}$. Ages are coded in colour. Stars marked as triangles do not have a significant IR excess (see Sect.\,\ref{ssec:targets:sel}). 
The horizontal dashed-dotted line marks the approximate boundary between the low-mass stellar and substellar regimes and the dashed line marks the approximate boundary between the planetary and brown-dwarf mass regimes. The dotted line marks 3\,M$_{\rm Jup}$.
}
\end{figure}
%%%%%%

The observations have already shown that, even with our careful target selection and the strategy for high-cadence observing campaigns, signatures of stellar activity and possible short-period companions are often not easy to disentangle and will require more data. Although we have not identified an HC in our sample from the high-cadence data, the data enable us to select those stars for dedicated follow-up observations that have a low-enough activity-related RV jitter to make searching for longer-period companions ($P\gtrsim$30\,days up to 5\,yrs) feasible.

Figure\,\ref{fig:Mp_vs_Teff} illustrates the feasibility of target selection for such long-term follow-up observations with adapted cadence. We compute for all targets the minimum mass of a hypothetical companion that might still be detectable, assuming that the effective measurement precision is limited by the total RV scatter in the high-cadence data. These masses are computed assuming a 1-year orbital period (as a proxy) and assuming that the required RV semi-amplitude is at least three times larger than the 'activity noise' (${\rm rms}_{\rm RV}(\tau$=14\,d)). 
While we use this simplified assessment to select the targets for the longer-term RV-monitoring, we will try to reduce the observed short-term jitter for the longer-term data by fitting and subtracting the rotational RV modulation where the rotation period is known from TESS and evident in the RV data, as well as by fitting with a gaussian process for correlated noise in the RVs.

Around eight stars, not counting the SB, we could only detect stellar-mass companions with periods of one year. With the exception of HD\,139664 (200\,Myr), all these stars are younger than 20\,Myr. These stars will not be followed up further for longer-period companions. We will also not follow up further the nine SB (Sect.\,\ref{ssec:res:sb}), nor the 18 stars that turned out to show no or only marginal debris disc emission (Sect.\,\ref{ssec:targets:sel}). All these stars are marked in Table\,\ref{tab:periods}. Around the remaining 84 stars, we could potentially detect longer-period brown dwarfs or even GPs (71 stars). While existing archival data allow us to derive constraints on longer-period companions for at least a subset of these 84 stars, the assessment of the longer-term RV variability, based both on archival and on our own FEROS data, will be subject of a subsequent paper.

%%%%%%%%%%%%%%%%%%%%%%%%%%%%%%%%%%%%

\subsection{Comparison with other RV surveys}
 \label{sec:dis:other}

Dedicated RV surveys for planets around stars younger than a few hundred Myr are rare, although various larger surveys do not explicitly exclude young stars. Two of the most recently published  relevant surveys by \mbox{\citet{Grandjean2020}} and \mbox{\citet{Grandjean2021}} targeted 143 young (median age 150\,Myr) nearby (median distance 28\,pc) A0--M5 stars with the HARPS and SOPHIE\footnote{Spectrographe pour l'Observation des Ph\'enom\`enes des Int\'eriours stellaires et des Exoplanetes.} instruments. 
\mbox{\citet{Grandjean2021}} also jointly analysed 120 of the stars in a statistical manner.
Most of their targets are part of the SPHERE GTO programme 'The SpHere INfrared survey for Exoplanets' (SHINE) survey sample \mbox{\citep{chauvin2017}}. Grandjean did not explicitly target stars with debris discs and their sample extends to earlier spectral types (A stars) than our survey, although their median mass (1.0\,M$_{\odot}$) is the same as ours (Sect.\,\ref{ssec:targets:list}). Correspondingly, 16 of their 120 targets are in common with our RVSPY target list.  
Their two surveys adopt slightly different observational strategies for late and early-type stars, but they also tried to sample both the short-term jitter as well as time baselines of several years. However, it is not well-described what the lengths and cadence of their high-cadence observing series is and how well stellar rotation periods and possible HCs are sampled.

\mbox{\citet{Grandjean2021}} measure a median short-term RV scatter of 50\,m/s in their combined surveys, which compares well with the value of 44\,m/s we measure for the younger half of our target list (Sect.\,\ref{ssec:res:rv}, see also Sect.\,\ref{ssec:dis:rmsrv-age}).
No HC with $P<10$\,days was discovered in their surveys, but they discover one longer-period multi-GP system around HD\,113337 \citep{borgniet2014,borgniet2019} and one additional long-period (P>1000\,days) sub-stellar companion candidate in the HD\,206893 planetary system \mbox{\citep[][]{Grandjean2019,romero2021}}. 
In addition, they report three new spectroscopic binaries and confirm the binary nature of 13 stars. Except for HD\,206893, none of their detections overlaps with our RVSPY target list.
From the combined analysis of their two surveys, \mbox{\citet{Grandjean2021}} derive upper limits on the occurrence rates of GPs and BDs with periods between 1 and 1000\,days around young A- to M-type stars of $0.9^{+2.2}_{-0.3}$\% and $0.9^{+2}_{-0.9}$\%, respectively. 
Also in accordance with their analysis is our finding that the RV jitter in several stars hotter than 6500\,K and more massive than 1.3\,M$_{\odot}$\ is dominated by pulsations, while the jitter in lower-mass stars is dominated by activity.

\mbox{\citet{Grandjean2021}} give a 90\% completeness for HC in their survey down to 2-3\,M$_{\rm Jup}$\ and 50\% completeness down to 0.2-0.4\,M$_{\rm Jup}$.
Combining their results with our RVSPY high-cadence survey, we can conclude that in a sample of 135 stars with ages between 10 and 400\,Myr, no HC was detected down a mean mass limit of $\sim$0.5\,M$_{\rm Jup}$, which corresponds to a 68.3\% confidence (1-\,$\sigma$) upper limit on their occurrence rate of 0.9\%. This is still consistent with the HC occurrence rate around older MS stars, which is in the range 0.4-1.2\% (Sect.\,\ref{sec:motivation}). Even the non-detection in all 210 targets of the combined surveys (without the SB), that is, ignoring age and assuming no evolution, would imply an upper limit occurrence rate of HC of 0.6\%, which is still consistent with the HC occurrence rate around older MS stars reported by other studies.

Several large RV surveys for planets around MS stars were carried out and published during the past decade \citep[e.g.][]{santerne2016,Borgniet2017,Borgniet2019b,Quirrenbach2020,Chontos2022}. Not all of these were unbiased, but the most unbiased surveys report occurrence rates of HC in the range 0.4-1.2\%, which if fully consistent with our finding of an 1\,$\sigma$\ upper limit of 0.9\%, that is, the same as for stars younger than 400\,Myr (Sect.\,\ref{ssec:res:rv}) since both age bins contain about half of our targets each.

\section{Summary and conclusions}
 \label{sec:summary}

This paper presents the survey strategy, target list with stellar properties, and the results of the high-cadence reconnaissance observations of our RVSPY. Our survey list contains 111 stars of spectral types between early F and late K at distances between 6 and 160\,pc (median 45\,pc). About half of the targets are also part of the NACO DI survey for planets around young stars \mbox{\citep[NACO-ISPY;][]{launhardt2020}}.

Phase\,1 of the RVSPY survey is dedicated to characterising the target stars and their activity, to search for HCs, and to select the targets for which searching for longer-period companions is feasible.
During the high-cadence survey, all stars are observed in one or two two-week long observing campaigns with one or two spectra each night. All observations are carried out with the FEROS spectrograph ($R=48000$) at the La Silla observatory, in 'object calibration' mode, and with integration times between 6 and 20\,mins. The main results for the stars observed during this first phase of our survey observations are summarised as follows:
\begin{itemize}
\item 
We achieve S/Ns >100 and a median single-measurement RV precision of 6\,m/s. The achievable RV precision strongly degrades with increasing $T_{\rm eff}$ and $v\sin(i)$\ and with decreasing age.
\item 
We derive the stellar parameters $T_{\rm eff}$, $[{\rm Fe/H}]$, $\log(g)$, and $v\sin(i)$\ from the FEROS spectra.
Values of $T_{\rm eff}$\ range from 3900\,K to 7300\,K with a median of 5900\,K. 
Values of $[{\rm Fe/H}]$\ cluster around solar (zero) with only eight stars having $|[{\rm Fe/H}]|>0.4$.
The distribution of surface gravities shows the main-sequence peak around \mbox{$\log(g)\sim4.4$} and a smaller secondary peak around \mbox{$\log(g)\sim3.9$}, which encompasses both the youngest PMS stars as well as a few old (sub-)giant stars.
Values of $v\sin(i)$\ range from <3\,km/s to 90\,km/s with a median of  7.7\,km/s.
The distribution of $v\sin(i)$\ is also bimodal, with the first peak between $v\sin(i)\sim0.5$\ and 20\,km/s encompassing those stars for which magnetic braking is efficient, and the second peak between $v\sin(i)\sim30$\ and 45\,km/s encompassing stars more massive than $\sim1.3$\,M$_{\odot}$\ for which magnetic braking does not act.
\item 
In addition, we derive stellar masses $M_{\ast}$\ and luminosities $L_{\ast}$\ from the SEDs.
Stellar masses range from 0.56 to 2.34\,M$_{\odot}$\ with a median 1.18\,M$_{\odot}$.
Luminosities range from 0.06 to 60\,L$_{\odot}$\ with a median 1.7\,L$_{\odot}$.
\item 
Stellar ages are derived via association with young moving groups (43 stars) and, for the remaining 68 field stars, from a combination of HRD isochronal fits and various literature ages derived with other methods. Our resulting ages range from 10\,Myr to 7.6\,Gyr with a median age of 400\,Myr.
\item We find a clear trend with $v\sin(i)$ decreasing with increasing age, and a bifurcation between stars with \mbox{$T_{\rm eff}>6000$\,K} having significantly larger and slower decreasing $v\sin(i)$ than cooler stars, owing to their ability for magnetic breaking.
\item The median short-term intrinsic RV scatter, ${\rm rms}_{\rm RV}(\tau$=14\,d), of our survey targets is 23\,m/s (44\,m/s for stars younger than 500\,Myr and 10\,m/s for older stars), with values ranging from about 2\,m/s to 1.5\,km/s. The RV scatter for the majority of our targets is caused by stellar activity and/or pulsations (in stars more massive than 1.3\,M$_{\odot}$\ or earlier than F5) and decays with age from >100\,m/s at <20\,Myr to <20\,m/s at >500\,Myr. 
\item 
We analyse time series periodograms of the high-cadence RV data and find  significant periods between 1.3 and 4.5\,d for 14 stars. However, all these RV periodicities are clearly caused by rotational modulation due to starspots and we do not detect an HC with $P<10$\,days in our high-cadence RV data down to a median mass detection limit of $\sim$1\,M$_{\rm Jup}$\ for stars younger than 500\,Myr (0.3\,M$_{\rm Jup}$\ for older stars). Combining our result with the surveys analysed by \mbox{\citet{Grandjean2021}}, we find an upper limit on the occurrence rate of HC around stars younger than 400\,Myr of 0.9\%, which is still consistent with the HC occurrence rate around older MS stars (0.4-1.2\%).
\item We confirm three spectroscopic binary stars: HD\,16673, HD\,27638\,B, HD\,141521 (which were overlooked in the target selection), and report six previously unreported new spectroscopic binary stars with orbital periods between 10 and 100\,days, but no orbit solutions yet. Three of these newly discovered companions have estimated minimum masses in the brown-dwarf regime (HD\,102902, MML\,43, and HD\,129590), the other three in the low-mass-stellar regime (HD\,20759, HD\,108857, and HD\,143811).
\item 
We also analyse the TESS photometric time series data for 91 of our target stars and find significant periodicities in nearly all of them. For 11 stars, the photometric periods are clearly detected also in the RV data. For 18 more stars, the photometric periods are marginally evident in the RV data.
\item
For 91 stars, we derive stellar rotation periods (59 confident and 32 tentative), mostly from TESS data. The majority of our targets have rotation periods between 1 and 10\,d (median 3.9\,d). 
\item From the intrinsic activity-related short-term RV jitter of our target stars, we derive the expected mass-detection thresholds for longer-period companions, and select 84 targets for the ongoing second phase of the survey.

\end{itemize}

The longer-term RV monitoring of our down-selected targets with individually adapted cadences, will go on for at least 2--3 more years.

%%%%%%%%%%%%%%%%%%%%%%%%%%%%%%%%%%%%

\begin{acknowledgements}
The authors thank Didier Queloz, Andreas Quirrenbach, Nestor Espinoza, Raffael Brahm, Damien Ségransan, Carlos Eiroa, Amelia Bayo, Daniela Paz Iglesias Vallejo, Andres Jordan, and Jan Eberhardt for helpful discussions and feedback. O.Z. acknowledges support  within the framework of the Ukraine aid package for individual grants of the Max-Planck Society 2022. A.M.\ acknowledges support of the DFG priority program SPP 1992 ``Exploring the Diversity of Extrasolar Planets” (MU 4172/1-1)''. T.H.\ acknowledges support from the European Research Council under the Horizon 2020 Framework Program via the ERC Advanced Grant Origins 832428. G.M.K. is supported by the Royal Society as a Royal Society University Research Fellow. T.T. acknowledges support by the DFG Research Unit FOR 2544 'Blue Planets around Red Stars' project No. KU 3625/2-1. T.T. further acknowledges support by the BNSF program 'VIHREN-2021' project No.KP-06-DV/5.%КП-06-ДВ/5
This work has made use of data from the European Space Agency (ESA) mission {\it Gaia} (\url{https://www.cosmos.esa.int/gaia}), processed by the {\it Gaia} Data Processing and Analysis Consortium (DPAC, \url{https://www.cosmos.esa.int/web/gaia/dpac/consortium}). Funding for the DPAC has been provided by national institutions, in particular the institutions participating in the {\it Gaia} Multilateral Agreement. This research has also made use of the SIMBAD database and the VizieR catalogue access tool, both operated at CDS, Strasbourg, France. The original description of the VizieR service was published in \mbox{\citet{vizier2000}}.
This paper includes data collected with the TESS mission, obtained from the MAST data archive at the Space Telescope Science Institute (STScI). Funding for the TESS mission is provided by the NASA Explorer Program. STScI is operated by the Association of Universities for Research in Astronomy, Inc., under NASA contract NAS 5–26555.
We also wish to thank the anonymous referee for constructive criticism that helped to improve the clarity of the paper.
\end{acknowledgements}

%%%%%%%%%%%%%%%%%%%%%%%%%%%%%%%%%%%%%%%%%%%%%%%%

\bibliography{RVSPY_introductoryPaper}
\bibliographystyle{aa}

%%%%%%%%%%%%%%%%%%%%%%%%%%%%%%%%%%%%%%%%%%%%%%%%

%\begin{sidewaystable}
\begin{appendix}

%--------------------------------------
%\section{Target list and basic stellar parameters}
%\label{sec:parametersTables}
%The tables~\ref{tab:RVspy_phys_pars} and \ref{tab:RVspy_phys_pars_nodeb} are available at the CDS.

\section{Activity analysis of HD\,38949}
\label{sec:activity:HD38949}

\begin{figure}[htb]
\includegraphics[width=0.48\textwidth]{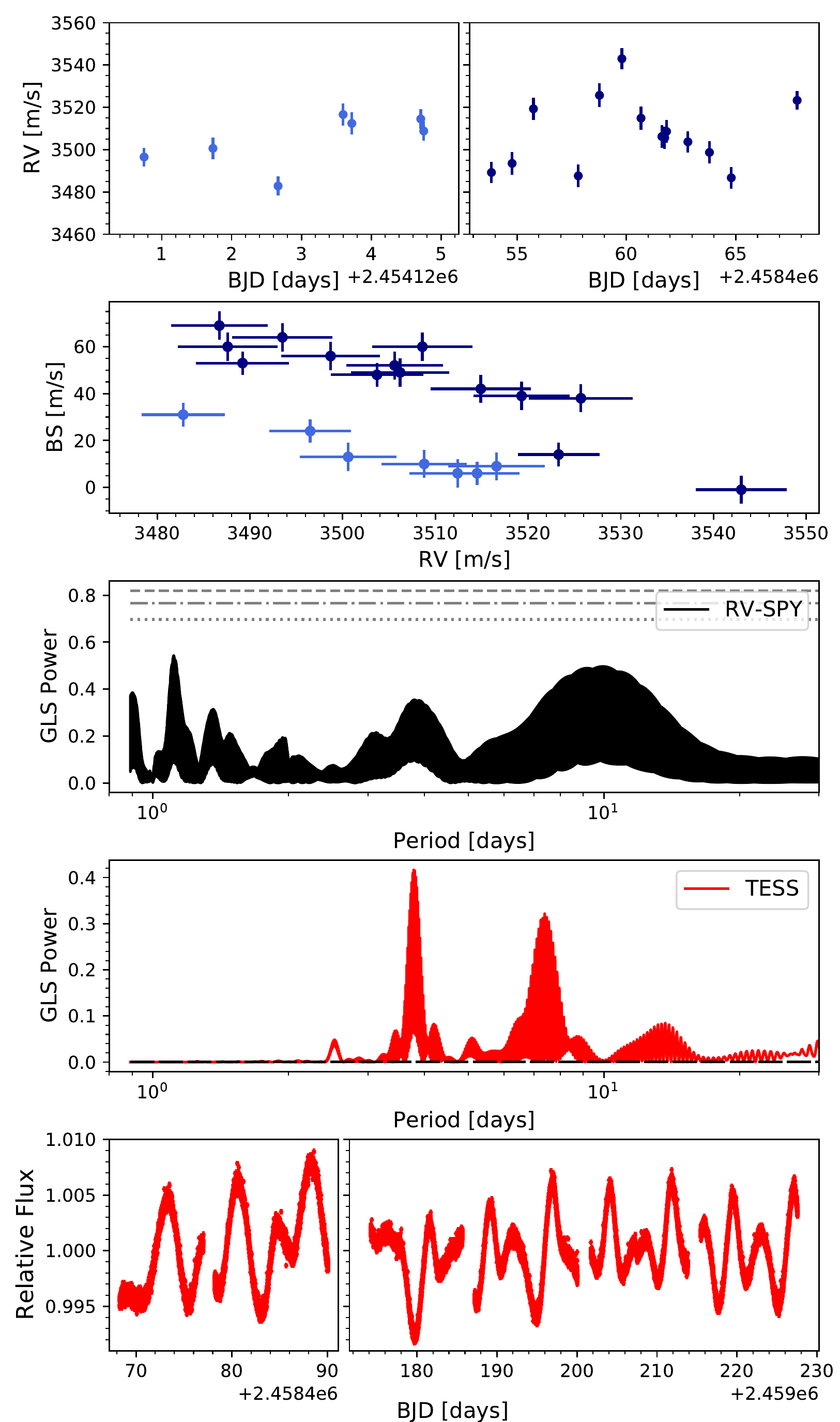}
\centering
\caption{\label{fig:fig_HD38949_RV_BS_TESS}
RVSPY and TESS data for HD\,38949. The panels show (from top to bottom): the RV\ time series, the relation between RV and BS, the GLS periodograms of the RVs and of the photometric data (measured by TESS), and the TESS time series. All data products of the RVSPY programme are shown in black, all TESS data products are shown in red. Horizontal lines in the GLS periodogram of RVs reflect the 0.1\%, 1\% and 10\% false alarm probabilities (from top to bottom). The corresponding false alarm probabilities of the TESS data GLS periodogram are all merged in the dashed line at the bottom of the periodogram, owing to the large number of the TESS data points. 
}
\end{figure}

Here we demonstrate our stellar activity analysis and characterisation (as described in Sect.\,\ref{ssec:red:activity}) which are exemplary for the star HD\,38949. Figure\,\ref{fig:fig_HD38949_RV_BS_TESS} shows (from top to bottom) the RV\ time series, the BS(RV) correlation, the GLS periodograms of the RVs and the photometric TESS\footnote{Obtained from \mbox{\url{https://dx.doi.org/10.17909/t9-h5bx-p296}.}} data, as well as the photometric data time series. The two top panels show the two sets of available high-cadence FEROS data: blue - archival data, navy - RVSPY data. A significant linear correlation ($r_{\rm P}=-0.92$) between the BS and the RV indicates that the observed periodic RV variations are caused by surface features on the rotating star. The GLS periodogram of all available TESS photometric data shows two significant peaks at 3.8\,d and 7.5\,d, of which we assign the latter to the stellar rotation period, and the shorter one to a second dominant spot group on the opposite side of the stellar surface. The rotation period of HD\,38949 was already estimated earlier from photometric measurements by \citet{2011ApJ...743...48W} to be $\approx$7.6\,d, which agrees well with the longer period of the TESS photometry (see Table\,\ref{tab:periods}). 
Two, albeit non-significant, periodogram peaks at similar periods are seen in the combined high-cadence RV data. The two additional and also non-significant peaks around 1\,d are their daily aliases.

\begin{figure*}[!ht]
\centering
\subfloat[]{\label{fig:fig_HD38949_2dglscak}\includegraphics[width=0.333\textwidth]{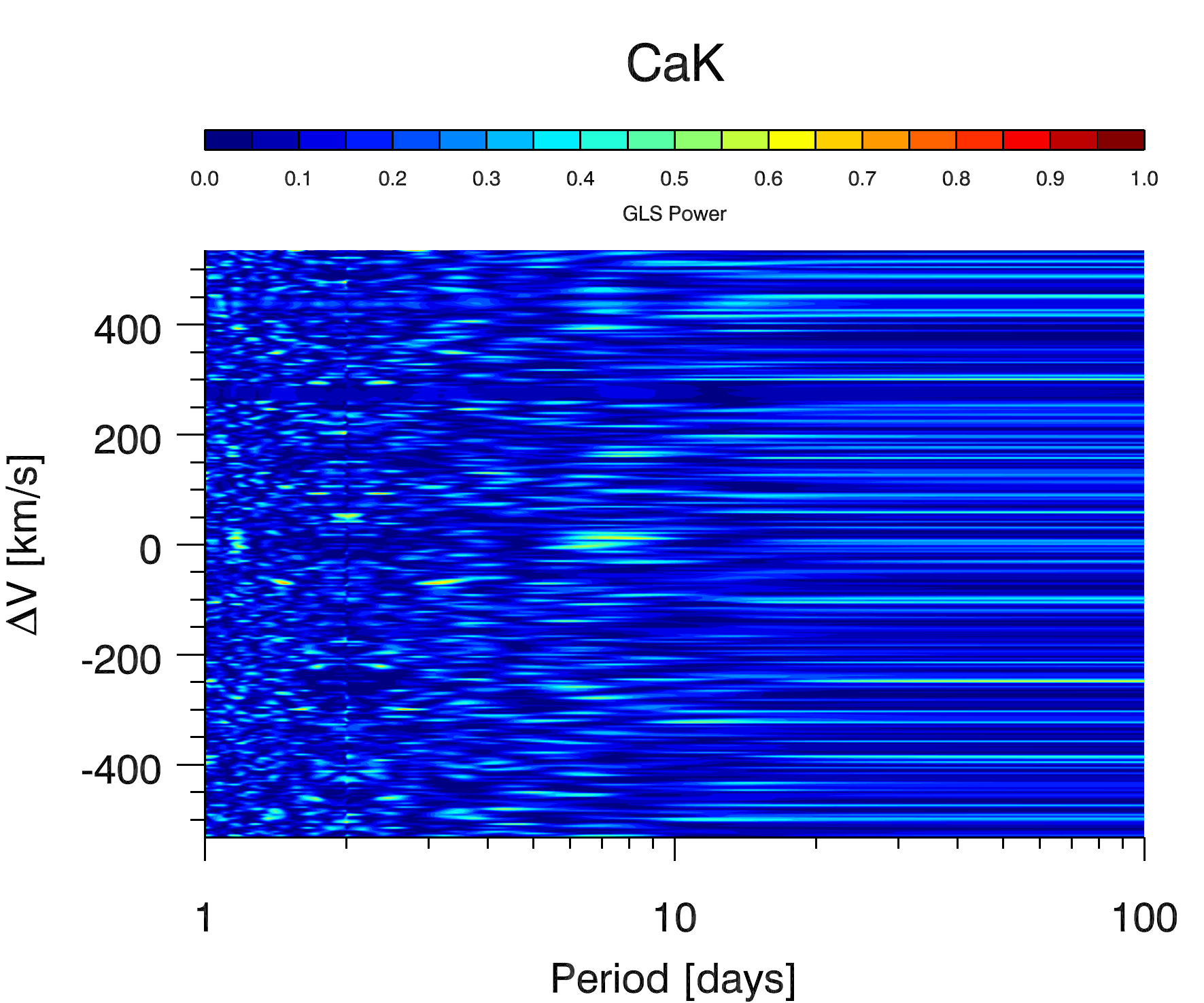}}
\subfloat[]{\label{fig:fig_HD38949_2dglscah}\includegraphics[width=0.333\textwidth]{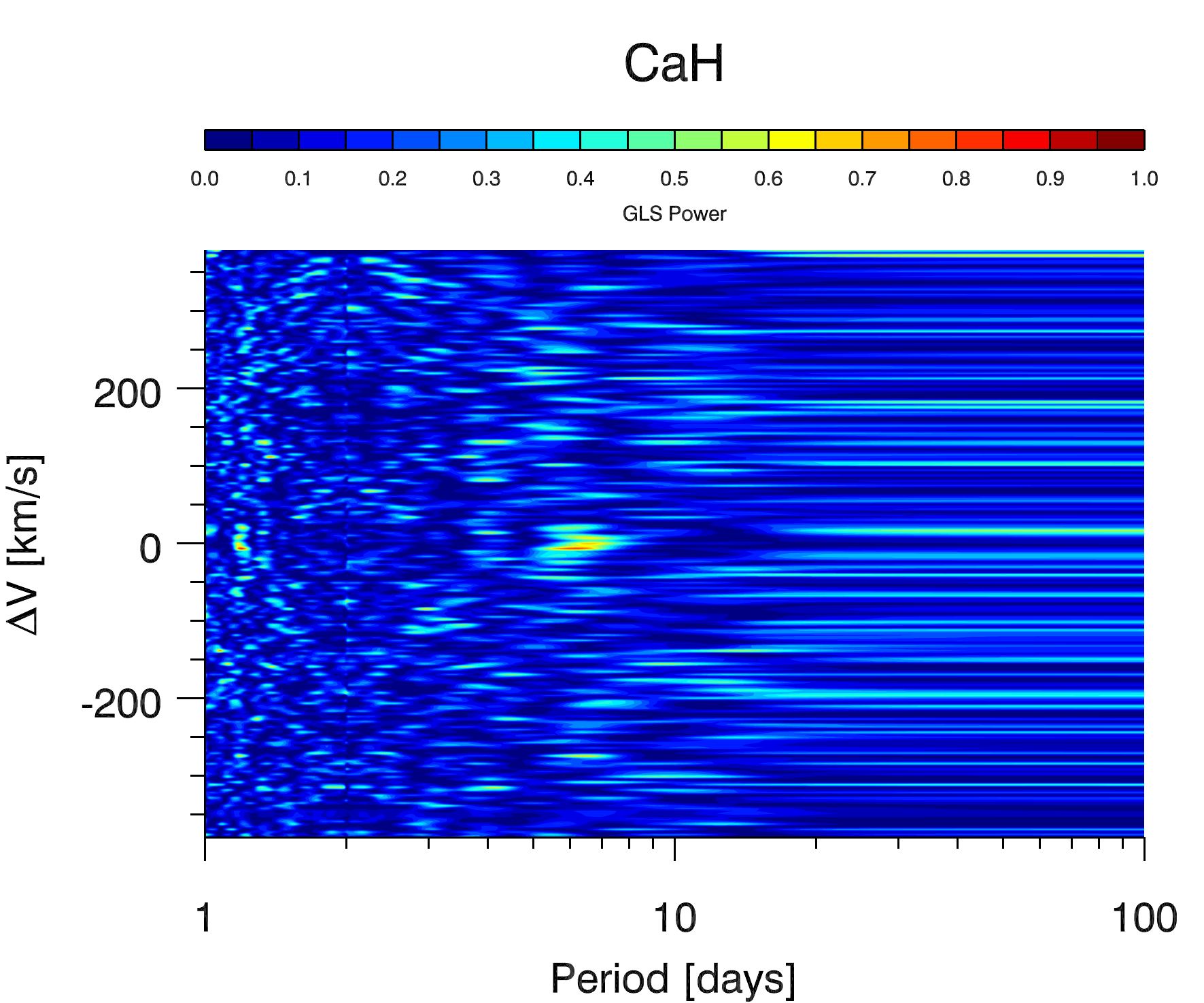}}
\subfloat[]{\label{fig:fig_HD38949_2dglsha}\includegraphics[width=0.333\textwidth]{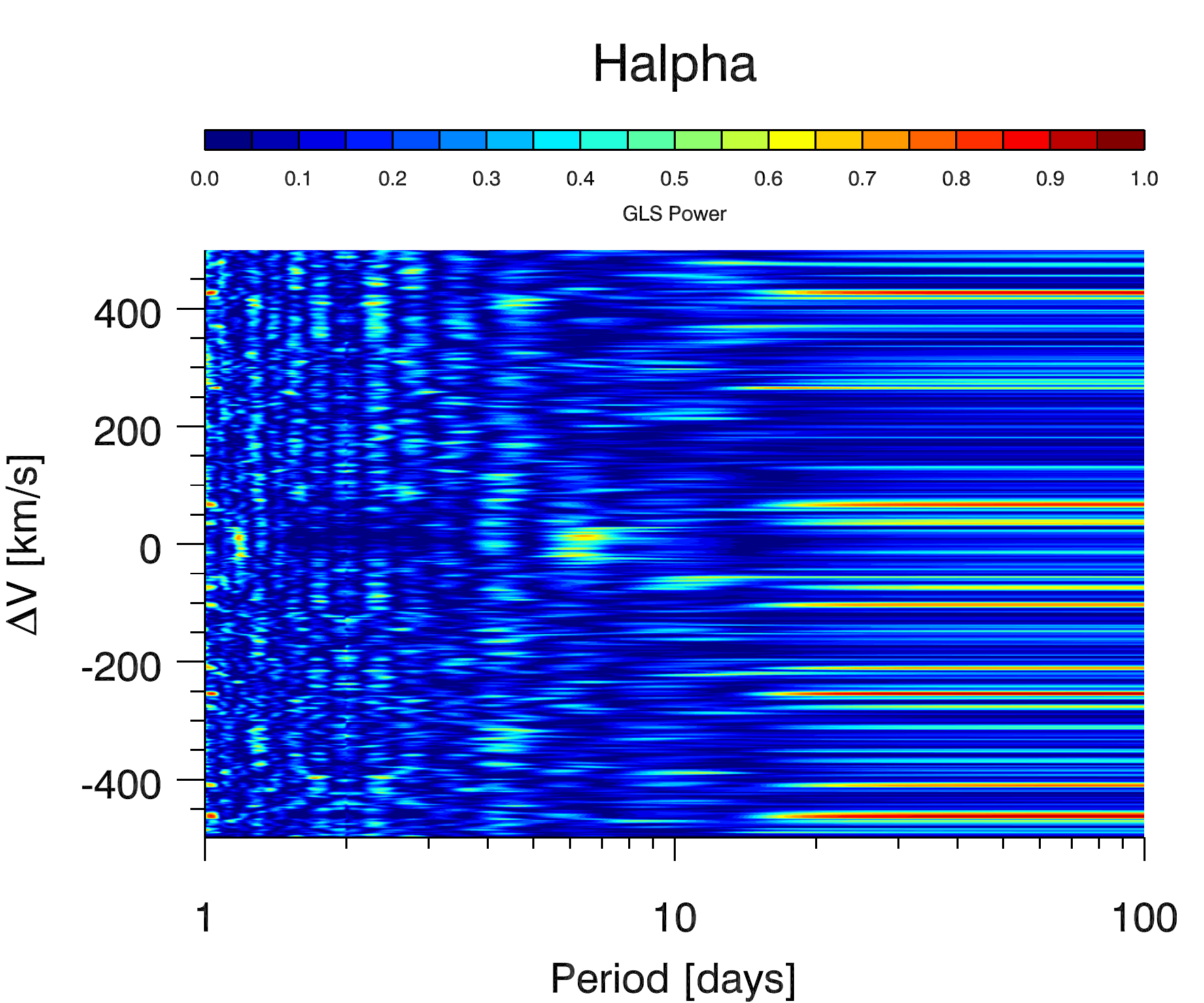}}
\quad
\subfloat[]{\label{fig:fig_HD38949_glswilson}\includegraphics[width=0.333\textwidth]{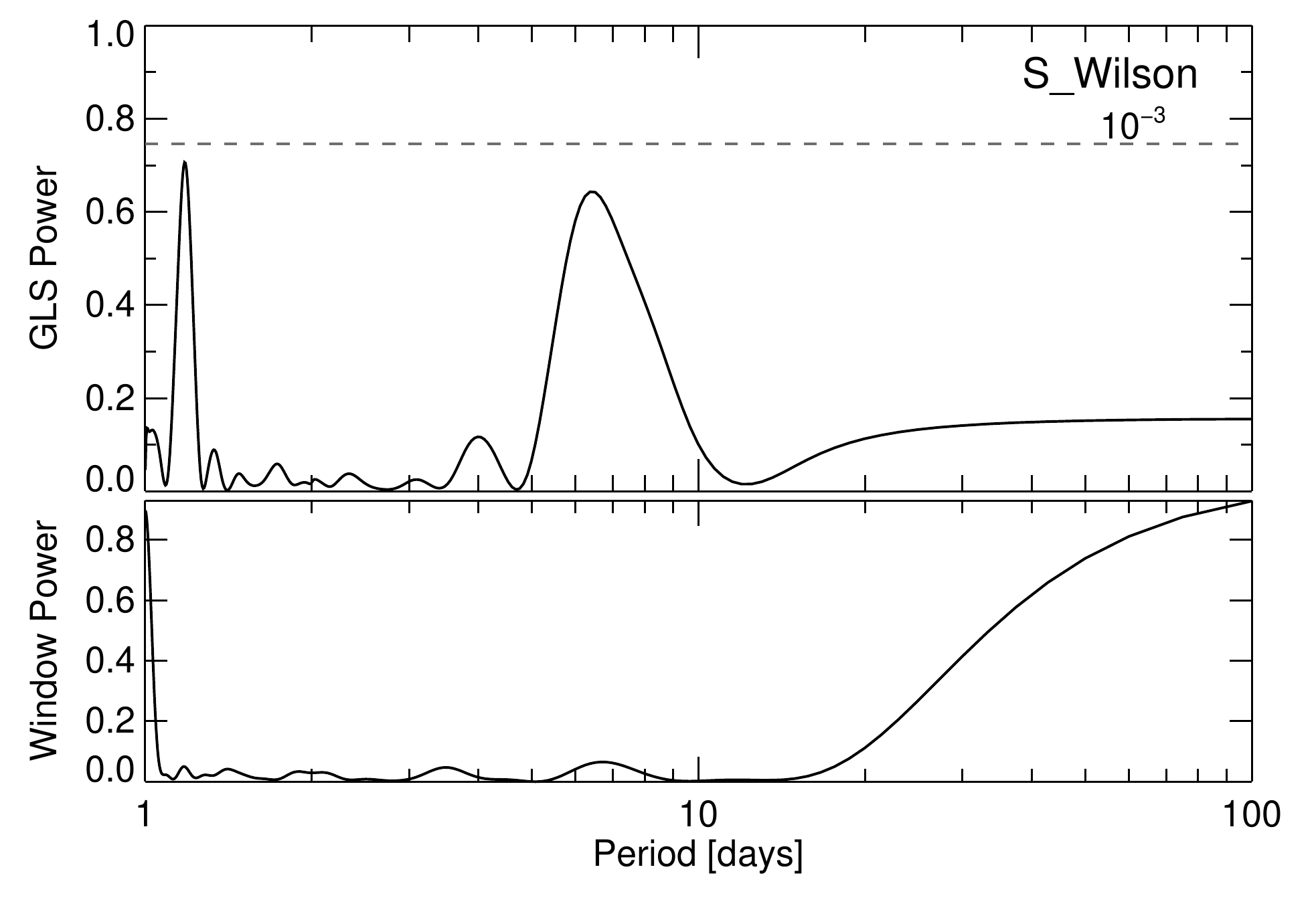}}
\subfloat[]{\label{fig:fig_HD38949_glshaindex}\includegraphics[width=0.333\textwidth]{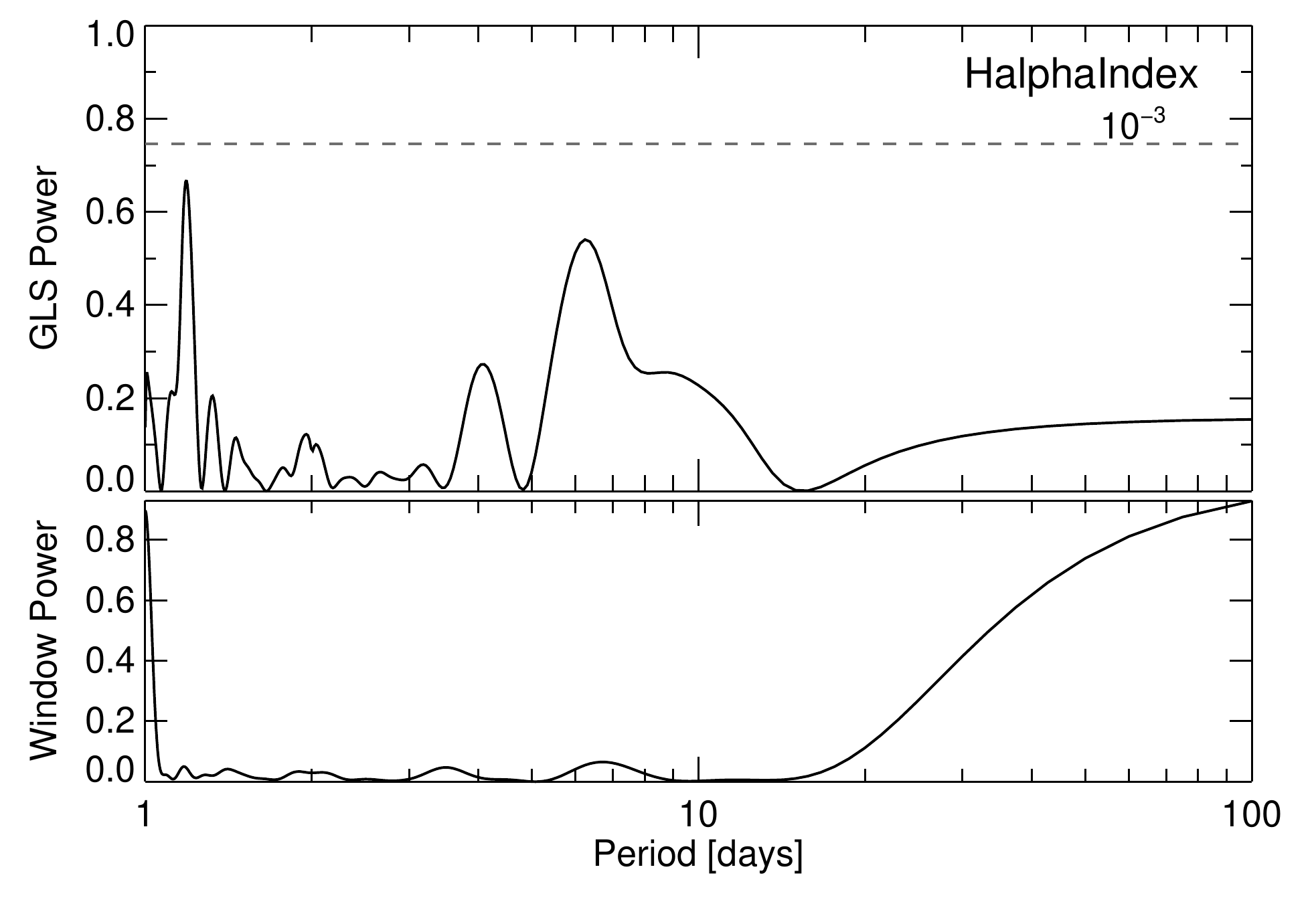}}
\caption{(a\,--\,c) 2D GLS periodograms of Ca\,II K, H, and H$\alpha$. (d) GLS periodogram of the $S$-index based on flux measurements of Ca\,II H\&K. (e) GLS periodogram of the H$\alpha$ index. All three lines display a clear variability of HD\,38949 on a time scale of 6--7\,days.}
\label{fig:2dgls}
\end{figure*}

To further verify the hypothesis that we observe a RV variation which is caused by rotational modulation due to starspots, we compute two-dimensional versions of GLS periodograms for the spectral lines Ca\,II H\&K and H$\alpha$ by shifting all spectra to their rest wavelength and computing a periodogram for each velocity channel  (Figs.\,\ref{fig:fig_HD38949_2dglscak} though \ref{fig:fig_HD38949_2dglsha}). These 2D periodograms show that all three line cores exhibit periodic variability on a time scale of about seven days. This subtle variability is not readily detected in the corresponding line EWs, which are integral quantities describing only the total intensity of the entire line. 
In addition, we also compute the $S$ index and the H$\alpha$ index (Sect.\,\ref{ssec:red:activity}) and show the corresponding GLS periodograms in Figs.\,\ref{fig:fig_HD38949_glswilson} and \ref{fig:fig_HD38949_glshaindex}. Similar to the line cores in the 2D periodograms, both quantities display significant  variability on a time scale of 6--8\,days, which agrees well with the longer of the two periods we find in both the TESS and the RV (albeit not significant) data. 
The additional 1\,d period peaks in both GLS periodograms are the daily aliases of these signals. 
We therefore conclude for HD\,38949 that the observed RV variations are dominated by rotational modulation due to starspots. This analysis also demonstrates the importance of using several indicators and methods simultaneously to identify and quantify processes related to stellar activity.

We perform a similar activity analysis for all stars in our sample in a semi-automatic fashion. Indeed, all 34 out of 40 stars with a clear anti-correlation between RV and BS ($r_{\rm P}\le-0.6$, Sect.\,\ref{ssec:res:rv}) that have TESS data available show significant periodicities in the TESS photometric data (Sect.\,\ref{ssec:res:phot}).

%--------------------------------------

\section{Activity analysis of CPD\,--72\,2713}
\label{sec:activity:CPD-722713}

\begin{figure}[htb]
\includegraphics[width=0.48\textwidth]{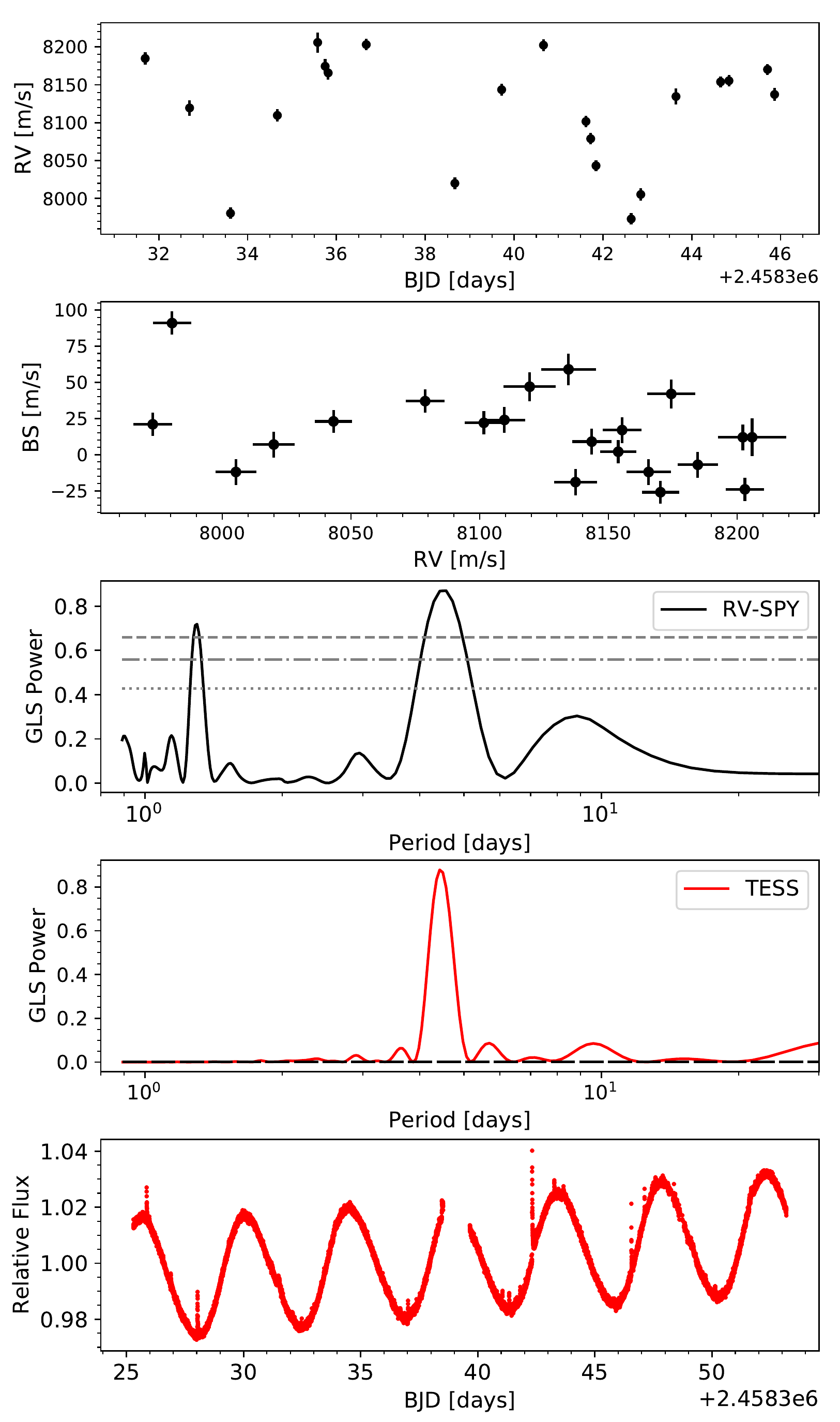}
\centering
\caption{\label{RV_FEROS_activity}
RVSPY and TESS data for CPD\,--72\,2713. The panels show (from top to bottom): the RV\ time series, the relation between RV and BS, the GLS periodograms of the RVs and of the photometric data (measured by TESS), and the TESS time series. All data products of the RVSPY programme are shown in black, all TESS data products are shown in red. Horizontal lines in the GLS periodogram of RVs reflect the 0.1\%, 1\% and 10\% false alarm probabilities (from top to bottom). The corresponding false alarm probabilities of the TESS data GLS periodogram are all merged in the dashed line at the bottom of the periodogram, owing to the large number of the TESS data points.
}
\end{figure}

A good example of a star with an interesting RV periodicity, that is not obviously doomed to be activity-related by a clear anti-correlation between RV and BS, is \mbox{CPD\,--72\,2713}. This star was observed in high cadence in August 2018 and was initially classified as a 'HC candidate' because of a significant RV signal at  \mbox{$P_{\rm RV} = 4.57\pm0.10$\,d} and the absence of a clear anti-correlation between RV and BS. However, a photometric study by \mbox{\citet{messina17}} reports a stellar rotational period of \mbox{$P_{\rm rot} = 4.46$\,d}, which is very close to the RV period we find. The TESS data (released in December 2018) also confirm the presence of a significant signal at \mbox{$P_{\rm phot} = 4.42\pm0.002$\,d}, which is clearly related to the rotational modulation caused by star spots. Hence, the periodic RV modulation is clearly related to the stellar rotation period and not to an orbiting companion. 
Figure\,\ref{RV_FEROS_activity} illustrates (from top to bottom) the RV\ time series, the BS(RV) (non-)correlation, the GLS periodograms of the RVs and the photometric TESS data, as well as the photometric data time series. The additional signal at a period of 1.3\,d in the GLS periodogram of the RV data is an alias of the real signal at 4.57\,d. 

The absence of a clear BS/RV correlation in \mbox{CPD\,--72\,2713} is most likely related to the fact that the inclination of its rotation axis, derived from $P_{\rm rot}$, $R_{\ast}$, and $v\sin(i)$, is $\sim$78$\pm$4\,deg, that is, the star is seen close to pole-on. At such high inclinations, the CCFs are just slightly shifted, but the CCF shape does not change with time, that is, the RVs are affected, but the bisectors are not \citep{desort2007}, which explains the absence of a clear BS/RV correlation.

\end{appendix}

%%%%%%%%%%%%%%%%%%%%%%%%%%%%%%%%%%%%%%%%%%%%%%%%%%%%%%%%%%%%%%%%%%%%%%%%
\onecolumn

%%%%%%%%%%%%%%%%%%%%%%%%%%% Table 2

\setcounter{table}{1}
\begin{landscape}
\begin{longtable}{lllllllllllllllll}
\caption{\label{tab:RVspy_phys_pars}
Target list and basic stellar parameters of debris disc stars}\\
\hline\hline
Name                   & 
dist.\tablefootmark{a} & 
$V$\tablefootmark{b}   & 
SpT\tablefootmark{c}   & 
$T_{\rm eff}^{\rm sp}$\tablefootmark{d} & 
$T_{\rm eff}^{\rm ph}$\tablefootmark{e} & 
$[{\rm Fe/H}]$\tablefootmark{d}   & 
$\log(g)$\tablefootmark{d}  & 
$M_{\ast}$\tablefootmark{f} & 
$L_{\ast}$\tablefootmark{e} & 
$v\sin(i)$\tablefootmark{d} & 
Assoc.\tablefootmark{g}     & 
log($age$)\tablefootmark{i} & 
Ref.                & 
Notes              \\ 
 & [pc] & [mag] & & [K] & [K] & & & [M$_{\odot}$] & [L$_{\odot}$] & [km\,s$^{-1}$] &  & [Myr] & & \\ 
\hline
\endfirsthead
\caption{-- continued from previous page}\\
\hline\hline
Name                & 
dist.\tablefootmark{a} & 
$V$\tablefootmark{b}   & 
SpT\tablefootmark{c}   & 
$T_{\rm eff}^{\rm sp}$\tablefootmark{d} & 
$T_{\rm eff}^{\rm ph}$\tablefootmark{e} & 
$[{\rm Fe/H}]$\tablefootmark{d}    & 
$\log(g)$\tablefootmark{d}   & 
$M_{\ast}$\tablefootmark{f}  & 
$L_{\ast}$ \tablefootmark{e} & 
$v\sin(i)$\tablefootmark{d}  & 
Assoc.\tablefootmark{g}      & 
log($age$)\tablefootmark{h}  & 
Ref.                & 
Notes              \\ 
 & [pc] & [mag] & & [K] & [K] & & & [M$_{\odot}$] & [L$_{\odot}$] & [km\,s$^{-1}$] &  & [Myr] & & \\ 
\hline
\endhead
\hline
\multicolumn{11}{r}{{Continued on next page}} \\
\endfoot
\hline
\endlastfoot
HD\,105     & 38.8  & 7.51 & G0\,V   & 6135 & 5950 &  0.02 & 4.47 & 1.13 & 1.26 & 14.7 & THA      & 1.65 (0.60) & 1 & \\ 
HD\,377     & 38.5  & 7.59 & G2\,V   & 5896 & 5780 &  0.05 & 4.43 & 1.07 & 1.17 & 14.5 & field    & 2.34 (1.70) & 2 & \\ 
HD\,870     & 20.6  & 7.23 & K0\,V   & 5455 & 5360 & -0.14 & 4.53 & 0.87 & 0.51 & <3.0 & field    & 3.48 (3.00) & 2 & \\ 
HD\,1466    & 42.9  & 7.46 & G0\,V   & 6191 & 6120 & -0.06 & 4.33 & 1.15 & 1.58 & 21.3 & THA      & 1.65 (0.60) & 1 & \\ 
HD\,3296    & 44.8  & 6.71 & F6\,V   & 6653 & 6480 &  0.14 & 4.31 & 1.28 & 3.36 & 14.2 & field    & 3.15 (2.70) & 3, 4 & \\ 
HD\,3670    & 77.4  & 8.21 & F5\,V   & 6496 & 6400 &  0.02 & 4.29 & 1.30 & 2.49 & 29.5 & COL\tablefootmark{2} & 1.62 (0.60) & 1 & \\ 
HD\,5133    & 14.0  & 7.17 & K2.5\,V & 4953 & 4900 & -0.23 & 4.52 & 0.78 & 0.29 & <3.0 & field    & 3.40 (3.00) & 2 & \\ 
HD\,7570    & 15.2  & 4.96 & F9\,V   & 6217 & 6220 &  0.20 & 4.28 & 1.25 & 1.95 &  5.1 & field    & 3.60 (3.18) & 2 & \\ 
HIP\,6276   & 35.3  & 8.43 & G9\,V   & 5452 & 5300 & -0.06 & 4.51 & 0.95 & 0.52 &  4.1 & ABDMG    & 2.17 (1.28) & 1 & \\ 
HD\,10008   & 24.0  & 7.66 & K0.5\,V & 5443 & 5280 & -0.06 & 4.57 & 0.88 & 0.47 &  0.9 & BPMG\tablefootmark{3} & 2.70 (2.48) & 1 & \\ 
HD\,13246   & 45.5  & 7.50 & F8\,V   & 6277 & 6160 &  0.04 & 4.38 & 1.19 & 1.72 & 33.3 & THA      & 1.65 (0.60) & 1 & \\ 
HD\,14082B  & 39.7  & 7.74 & G2\,V   & 5898 & 5800 &  0.04 & 4.42 & 0.98 & 1.08 &  8.1 & BPMG     & 1.38 (0.48) & 1 & PMaG2 (3.8\,$\sigma$) \\ 
HD\,16673\tablefootmark{1}   & 21.9  & 5.79 & F8\,V   & 6413 & 6230 &  0.06 & 4.38 & 1.16 & 1.93 &  7.7 & field    & 3.18 (3.00) & 5, 6 & SB1\\ 
HD\,17925   & 10.4  & 6.05 & K1\,V   & 5259 & 5140 &  0.02 & 4.54 & 0.89 & 0.40 &  4.1 & field    & 2.30 (2.00) & 2 & \\ 
HD\,19668   & 38.7  & 8.48 & G8\,V   & 5498 & 5420 & -0.08 & 4.47 & 0.94 & 0.58 &  6.8 & ABDMG    & 2.17 (1.28) & 1 & \\ 
HD\,22484   & 14.0  & 4.30 & F9\,V   & 6031 & 6000 & -0.08 & 3.89 & 1.30 & 3.17 &  4.9 & field    & 3.70 (3.40) & 2 & \\ 
HD\,23356   & 13.9  & 7.10 & K2\,V   & 4986 & 4980 & -0.15 & 4.53 & 0.79 & 0.30 & <3.0 & field    & 3.18 (3.00) & 2 & \\ 
HD\,23484   & 16.2  & 7.00 & K2\,V   & 5278 & 5130 &  0.02 & 4.57 & 0.89 & 0.41 &  0.6 & field    & 3.54 (3.40) & 2 & \\ 
BD+23\,551  & 140.4 & 10.1 & F9\,V   & 6255 & 5960 & -0.02 & 4.39 & 1.21 & 1.49 & 20.8 & PLE      & 2.05 (0.70) & 7 & \\ 
HD\,24649   & 41.5  & 7.22 & K2\,V   & 6271 & 6210 & -0.04 & 4.33 & 1.05 & 1.83 & 31.5 & field    & 3.30 (3.00) & 6 & \\ 
HD\,27638\,B\tablefootmark{1}& 86.5  & 8.47 & F8\,V   & 6096 & 5770 &  0.04 & 4.36 & 1.34 & 2.87 &  2.6 & field    & 2.30 (2.00) & 8 & SB2 \\ 
HD\,28069   & 47.8  & 7.35 & F7\,V   & 6390 & 6210 &  0.17 & 4.37 & 1.18 & 2.18 & 24.5 & field    & 3.30 (3.00) & 6 & \\ 
HD\,28447   & 40.5  & 6.51 & G5\,V   & 5591 & 5490 & -0.15 & 3.84 & 1.50 & 3.69 & <3.0 & field    & 3.86 (3.18) & 6, 14 & PMaG2 (29\,$\sigma$) \\ 
HD\,31392   & 25.7  & 7.61 & G9\,V   & 5479 & 5310 &  0.02 & 4.52 & 0.88 & 0.56 & <3.0 & field    & 3.57 (3.00) & 3, 14 & \\ 
HD\,33081   & 53.8  & 7.04 & F7\,V   & 6378 & 6340 & -0.06 & 3.93 & 1.21 & 3.62 &  7.6 & field    & 3.40 (2.70) & 14 & \\ 
SAO\,150676 & 73.0  & 8.96 & G2\,V   & 5838 & 5630 &  0.00 & 4.38 & 1.20 & 1.15 & 27.7 & COL      & 1.62 (0.60) & 1 & \\ 
HD\,38397   & 53.7  & 8.14 & G0\,V   & 6295 & 5960 &  0.24 & 4.33 & 1.13 & 1.36 & 16.2 & COL      & 1.62 (0.60) & 1 & \\ 
HD\,40136   & 14.5  & 3.71 & F1\,V   & 6992 & 7030 &  0.02 & 3.90 & 1.42 & 5.54 & 17.0 & field    & 3.18 (2.85) & 2 & \\ 
HD\,43989   & 51.9  & 7.95 & G0\,V   & 6163 & 5930 &  0.06 & 4.33 & 1.23 & 1.57 & 41.5 & COL      & 1.62 (0.60) & 1 & \\ 
HD\,48370   & 36.0  & 7.91 & G5\,V   & 5646 & 5580 &  0.10 & 4.39 & 0.96 & 0.78 & 10.1 & field    & 1.78 (1.30) & 2 & \\ 
HD\,50571   & 34.0  & 6.10 & F5\,V   & 6647 & 6560 &  0.24 & 4.34 & 1.32 & 3.36 & 42.8 & field    & 3.30 (3.00) & 9 & \\ 
HD\,53143   & 18.4  & 6.81 & G9\,V   & 5450 & 5410 &  0.10 & 4.48 & 0.99 & 0.59 &  5.2 & field    & 3.00 (2.70) & 2 & \\ 
HD\,57703   & 43.8  & 6.77 & F2\,E   & 6474 & 6560 & -0.06 & 3.89 & 1.25 & 3.02 & 36.0 & field    & 2.78 (2.08) & 14 & \\ 
HD\,59659   & 87.4  & 8.79 & F7\,V   & 6295 & 6150 &  0.06 & 4.32 & 1.27 & 1.92 & 33.4 & ARG      & 1.65 (0.70) & 10 & \\ 
HD\,59967   & 21.8  & 6.64 & G3\,V   & 5896 & 5790 & -0.08 & 4.44 & 0.98 & 0.89 &  4.3 & field    & 2.65 (2.40) & 2 & \\ 
HD\,72687   & 45.4  & 8.26 & G5\,V   & 5895 & 5660 & -0.02 & 4.53 & 0.93 & 0.89 &  7.3 & field    & 3.48 (3.00) & 14 & \\ 
HD\,76151   & 16.8  & 6.00 & G3\,V   & 5815 & 5830 &  0.06 & 4.39 & 1.07 & 0.97 & <3.0 & field    & 3.30 (2.90) & 2 & \\ 
HD\,76653   & 24.3  & 5.70 & F6\,V   & 6647 & 6430 &  0.20 & 4.42 & 1.27 & 2.50 & 11.1 & field    & 2.85 (2.30) & 14 & \\ 
CD-49\,3972 & 139.0 & 10.2 & F5\,E   & 6072 & 5720 &  0.02 & 4.37 & 1.24 & 1.36 &  3.7 & ARG      & 1.65 (0.70) & 10 & \\ 
HD\,84075   & 64.0  & 8.59 & G2\,V   & 6051 & 5840 &  0.04 & 4.42 & 1.18 & 1.32 & 18.3 & ARG      & 1.65 (0.70) & 10 & \\ 
HD\,90905   & 31.0  & 6.88 & G1\,V   & 6192 & 6000 &  0.02 & 4.43 & 1.13 & 1.43 &  9.4 & field    & 2.54 (2.18) & 2 & \\ 
HD\,92945   & 21.5  & 7.72 & K1\,V   & 5249 & 5150 & -0.04 & 4.53 & 0.87 & 0.37 &  4.8 & field    & 2.40 (2.30) & 3 ,8, 11, 14, 15 & \\ 
HD\,102458  & 113.0 & 9.07 & G5\,V   & 5804 & 5680 &  0.00 & 4.12 & 1.38 & 2.69 & 29.2 & LCC      & 1.18 (0.48) & 12 & \\ 
HD\,104231  & 102.4 & 8.54 & F5\,V   & 6495 & 6310 &  0.14 & 3.93 & 1.35 & 3.21 & 33.8 & LCC      & 1.18 (0.48) & 12 & \\ 
HD\,105912  & 48.2  & 6.94 & F5\,V   & 6635 & 6700 &  0.04 & 4.29 & 1.32 & 3.08 & 41.2 & field    & 3.30 (3.00) & 9 & \\ 
MML\,8      & 112.9 & 10.5 & K0\,V   & 5055 & 4720 & -0.08 & 4.05 & 1.25 & 0.91 & 27.5 & LCC      & 1.18 (0.48) & 12 & \\ 
HD\,107146  & 27.5  & 7.04 & G2\,V   & 5997 & 5830 & -0.04 & 4.48 & 1.03 & 1.00 &  5.1 & field    & 2.18 (1.70) & 2 & PMaG2 (3.4\,$\sigma$) \\ 
HD\,107649  & 108.0 & 8.78 & F5\,V   & 6405 & 6290 &  0.04 & 4.03 & 1.38 & 2.83 & 69.4 & LCC      & 1.18 (0.48) & 12 & \\ 
HD\,108857\tablefootmark{1}  & 104.2 & 8.60 & F7\,V   & 6271 & 5800 & -0.05 & 4.20 & 1.49 & 3.58 &  4.7 & LCC      & 1.18 (0.48) & 12 & SB1, PMaG2 (3.5\,$\sigma$) \\ 
HD\,109832  & 108.0 & 8.09 & A9\,V   & 6879 & 6930 &  0.02 & 3.83 & 1.55 & 5.34 & 46.3 & LCC      & 1.18 (0.48) & 12 & \\ 
HD\,111520  & 108.6 & 8.87 & F5\,V   & 6575 & 6140 &  0.05 & 4.31 & 1.32 & 2.70 & 37.8 & LCC      & 1.18 (0.48) & 12 & \\ 
HD\,111631  & 10.6  & 8.49 & M1\,V   & 4100 & 3985 & -0.18 & 4.83 & 0.64 & 0.10 & <3.0 & field    & 2.78 (2.48) & 2 & \\ 
HD\,114082  & 95.4  & 8.21 & F3\,V   & 6600 & 6000 &  0.01 & 4.17 & 1.47 & 3.80 & 39.2 & LCC      & 1.18 (0.48) & 12 & \\ 
HD\,115820  & 115.5 & 7.95 & A7\,V   & 6807 & 7480 & -0.33 & 2.49 & 1.64 & 6.97 & 92.3 & LCC      & 1.18 (0.48) & 12 & \\ 
HD\,117214  & 107.3 & 8.06 & F6\,V   & 6417 & 6360 &  0.00 & 3.66 & 1.48 & 5.57 & 37.1 & LCC      & 1.18 (0.48) & 12 & \\ 
MML\,36     & 99.5  & 10.1 & K0\,IV  & 5251 & 4980 &  0.02 & 4.23 & 1.21 & 0.99 & 13.8 & UCL      & 1.20 (0.30) & 12 & \\ 
HD\,118972  & 15.7  & 6.92 & K0\,V   & 5382 & 5170 &  0.38 & 4.57 & 0.85 & 0.41 &  5.6 & field    & 2.60 (2.00) & 14, 15 & \\ 
CD-29\,10609& 110.9 & 10.4 & G5      & 5595 & 5440 & -0.04 & 4.39 & 0.97 & 0.86 &  9.2 & UCL      & 1.20 (0.30) & 12 & \\ 
HD\,125451  & 26.2  & 5.41 & F5\,IV  & 6820 & 6720 &  0.12 & 4.33 & 1.33 & 3.81 & 38.1 & field    & 3.15 (2.81) & 2 & \\ 
MML\,43\tablefootmark{1}     & 120.0 & 10.6 & G9\,V   & 4895 & 5030 & -0.15 & 3.93 & 1.11 & 0.84 & 10.4 & UCL      & 1.20 (0.30) & 12 & SB1 \\ 
HD\,129590\tablefootmark{1}  & 135.5 & 9.33 & G3\,V   & 6483 & 5820 & -0.09 & 4.37 & 1.37 & 2.98 & 19.1 & UCL      & 1.20 (0.30) & 12 & SB2 \\ 
HD\,134910  & 143.3 & 9.53 & G0\,V   & 6249 & 6060 &  0.00 & 4.24 & 1.23 & 2.62 & 14.1 & UCL      & 1.20 (0.30) & 12 & \\ 
HD\,135953  & 130.0 & 9.36 & F5\,V   & 6382 & 6120 &  0.04 & 4.29 & 1.21 & 2.53 & 31.3 & UCL      & 1.20 (0.30) & 12 & \\ 
HD\,139664  & 17.4  & 4.64 & F5\,IV  & 6666 & 6710 & -0.01 & 4.28 & 1.25 & 3.37 & 69.1 & field    & 2.30 (2.11) & 2 & \\ 
HD\,141011  & 128.1 & 8.97 & F5\,V   & 6409 & 6410 &  0.00 & 3.84 & 1.33 & 3.47 & 77.5 & UCL      & 1.20 (0.30) & 12 & \\ 
HD\,143811\tablefootmark{1}  & 136.6 & 8.91 & F5\,V   & 6632 & 6220 & -0.12 & 4.23 & 1.39 & 4.22 & 12.3 & USCO\tablefootmark{2} & 1.00 (0.48) & 12 & SB2 \\ 
HD\,145229  & 33.7  & 7.44 & G0\,D   & 6044 & 5920 & -0.15 & 4.42 & 0.97 & 1.04 &  4.8 & field    & 3.48 (3.00) & 14 & \\ 
HD\,145560  & 120.0 & 8.90 & F5\,V   & 6408 & 6330 &  0.02 & 4.03 & 1.30 & 3.23 & 40.0 & UCL      & 1.20 (0.30) & 12 & \\ 
HD\,145972  & 125.4 & 8.40 & F0\,V   & 6888 & 6990 &  0.16 & 3.82 & 1.56 & 5.33 & 56.3 & UCL      & 1.20 (0.30) & 12 & \\ 
HD\,146181  & 124.5 & 9.16 & F6\,V   & 6514 & 6380 &  0.00 & 4.28 & 1.28 & 2.70 & 33.6 & UCL      & 1.20 (0.30) & 12 & \\ 
HD\,147594  & 133.8 & 9.27 & G3\,IV  & 5695 & 5590 &  0.00 & 4.02 & 1.52 & 3.07 & 32.0 & UCL/USCO & 1.11 (0.70) & 12 & \\ 
HD\,166348  & 13.2  & 8.38 & K7\,V   & 4257 & 4300 & -0.26 & 4.68 & 0.72 & 0.13 & <3.0 & field    & 3.54 (3.40) & 2 & \\ 
HD\,170773  & 37.0  & 6.22 & F5\,V   & 6644 & 6590 &  0.22 & 3.96 & 1.29 & 3.58 & 46.7 & field    & 3.08 (2.70) & 9, 14 & \\ 
HD\,180134  & 45.6  & 6.36 & F7\,V   & 6249 & 6230 & -0.15 & 3.88 & 1.39 & 4.93 &  6.9 & field    & 3.45 (2.78) & 14 & \\ 
HD\,181327  & 48.2  & 7.04 & F6\,V   & 6698 & 6430 &  0.20 & 4.38 & 1.41 & 2.87 & 18.5 & BPMG     & 1.38 (0.48) & 1 & \\ 
HD\,183216  & 35.8  & 7.14 & G2\,V   & 6171 & 6040 &  0.20 & 4.42 & 1.18 & 1.50 &  6.4 & field    & 3.23 (2.60) & 14 & \\ 
HD\,187897  & 32.3  & 7.13 & G5\,D   & 5982 & 5880 &  0.10 & 4.38 & 1.09 & 1.23 &  4.9 & field    & 3.65 (3.00) & 14 & \\ 
HD\,190470  & 22.1  & 7.77 & K3\,V   & 5096 & 5030 &  0.04 & 4.52 & 0.80 & 0.38 & <3.0 & field    & 2.95 (2.30) & 5, 16 & \\ 
HD\,191089  & 50.1  & 7.18 & F5\,V   & 6632 & 6460 &  0.08 & 4.35 & 1.20 & 2.74 & 36.8 & BPMG     & 1.38 (0.48) & 1 & \\ 
HD\,191849  & 6.2   & 7.97 & M0\,V   & 4000 & 3832 & -0.52 & 4.79 & 0.56 & 0.06 & <3.0 & field    & 2.95 (2.60) & 5, 13 & \\ 
HD\,199260  & 21.3  & 5.70 & F6\,V   & 6424 & 6280 & -0.04 & 4.33 & 1.14 & 1.97 & 14.1 & field    & 2.60 (2.00) & 2 & PMaG2 (3.6\,$\sigma$) \\ 
HD\,201219  & 37.9  & 8.00 & G5\,D   & 5661 & 5600 &  0.06 & 4.43 & 1.03 & 0.79 & <3.0 & field    & 3.48 (3.00) & 14 & \\ 
HD\,202917  & 46.8  & 8.65 & G7\,V   & 5496 & 5440 & -0.08 & 4.34 & 1.00 & 0.69 & 15.0 & THA      & 1.65 (0.60) & 1 & \\ 
HD\,206893  & 40.8  & 6.67 & F5\,V   & 6680 & 6550 &  0.08 & 4.34 & 1.26 & 2.85 & 32.7 & field    & 3.18 (2.70) & 9 & PMaG2 (3.4\,$\sigma$), planets! \\ 
HD\,209253  & 31.4  & 6.63 & F6.5\,V & 6399 & 6220 &  0.00 & 4.44 & 1.12 & 1.70 & 16.5 & field    & 2.78 (2.60) & 2 & \\ 
HD\,212695  & 48.1  & 6.95 & F3\,V   & 6701 & 6500 &  0.02 & 4.27 & 1.37 & 3.13 & 18.2 & field    & 3.18 (3.00) & 6, 9 & PMaG2 (3.4\,$\sigma$) \\ 
CPD-72\,2713& 36.6  & 10.6 & K7\,V   & 4144 & 3910 & -0.47 & 3.41 & 0.80 & 0.19 &  7.3 & BPMG     & 1.38 (0.48) & 1 & \\ 
HD\,218340  & 56.1  & 8.44 & G3\,V   & 5981 & 5840 &  0.08 & 4.38 & 1.09 & 1.16 &  3.4 & field    & 3.04 (3.00) & 2 & \\ 
HD\,218511  & 14.8  & 8.29 & K5\,V   & 4492 & 4380 & -0.02 & 4.52 & 0.69 & 0.16 & <3.0 & field    & 3.00 (2.60) & 2 & \\ 
HD\,219498  & 57.0  & 9.05 & G5\,D   & 5646 & 5540 & -0.11 & 4.44 & 0.98 & 0.70 &  8.2 & field    & 2.48 (2.00) & 2 & \\ 
HD\,219482  & 20.5  & 5.65 & F6\,V   & 6416 & 6230 &  0.06 & 4.43 & 1.11 & 1.90 &  8.8 & field    & 2.63 (2.30) & 2 & \\ 
HD\,223340  & 44.2  & 9.28 & K1\,V   & 5408 & 5240 & -0.02 & 4.59 & 1.39 & 0.44 & <3.0 & ABDMG    & 2.17 (1.28) & 1 & \\ 
CD-44\,15399& 159.9 & 10.5 & F9      & 6175 & 5920 & -0.24 & 4.38 & 1.11 & 1.43 & <3.0 & field    & 3.28 (3.18) & Iso & \\ 
\end{longtable}
\noindent 
\tablefoottext{a}{Distances are inferred from {\it Gaia}\,DR2 parallaxes with the method described by \mbox{\citet{bailer2018}}.}
\tablefoottext{b}{Visual magnitudes are taken from The Hipparcos and Tycho Catalogues and from SIMBAD.}
\tablefoottext{c}{Spectral types are taken from SIMBAD.}
\tablefoottext{d}{Spectroscopic stellar $T_{\rm eff}$, metallicities $[{\rm Fe/H}]$, surface gravity $\log(g)$, and Values of $v\sin(i)$\ are derived from FEROS spectra with the ZASPE pipeline \mbox{\citep{Brahm2017b}}.}
\tablefoottext{d}{Photometric stellar $T_{\rm eff}$\ and luminosities $L_{\ast}$\ are derived by fitting stellar atmosphere models  \mbox{\citep[PHOENIX;][]{husser2013}} to observed photometry compiled from various data bases (see Section\,\ref{ssec:targets:list}).}
\tablefoottext{f}{Stellar masses $M_{\ast}$\ are derived HRD isochrone fits as described in Section\,\ref{ssec:targets:list}. Typical (median) 1$\sigma$\ uncertainties are 0.035\,M$_{\odot}$.}
\tablefoottext{g}{If the membership  probability is >\,80\% according to  BANYAN\,$\Sigma$\ \citep[for association abbreviations see][]{gagne2018}.}
\tablefoottext{h}{Ages and their uncertainties, where available, are compiled from the literature with references given in the next column.}\newline
\noindent 
\tablefoottext{1}{Spectroscopic binary, both photometrically and spectroscopically derived quantities may be affected and thus inaccurate.}
\tablefoottext{2}{Membership probabilities <\,80\% (64\,--\,70\%).}
\tablefoottext{3}{BPMG member based on kinematic properties only \citep[Banyan\,$\Sigma$: $>90\%$;][]{gagne2018}, but low $v\sin(i))$\ and chemical age \citep{delgado2019} both contradict the young age and BPMG membership.}
\tablebib{
(1)~\mbox{\citet{bell2015}};
(2)~\mbox{\citet{pearce2022}};
(3)~\mbox{\citet{holland2017}};
(4)~\mbox{\citet{stone2018}};
(5)~\mbox{\citet{Vic12}}; 
(6)~\mbox{\citet{holmberg2009}}; 
(7)~\mbox{\citet{dahm2015}};  
(8)~\mbox{\citet{torres2006}}; 
(9)~\mbox{\citet{DH2015}};
(10)~\mbox{\citet{zuc2019}};  
(11)~\mbox{\citet{baffles2020}}; 
(12)~\mbox{\citet{pm2016}}; 
(13)~\mbox{\citet{gaspar2013}}; 
(14)~\mbox{\citet{Casa11}}; 
(15)~\mbox{\citet{MH08}}; 
(16)~\mbox{\citet{chen2014}}; 
(17)~\mbox{\citet{delgado2019}}; 
(18)~\mbox{\citet{tetzlaff2011}};
(Iso)~Isochrone fit, this paper (Section\,\ref{ssec:targets:list}).
}
\end{landscape}

%%%%%%%%%%%%%%%%%%%%%%%%%%% Table 3
\clearpage

\begin{landscape}
\begin{table*}[htb]
\caption{Target list and basic stellar parameters of stars without disc excess}             
\label{tab:RVspy_phys_pars_nodeb}
%\centering       
\begin{tabular}{lllllllllllllllll}
\hline\hline       
Name                   & 
dist.\tablefootmark{a} & 
$V$\tablefootmark{b}   & 
SpT\tablefootmark{c}   & 
$T_{\rm eff}^{\rm sp}$\tablefootmark{d} & 
$T_{\rm eff}^{\rm ph}$\tablefootmark{e} & 
$[{\rm Fe/H}]$\tablefootmark{d}    & 
$\log(g)$\tablefootmark{d} & 
$M_{\ast}$\tablefootmark{f} & 
$L_{\ast}$\tablefootmark{e} & 
$v\sin(i)$\tablefootmark{d} & 
Assoc.\tablefootmark{g}     & 
log($age$)\tablefootmark{h} & 
Ref.\tablefootmark{i}       & 
Notes              \\ 
 & [pc] & [mag] & & [K] & [K] & & & [M$_{\odot}$] & [L$_{\odot}$] & [km\,s$^{-1}$] &  & [Myr] & & \\ 
\hline
HD\,1835   & 21.3  & 6.39 & G3\,V   & 5896 & 5800 &  0.20 & 4.44 & 1.12 & 1.09 & 7.0  & field & 2.70 (2.00) & 5 & \\ 
HD\,5349   & 59.2  & 7.91 & K0\,IV  & 5097 & 5100 &  0.49 & 3.66 & 1.12 & 2.58 & <3.0 & field & 4.00 (3.30) & 14, 17 & \\ 
HD\,15060  & 79.0  & 7.02 & F5\,V   & 6299 & 6280 & -0.14 & 3.73 & 1.66 & 8.10 & 5.8  & field & 3.55 (3.13) & 9 & \\ 
HD\,20759\tablefootmark{1}  & 86.0  & 7.71 & F5\,V   & 6432 & 6240 & -0.48 & 3.80 & 1.36 & 5.00 & 7.2 & field & 3.49 (2.78) & 6, 14 & SB1, PMaG2 (7.6\,$\sigma$) \\ 
HD\,38949  & 44.6  & 7.80 & G1\,V   & 6197 & 6030 &  0.03 & 4.38 & 1.09 & 1.28 & 7.4  & field & 3.40 (2.70) & 14 & \\ 
HD\,76748  & 48.9  & 9.44 & K0\,V   & 5163 & 5130 & -0.32 & 4.48 & 0.85 & 0.40 & <3.0 & field & 3.60 (3.00) & 14 & \\ 
HD\,93932  & 50.8  & 7.53 & G3\,V   & 5990 & 5950 &  0.06 & 4.08 & 1.12 & 2.16 & 4.1  & field & 3.65 (3.00) & 14 & \\ 
HD\,101259 & 67.6  & 6.42 & G7\,V   & 4968 & 4990 & -0.80 & 2.96 & 1.90 & 13.0 & <3.0 & field & 3.40 (2.70) & 14 & \\ 
HD\,102902\tablefootmark{1} & 89.1  & 7.37 & G3\,V   & 5658 & 5390 & -0.68 & 4.81 & 2.00 & 8.61 & 0.5 & field & 3.36 (2.78) & 14 & SB2, PMaG2 (4.8\,$\sigma$) \\ 
HD\,117524 & 129   & 9.89 & G8\,V   & 5421 & 5090 & -0.08 & 3.93 & 1.50 & 1.93 & 32   & LCC   & 1.18 (0.48) & 12 & PMaG2 (20\,$\sigma$) \\ 
HD\,122948 & 44.0  & 8.51 & G5\,E   & 5785 & 5650 & -0.20 & 4.53 & 0.95 & 0.67 & <3.0 & field & 2.70 (2.60) & 18, Iso & \\ 
HD\,131156 & 6.7   & 4.54 & G8\,V   & 5548 & 5280 & -0.15 & 4.48 & 0.90 & 0.65 & 4.4  & field & 2.48 (2.00) & 5 & PMaG2 (25\,$\sigma$) \\ 
HD\,132950 & 29.0  & 9.35 & G3\,V   & 4695 & 4690 & -0.15 & 4.49 & 0.79 & 0.26 & <3.0 & field & 2.78 (2.30) & 5, 16 & \\ 
HD\,138398 & 334.6 & 8.29 & G6\,III & 5008 & 4950 & -0.30 & 3.01 & 2.32 & 60.0 & 3.9  & field & 3.87 (3.30) & 16 & \\ 
HD\,141521\tablefootmark{1} & 131.3 & 10.3 & G8\,IV  & 5381 & 5240 & -0.76 & 3.83 & 1.34 & 2.44 & 8.1 & UCL   & 1.20 (0.30) & 12 & SB2 \\ 
HD\,204277 & 33.1  & 6.72 & F8\,V   & 6393 & 6170 &  0.04 & 4.42 & 1.12 & 1.87 & 6.1 & field & 3.40 (3.18) & 14 & \\ 
HD\,208038 & 23.3  & 8.15 & K2.5\,V & 5050 & 5010 & -0.15 & 4.58 & 0.82 & 0.31 & <3.0 & field & 2.93 (2.85) & Iso & \\ 
HD\,213941 & 31.6  & 7.58 & G8\,V   & 5496 & 5570 & -0.48 & 3.84 & 0.90 & 0.84 & <3.0 & field & 3.70 (3.54) & 14 & PMaG2 (3.3\,$\sigma$) \\ 
\hline                  
\end{tabular}
\end{table*}
\noindent 
\tablefoottext{a}{Distances are inferred from {\it Gaia}\,DR2 parallaxes with the method described by \mbox{\citet{bailer2018}}.}
\tablefoottext{b}{Visual magnitudes are taken from The Hipparcos and Tycho Catalogues and from SIMBAD.}
\tablefoottext{c}{Spectral types are taken from SIMBAD.}
\tablefoottext{d}{Spectroscopic stellar $T_{\rm eff}$, metallicities $[{\rm Fe/H}]$, surface gravity $\log(g)$, and Values of $v\sin(i)$\ are derived from FEROS spectra with the ZASPE pipeline \mbox{\citep{Brahm2017b}}.}
\tablefoottext{d}{Photometric stellar $T_{\rm eff}$\ and luminosities $L_{\ast}$\ are derived by fitting stellar atmosphere models  \mbox{\citep[PHOENIX;][]{husser2013}} to observed photometry compiled from various data bases (see Section\,\ref{ssec:targets:list}).}
\tablefoottext{f}{Stellar masses $M_{\ast}$\ are derived HRD isochrone fits as described in Section\,\ref{ssec:targets:list}. Typical (median) 1$\sigma$\ uncertainties are 0.035\,M$_{\odot}$.}
\tablefoottext{g}{${\rm rms}_{\rm RV}(\tau\,14\,{\rm d}$\ as derived from our high-cadence data.}
\tablefoottext{h}{If the membership  probability is >\,80\% according to  banyan\,$\Sigma$\ \citep[for association abbreviations see][]{gagne2018}.}
\tablefoottext{i}{Ages and their uncertainties, where available, are compiled from the literature with references given in the next column. Age reference numbers refer to the references listed below Table\,\ref{tab:RVspy_phys_pars}).}\newline
\noindent 
\tablefoottext{1}{Spectroscopic binary, both photometrically and spectroscopically derived quantities may be affected and thus inaccurate.}
\end{landscape}

%%%%%%%%%%%%%%%%%%%%%%%%%%% Table 4
\clearpage
\begin{landscape}
\begin{longtable}{llllllclllll}
\caption{\label{tab:periods}
RV variability and periodicities in RV and TESS data}\\
\hline\hline
Name & N(RV)\tablefootmark{a}   & 
${\rm rms}_{\rm RV}$\tablefootmark{b} & 
$M^{\prime}_{\rm 10d}$\tablefootmark{k} &  
P1(RV)\tablefootmark{c}         & 
P2(RV)                          & 
N(TESS)                         & 
P1(TESS)\tablefootmark{d}       & 
P2(TESS)                        & 
P3(TESS)                        & 
$P_{\rm rot}$\tablefootmark{e}  & 
Notes\tablefootmark{g}          \\
 & & [m\,s$^{-1}$] & [M$_{\rm Jup}$] & [d (FAP)] & [d (FAP)] & Sectors & [d (power)] & [d (power)] & [d (power)] & [d] & \\
\hline
\endfirsthead
\multicolumn{7}{c}%
{\tablename\ \thetable\ -- continued from previous page} \\
\hline\hline
Name & N(RV)\tablefootmark{a}   & 
${\rm rms}_{\rm RV}$\tablefootmark{b} &  
$M^{\prime}_{\rm 10d}$\tablefootmark{k} &  
P1(RV)\tablefootmark{c}         & 
P2(RV)                          & 
N(TESS)                         & 
P1(TESS)\tablefootmark{d}       & 
P2(TESS)                        & 
P3(TESS)                        & 
$P_{\rm rot}$\tablefootmark{e}  & 
Notes\tablefootmark{g}          \\
     & & [m\,s$^{-1}$] & [M$_{\rm Jup}$] & [d (FAP)] & [d (FAP)] & Sectors & [d (power)] & [d (power)] & [d (power)] & [d] &\\
\hline
\endhead
\hline \multicolumn{7}{c}{Continued on next page} \\
\endfoot
\hline
\endlastfoot

HD\,105     & 16 & 40.7  & 1.7 & 1.5 ($<$0.001) & 2.9 (0.005) & 2 & 2.95 (0.82) & -- & -- & 2.95$\pm$0.1  & 1, 10 \\ 
HD\,377\tablefootmark{i}     & 10 & 62.9\tablefootmark{f}  & 2.4 & (1.8 (0.3)) & (3.8 (0.5)) & 2 & 1.80 (0.46) & 3.61 (0.28) & -- & 3.6$\pm$0.1 & 2, 11 \\ 
HD\,870\tablefootmark{i}     & 16 & 5.8  & 0.2 &  -- & -- & 4 & 5.7 (0.7) & 9.7 (0.44) & 3.8 (0.08) & 10$\pm$1 & 2 \\ 
HD\,1466\tablefootmark{i}    & 16 & 34.8\tablefootmark{f}  & 1.6 & (2.7 (0.03)) & -- & 3 & 2.3 (0.54) & 2.6 (0.42) & -- & 2.45$\pm$0.2 & 3, 11 \\ 
HD\,1835    & 14 & 25.9  & 1.0 & (3.9 (0.04)) & -- & 1 & 7.8 (0.60) & 3.8 (0.28) & -- & 7.8$\pm$0.2 & 2, 11, 22 \\ 
HD\,3296    & 19 & 7.1   & 0.5 & -- & -- & 4 & 2.4--8 (0.3) & ... & ... & (6$\pm$3) & 4 \\ 
HD\,3670\tablefootmark{i}    & 24 & 27.2  & 2.1 & -- & -- & 1 & 0.73 (0.32) & 1.44 (0.11) & -- & 1.44$\pm$0.1 & 2 \\ 
HD\,5133\tablefootmark{i}    & 26 & 6.8   & 0.2 & -- & -- & 2 & 9.9 (0.16) & 5.4 (0.14) & -- & 9.9$\pm$0.5 & 2 \\ 
HD\,5349    & 12 & 2.4   & 0.2 & -- & -- & 2 & 6.0 ($<$0.01) & 3.0 ($<$0.01) &  & (5$\pm$2) & 5, 22 \\ 
HD\,7570\tablefootmark{i}    & 28 & 6.1   & 0.3 & -- & -- & 3 & 5.5 (0.01) & 11 (0.01) & -- & (11$\pm$2) & 2 \\  
HIP\,6276\tablefootmark{i}   & 18 & 27.3\tablefootmark{f}  & 0.9  & (6.0 (0.5)) & -- & 1 & 5.96 (0.96) & -- & -- & 5.96$\pm$0.1 & 1, 11 \\ 
HD\,10008\tablefootmark{i}   & 17 & 10.4  & 0.3 & -- & -- & 2 & 7.1 (0.27) & 3.5 (0.26) & 12.3 (0.1) & (7 or 12) & 7 \\ 
HD\,13246\tablefootmark{i}   & 21 & 19.3  & 1.4 & -- & -- & 1 & 1.71 (0.63) & -- & -- & 1.71$\pm$0.1 & 1 \\ 
HD\,14082\,B & 16 & 20.8\tablefootmark{f} & 0.7 & -- & -- & 1 & 4.09 (0.22) & 0.79 (0.18) & --  & (4.1$\pm$0.3) & 8 \\ 
HD\,15060   & 12 & 9.3   & 0.6 & -- & -- & 4 & 5.1 (0.04) & 8.6 (0.03) & -- & (8$\pm$3) & 2, 22 \\ 
HD\,16673   & 13 & 3208\tablefootmark{l}  & ... & -- & -- & 2 & 5.0 (0.45) & 12.2 (0.14) & 7.9 (0.11) & (5 or 12) & 7, 20 \\ 
HD\,17925\tablefootmark{i}   & 11 & 18.7\tablefootmark{f}  & 0.6 & -- & -- & 2 & 6.9 (0.52) & 4.9 (0.30) & 11.9 (0.29) & (7 or 12) & 7 \\ 
HD\,19668\tablefootmark{i}   & 14 & 47.6\tablefootmark{f}  & 1.6 & 2.8 ($<$0.001) & -- & 1 & 5.68 (0.59) & 2.69 (0.38) & 4.26 (0.19) & 5.7$\pm$2 & 2, 10 \\ 
HD\,20759   & 17 & 5002\tablefootmark{l}  & ... & -- & -- & 3 & 3--10 (0.01) & ... & ... & (7$\pm$3) & 4, 20, 22 \\ 
HD\,22484\tablefootmark{i}   & 11 & 2.9   & 0.2 & -- & -- & 1 & 4.43 (0.05) & -- & -- & (4.4$\pm$0.5) & 1a \\ 
HD\,23356\tablefootmark{i}   & 11 & 4.6   & 0.1 & -- & -- & 2 & 6.7 (0.06) & 11.9 (0.06) & -- & (12$\pm$1) & 2 \\ 
HD\,23484\tablefootmark{i}   & 44 & 12.7  & 0.4 & -- & -- & 1 & 4.1 (0.22) & 9.0 (0.21) & 6.3 (0.15) & (9$\pm$1) & 8 \\ 
BD\,+23\,551 & 19 & 41.4 & 2.3 & -- & -- & 3 & 2.4 (0.59) & 1.2 (0.20) & -- & 2.4$\pm$0.1 & 2 \\ 
HD\,24649   & 14 & 79.5\tablefootmark{f}  & 3.2 & (1.7 (0.8)) & -- & 2 & 1.7 (0.62) & 1.88 (0.42) & -- & 1.8$\pm$0.2 & 3, 11 \\ 
HD\,27638\,B & 16 & 26190\tablefootmark{l} & ... & -- & -- & 0 & ... & ... & ... & ... & 6, 20 \\ 
HD\,28069   & 22 & 78.5  & 3.4 & -- & -- & 1 & 2.3 (0.78) & -- & -- & 2.3$\pm$0.1 & 1 \\ 
HD\,28447   & 15 & 4.2   & 0.3 & -- & -- & 3 & 2.6--7 (0.01) & ... & ... & (5$\pm$2) & 4 \\ 
HD\,31392   & 16 & 10.1  & 0.3 & -- & -- & 3 & 6.0 (0.19) & 12.6 (0.15) & -- & (12$\pm$2) & 2 \\ 
HD\,33081   & 45 & 2.9   & 0.2 & -- & -- & 2 & 3.5-10.5 (0.03) & ... & ... & (8$\pm$3) & 4 \\ 
SAO\,150676\tablefootmark{i} & 19 & 211   & 9.1 & (1.8 (0.7)) & -- & 3 & 1.75 (0.8) & -- & -- & 1.75$\pm$0.1 & 1, 11 \\ 
HD\,38397\tablefootmark{i}   & 17 & 15.8  & 0.8 & -- & -- & 4 & 2.27 (0.84) & -- & -- & 2.27$\pm$0.1 & 1 \\ 
HD\,38949   & 14 & 14.9\tablefootmark{f}  & 0.6 & (3.8 (0.7)) & (9 (0.5)) & 3 & 3.8 (0.46) & 7.5 (0.38) & -- & 7.5$\pm$0.3 & 2, 11, 22 \\ 
HD\,40136\tablefootmark{i}   & 13 & 18.6\tablefootmark{f}  & 1.1 & -- & -- & 2 & 0.96 (0.11) & -- & -- & 0.96$\pm$0.1 & 1 \\ 
HD\,43989\tablefootmark{i}   & 17 & 143\tablefootmark{f}   & 6.6 & -- & -- & 2 & 1.36 (0.67) & -- & -- & 1.36$\pm$0.1 & 1 \\ 
HD\,48370\tablefootmark{i}   & 20 & 47.3\tablefootmark{f}  & 1.6 & -- & -- & 1 & 5.18 (0.87) & -- & -- & 5.18$\pm$0.1 & 1 \\ 
HD\,50571   & 37 & 188\tablefootmark{f}   & 9.1 & -- & -- & 24 & 1.69 (0.12) & ... & ... & 1.69$\pm$0.1 & 1 \\ 
HD\,53143   & 21 & 15.4\tablefootmark{f}  & 0.5 & (4.8 (0.1)) & -- & 25 & 9.6 (0.6) & 4.9 (0.27) & -- & 9.6$\pm$0.3 & 2, 11 \\ 
HD\,57703   & 20 & 30.4  & 2.1 & -- & -- & 3 & 1.69 (0.13) & 1.9 (0.08) & -- & (1.8$\pm$0.2) & 3 \\ 
HD\,59659   & 44 & 48.4  & 3.2 & -- & -- & 5 & 1.64 (0.73) & -- & -- & 1.64$\pm$0.1 & 1 \\ 
HD\,59967\tablefootmark{i}   & 28 & 14.4  & 0.5 & -- & -- & 3 & 5.2 (0.58) & 7.9 (0.17) & -- & (6$\pm$1.5) & 3 \\ 
HD\,72687   & 21 & 47.9\tablefootmark{f}  & 1.6 & 3.8 (0.05) & -- & 3 & 3.8 (0.6) & 5.1 (0.27) & 1.9 (0.15) & (3.8 or 5.1) & 7, 10 \\ 
HD\,76151\tablefootmark{i}   & 34 & 4.6   & 0.2 & -- & -- & 2 & 3.2--14 (0.15) & ... & ... & (8$\pm$3) & 4 \\ 
HD\,76748   & 22 & 7.7   & 0.3 & -- & -- & 2 & 1.4--3 ($<$0.01) & -- & -- & ... & 5, 22 \\ 
HD\,76653   & 16 & 16.2\tablefootmark{f}  & 0.8 & 2.0 ($<$0.001) &  & 4 & 2.1 (0.5) & -- & -- & 2.1$\pm$0.2 & 1, 10 \\ 
CD\,-49\,3972 & 35 & 9.5 & 0.5 & -- & -- & 4 & 3.1--11.5 (0.1) & ... & ... & (7$\pm$3) & 4 \\ 
HD\,84075   & 16 & 92.9\tablefootmark{f}  & 3.9 & (2.4 (0.05)) &  & 5 & 2.44 (0.76) & -- & -- & 2.44$\pm$0.1 & 1, 11 \\ 
HD\,90905\tablefootmark{i}   & 19 & 9.7   & 0.4 & -- & -- & 3 & 2.55 (0.66) & -- & -- & 2.55$\pm$0.1 & 1 \\ 
HD\,92945\tablefootmark{i}   & 14 & 27.5\tablefootmark{f}  & 0.9 & (3.7 (0.4)) & (7.6 (0.5)) & 2 & 3.4 (0.45) & 7.4 (0.38) & 13.5 (0.21) & (7.4 or 13.5) & 7, 11 \\ 
HD\,93932   & 40 & 6.3   & 0.3 & -- & -- & 2 & 4.3 ($<$0.01) & 5.8 ($<$0.01) & -- & (5$\pm$1) & 3, 22 \\ 
HD\,101259  & 15 & 9.6   & 0.7 & -- & -- & 2 & 0.75--5.5 (0.02) & ... & ... & ... & 5, 22  \\ 
HD\,102458\tablefootmark{i}  & 14 & 504\tablefootmark{f}   & 25 & 2.0 ($<$0.001) & -- & 2 & 2.0 (0.53) & 0.85 (0.2) & -- & 2.0$\pm$0.1 & 2, 10 \\ 
HD\,102902  & 6  & 911\tablefootmark{l}   & ... &  -- & -- & 3 & 0.7 (0.13) & 1.5 (0.07) & -- & 1.5$\pm$0.2 & 2, 20, 22 \\ 
HD\,104231\tablefootmark{i}  & 27 & 26.0  & 2.1 & -- & -- & 3 & 0.55 (0.11) & 1.04 (0.12) & -- & 1.0$\pm$0.2 & 2 \\ 
HD\,105912  & 27 & 145\tablefootmark{f}   & 7.4 & -- & -- & 1 & 0.7 (0.13) & 1.5 (0.08) & -- & 1.5$\pm$0.2 & 2 \\ 
MML\,8\tablefootmark{i}      & 8  & 767\tablefootmark{f}   & 34 & 2.4 (0.01) & -- & 2 & 2.41 (0.97) & -- & -- & 2.4$\pm$0.1 & 1, 10, 21 \\ 
HD\,107146\tablefootmark{i}  & 13 & 33.1\tablefootmark{f}  & 1.2 & 3.6 (0.001) & -- & 0 & ... & ... & ... & 3.6$\pm$0.3 & 6, 12, 21 \\ 
HD\,107649\tablefootmark{i}  & 29 & 577   & 30 & -- & -- & 3 & 0.94 (0.58) & -- & -- & 0.94$\pm$0.1 & 1 \\ 
HD\,108857  & 5  & 4851\tablefootmark{l}  & ... & -- & -- & 3 & 11.4 ($<$0.01) & -- & -- & (11$\pm$2) & 1a, 20 \\ 
HD\,109832\tablefootmark{i}  & 59 & 804   & 45 & 1.3 (0.001) & -- & 2 & 0.79 (0.30) & 1.2 (0.26) & -- & 1.2$\pm$0.2 & 2, 21 \\ 
HD\,111520\tablefootmark{i}  & 32 & 219\tablefootmark{f}   & 10.9 & -- & -- & 3 & 4.7 (0.27) & 1.5 (0.1) & -- & (5$\pm$1) & 8 \\ 
HD\,111631\tablefootmark{i}  & 17 & 7.9   & 0.2 & -- & -- & 0 & ... & ... & ... & ... & 6 \\ 
HD\,114082\tablefootmark{i}  & 26 & 75.6\tablefootmark{f}  & 5.1 & -- & -- & 1 & 1.5--4.2 (0.05) & 8 (0.02) & -- & (3.5$\pm$2) & 7 \\ 
HD\,115820\tablefootmark{i}  & 16 & 6282\tablefootmark{h}  & 368 & -- & -- & 3 & -- & -- & -- & ... & 5, 21  \\ 
HD\,117214  & 20 & 87.8\tablefootmark{f}  & 5.5 & -- & -- & 2 & 2.7 (0.06) & 0.55 (0.04) & -- & 2.7$\pm$0.5 & 8 \\ 
HD\,117524  & 19 & 617\tablefootmark{f}   & 33 & 1.8 (0.005) & -- & 1 & 1.75 (0.92) & -- & -- & 1.75$\pm$0.1 & 1, 10, 21, 22 \\ 
MML\,36\tablefootmark{i}     & 21 & 150\tablefootmark{f}   & 6.5 & 4.5 (0.005) & -- & 1 & 4.73 (0.94) & -- & -- & 4.73$\pm$0.1 & 1, 10 \\ 
HD\,118972  & 22 & 20.3  & 0.6 & -- & -- & 2 & 4.7 (0.6) & 9 (0.25) & -- & 9$\pm$1 & 2 \\ 
CD\,-29\,10609 & 40 & 37.7\tablefootmark{f} & 1.3 & -- & -- & 0 & ... & ... & ... & ... & 6 \\ 
HD\,122948  & 24 & 8.0   & 0.3 & -- & -- & 0 & ... & ... & ... & ... & 6, 22 \\ 
HD\,125451\tablefootmark{i}  & 16 & 96.1  & 5.0 & -- & -- & 0 & ... & ... & ... & ... & 6 \\ 
MML\,43     & 6  & 2659\tablefootmark{l}  & ... & -- & -- & 2 & 4.2 (0.91) & -- & -- & 4.2$\pm$0.1 & 1, 20 \\ 
HD\,129590\tablefootmark{i}  & 6  & 11426\tablefootmark{l} & ... & -- & -- & 2 & 4.5 (0.7) & -- & -- & 4.5$\pm$0.1 & 1, 20 \\ 
HD\,131156  & 18 & 13.2\tablefootmark{f}  & 0.4 & (3.1 (0.05)) & -- & 0 & ... & ... & ... & (3.1$\pm$0.5) & 6, 13, 22 \\ 
HD\,132950  & 34 & 8.3   & 0.3 & -- & -- & 0 & ... & ... & ... & ... & 6, 22 \\ 
HD\,134910  & 22 & 49.7  & 2.3 & -- & -- & 1 & 3.4 (0.55) & -- & -- & 3.4$\pm$0.1 & 1 \\ 
HD\,135953\tablefootmark{i}  & 20 & 68.0  & 3.8 & -- & -- & 1 & 1.9 (0.41) & 0.94 (0.10) & -- & 1.9$\pm$0.1 & 2 \\ 
HD\,138398  & 19 & 11.0  & 1.0 & -- & -- & 0 & ... & ... & ... & ... & 6, 22 \\ 
HD\,139664\tablefootmark{i}  & 23 & 765   & 34 & -- & -- & 2 & 0.88 (0.23) & 1.04 (0.1) & -- & 0.95$\pm$0.1 & 3, 21 \\ 
HD\,141011  & 28 & 1712\tablefootmark{f}  & 82 & -- & -- & 2 & 0.96 (0.15) & -- & -- & 0.96$\pm$0.1 & 1, 21 \\ 
HD\,141521\tablefootmark{i}  & 22 & 127123\tablefootmark{l} & ... & (7.9 (0.2)) & -- & 1 & 7.1 (0.95) & 13.8 (0.35) & 5.0 (0.25) & 7.1$\pm$0.2 & 1, 11, 20, 22 \\ 
HD\,143811  & 15 & 11469\tablefootmark{l} & ... & -- & -- & 1 & 4.3 (0.028) & 6.4 (0.022) & 3.4 (0.019) & (5$\pm$2) & 7, 20  \\ 
HD\,145229  & 25 & 12.5  & 0.5 & -- & -- & 0 & ... & ... & ... & ... & 6, 20 \\ 
HD\,145560  & 30 & 152   & 7.8 & -- & -- & 2 & 1.4 (0.54) & 0.7 (0.1) & -- & 1.4$\pm$0.1 & 2 \\ 
HD\,145972\tablefootmark{i}  & 19 & 1345  & 75 & (1.5 (0.8)) & -- & 2 & 1.41 (0.41) & 1.54 (0.38) & -- & 1.48$\pm$0.1 & 3, 11, 21 \\ 
HD\,146181\tablefootmark{i}  & 20 & 73.2  & 4.4 & -- & -- & 1 & 1.95 (0.26) & 0.99 (0.14) & -- & 1.95$\pm$0.1 & 2 \\ 
HD\,147594  & 23 & 198   & 11 & (2.8 (0.9)) & -- & 2 & 2.98 (0.8) & -- & -- & 2.98$\pm$0.1 & 1, 11 \\ 
HD\,166348\tablefootmark{i}  & 26 & 5.2   & 0.2 & -- & -- & 1 & 11.6 (0.61) & 6.5 (0.17) & -- & 11.6$\pm$0.1 & 2 \\ 
HD\,170773  & 16 & 198   & 9.5 & -- & -- & 1 & 0.8 (0.09) & -- & -- & (0.8$\pm$0.2) & 1 \\ 
HD\,180134  & 32 & 8.6   & 0.5 & -- & -- & 2 & 4.5 (0.005) & 12 (0.004) & 2.3 (0.003) & ... & 9 \\ 
HD\,181327\tablefootmark{i}  & 16 & 6.0\tablefootmark{f}   & 0.5 & -- & -- & 2 & 1.57 (0.3) & 0.75 (0.1) & -- & 1.6$\pm$0.2 & 2 \\ 
HD\,183216  & 31 & 15.7  & 0.7 & -- & -- & 2 & 9.5 (0.3) & 5.1 (0.2) & -- & (10$\pm$2) & 2 \\ 
HD\,187897  & 32 & 10.9  & 0.4 & -- & -- & 0 & ... & ... & ... & ... & 6 \\ 
HD\,190470  & 30 & 8.9\tablefootmark{f}   & 0.3 & -- & -- & 2 & 11.3 (0.12) & 5.9 (0.08) & -- & 11.3$\pm$0.5 & 2 \\ 
HD\,191089\tablefootmark{i}  & 15 & 25.6  & 1.8 & -- & -- & 0 & ... & ... & ... & ... & 6 \\ 
HD\,191849\tablefootmark{i}  & 23 & 8.7   & 0.2 & -- & -- & 0 & ... & ... & ... & ... & 6 \\ 
HD\,199260\tablefootmark{i}  & 23 & 31.9\tablefootmark{f}  & 1.3 & 4.1 (0.01) & -- & 2 & 4.08 (0.85) & -- & -- & 4.08$\pm$0.1 & 1, 10 \\ 
HD\,201219  & 29 & 14.4\tablefootmark{f}  & 0.5 & -- & -- & 0 & ... & ... & ... & ... & 6 \\ 
HD\,202917  & 7  & 178\tablefootmark{f}   & 6.3 & (2.8 (0.3)) & -- & 2 & 3.5 (0.8) & -- & -- & 3.5$\pm$0.1 & 1, 11 \\ 
HD\,204277  & 7  & 12.8  & 0.5 & -- & -- & 0 & ... & ... & ... & ... & 6, 22 \\ 
HD\,206893  & 23 & 84.6  & 4.1 & -- & -- & 0 & ... & ... & ... & ... & 6 \\ 
HD\,208038  & 5  & 8.6   & 0.3 & -- & -- & 0 & ... & ... & ... & ... & 6, 22 \\ 
HD\,209253\tablefootmark{i}  & 16 & 33.9\tablefootmark{f}  & 1.4 & (3.0 (0.6)) & -- & 2 & 2.95 (0.60) & 1.5 (0.18) & -- & 2.95$\pm$0.1 & 2, 11 \\ 
HD\,212695\tablefootmark{i}  & 21 & 15.5  & 1.1 & -- & -- & 1 & 1.5 (0.18) & 2.8 (0.1) & -- & 2.8$\pm$0.2 & 2 \\ 
HD\,213941  & 14 & 5.2   & 0.2 & -- & -- & 0 & ... & ... & ... & ... & 6, 22 \\ 
CPD\,-72\,2713\tablefootmark{i} & 21 & 70.2 & 2.0 & 4.4 ($<$0.001) & -- & 4 & 4.45 (0.95) & -- & -- & 4.45$\pm$0.1 & 1, 10 \\ 
HD\,218340\tablefootmark{i}  & 21 & 8.7   & 0.4 & -- & -- & 1 & 7.7 (0.12) & -- & -- & 7.7$\pm$0.1 & 1 \\ 
HD\,218511\tablefootmark{i}  & 41 & 7.2   & 0.2 & -- & -- & 3 & 8.8 (0.22) & 5.8 (0.22) & -- & (9$\pm$1) & 2 \\ 
HD\,219498\tablefootmark{i}  & 22 & 32.3\tablefootmark{f}  & 1.2 & 2.8 (0.005) & 1.5 (0.008) & 0 & ... & ... & ... & 2.8$\pm$0.3 & 6, 12\\ 
HD\,219482\tablefootmark{i}  & 16 & 14.9\tablefootmark{f}  & 0.6 & 2.2 ($<$0.001) & -- & 2 & 2.1 (0.5) & -- & -- & 2.1$\pm$0.1 & 1, 10 \\ 
HD\,223340\tablefootmark{i}  & 13 & 24.5\tablefootmark{f}  & 1.2 & (6.3 (0.1)) & -- & 0 & ... & ... & ... & (6$\pm$1) & 6, 13 \\ 
CD\,-44\,15399 & 21 & 22.5 & 0.9 & -- & -- & 1 & 5.1 ($<$0.01) & -- & -- & (5.1$\pm$0.3) & 1a \\ 
\hline     
\end{longtable}
\noindent 
\looseness=-5
\tablefoottext{a}{Number of usable spectra in high-cadence RV times series.}
\tablefoottext{b}{${\rm rms}_{\rm RV}(\tau\,14\,{\rm d})$\ as derived from our high-cadence data.}
\tablefoottext{c}{Periodicities in the RV data and their false alarm probability (FAP). Non-significant periodicities are listed in brackets.}
\tablefoottext{d}{Periodicities in photometric TESS data and power in the GLS periodogram. All listed periods are significant.}
\tablefoottext{e}{Most likely stellar rotation period $P_{\rm rot}$. Uncertain guesses are listed in brackets.}
\tablefoottext{f}{Strong anti-correlation between BS and RV with $r_P\le-0.6$\ suggests RV variability is caused by rotational modulation due to starspots.}
\tablefoottext{g}{{bf Notes regarding $P_{\rm rot}$\ and selection for longer-period survey}: 
(1) Single dominant period in the photometric TESS data.
(1a) Single dominant period in the photometric TESS data, but only marginally significant.
(2) An approximate 2:1 ratio of two dominant photometric periods and alternating strengths and shapes in the light curve is indicative of two dominant spot groups and the longer of the two periods representing $P_{\rm rot}$.
(3) Two periods close together in the photometric TESS data, we adopt the mean.
(4) Multiple frequencies in the photometric TESS data, shifting between sectors, but periodogram power is limited to the given range of periods.
(5) No significant periodicity in the photometric TESS data.
(6) No TESS data available (yet).
(7) Not clear which of the periods represents $P_{\rm rot}$.
(8) The longest period most likely represents $P_{\rm rot}$.
(9) Marginally significant periods, shifting between sectors, not clear indication of $P_{\rm rot}$.
(10) Photometric period(s) clearly detected in RV.
(11) Photometric period(s) marginally evident in RV.
(12) Strong BS(RV) anti-correlation ($r_P\leq-0.8$) indicates that the significant RV period is caused by rotational modulation and most likley represents $P_{\rm rot}$.
(13) Strong BS(RV) anti-correlation ($r_P\leq-0.7$) suggests that the marginally significant RV period is caused by rotational modulation and may represent $P_{\rm rot}$.
(20) SB.
(21) $M^{\prime}_{\rm 1yr}>80\,{\rm M}_{\rm Jup}$.
(22) No significant debris disc signal.}
\tablefoottext{h}{HD\,115820 turns out to be a $\delta$\,Scuti varaible. The GLS periodigram of the TESS photometry shows several g-mode pulsation periods between 42 and 68\,min, which explains the large RV scatter.}
\tablefoottext{i}{Observed with NaCo under the ISPY programme.}
\tablefoottext{k}{Mass detection limits for $P=10$\,d corresponding to 3$\times{\rm rms}_{\rm RV}(\tau\,14\,{\rm d})$. The corresponding mass-detection limit for $P=1$\,yr is $M^{\prime}_{\rm 1yr} = M^{\prime}_{\rm 10d}\times(365/10)^(1/3)$.}
\tablefoottext{l}{Spectroscopic binary (Sect.\,\ref{ssec:res:sb}).}

\end{landscape}

%%%%%%%%%%%%%%%%%%%%%%%%%%%%%%%%

\end{document}